\documentclass[sigconf,balance=false]{acmart}

\usepackage{popets}
\usepackage{epsfig,endnotes}
\usepackage{pifont}
\usepackage{tikz}
\usepackage{multirow}
\usepackage{colortbl}
\usepackage{rotating}
\usepackage{pgfplots,pgfplotstable}
\usepackage{scalefnt}
\usepackage{algorithm}
\usepackage{enumitem}
\usepackage{caption}
\usepackage{subcaption}
\usepackage{tabularx}
\pgfplotsset{compat=1.17} 
\usepackage{graphicx}
\usepackage{booktabs}
\usepackage{afterpage}
\usepackage{algpseudocode}
\usepackage{siunitx}
\usepackage{float}

\makeatletter
\usepgfplotslibrary{groupplots}
\setcopyright{popets}
\copyrightyear{YYYY}
\usepackage{xcolor}
\usepackage[normalem]{ulem} 

\newif\ifshowdiff
\showdifffalse  

\ifshowdiff
  \newenvironment{addedenv}{\color{blue}}{}
  \newenvironment{deletedenv}{\color{red}\sout}{}
  \newcommand{\added}[1]{\textcolor{blue}{#1}}
  \newcommand{\deleted}[1]{\textcolor{red}{\sout{#1}}}
\else
  \newenvironment{addedenv}{}{}
  
  \newcommand{\added}[1]{#1} 
  \newcommand{\deleted}[1]{}  
\fi

\acmYear{YYYY}
\acmVolume{YYYY}
\acmNumber{X}
\acmDOI{XXXXXXX.XXXXXXX}
\acmISBN{}
\acmConference{Proceedings on Privacy Enhancing Technologies}
\makeatletter
\renewcommand{\subsubsection}{%
  \@startsection{subsubsection}{3}{\z@}%
  {3.25ex \@plus 1ex \@minus .2ex}%
  {1.5ex \@plus .2ex}%
  {\normalfont\normalsize\bfseries}%
}%
\makeatother
\begin{document}

\title{GNNBleed: Inference Attacks to Unveil Private Edges in Graphs with Realistic Access to GNN Models}

\author{Zeyu Song}
\affiliation{%
  \institution{Penn State University}
  \city{} 
  \state{} 
  \country{} 
}
\email{zysong@psu.edu}

\author{Ehsanul Kabir}
\affiliation{%
  \institution{Penn State University}
  \city{} 
  \state{} 
  \country{} 
}
\email{ekabir@psu.edu}

\author{Shagufta Mehnaz}
\affiliation{%
  \institution{Penn State University}
  \city{} 
  \state{} 
  \country{} 
}
\email{smehnaz@psu.edu}

\begin{abstract}
Graph Neural Networks (GNNs) have become indispensable tools for learning from graph-structured data, catering to various applications such as social network analysis and fraud detection for financial services. 
At the heart of these networks are the edges, which are crucial in guiding GNN models' predictions. In many scenarios, these edges represent sensitive information, such as personal associations or financial dealings, which require privacy assurance. 
However, their contributions to GNN model predictions may, in turn, be exploited by the adversary to compromise their privacy. 
Motivated by these conflicting requirements, this paper investigates edge privacy in contexts where adversaries possess only black-box access to the target GNN model, restricted further by access controls, preventing direct insights into arbitrary node outputs.
Moreover, we are the first to extensively examine situations where the target graph continuously evolves—a common trait of many real-world graphs. 
In this setting, we present a range of attacks that leverage the message-passing mechanism of GNNs. 
We evaluated the effectiveness of our attacks using nine real-world datasets, encompassing both static and dynamic graphs, across four different GNN architectures. 
The results demonstrate that our attack outperforms existing methods across various GNN architectures, consistently achieving an F1 score of at least 0.8 in static scenarios. 
Furthermore, our attack retains robustness in dynamic graph scenarios, maintaining F1 scores up to 0.8, unlike previous methods that only achieve F1 scores around 0.2.

\end{abstract}
\maketitle
 

\section{Introduction}
Graph Neural Networks (GNNs) are deep learning models specifically designed to handle graph-structured data \cite{model:gat, model:gcn, model:gin, model:sage}. Unlike traditional neural networks that assume data is structured in a grid-like manner (e.g., images) or in sequences (e.g., text), GNNs are designed for data where entities (nodes) and their relationships (edges) are explicitly represented as a graph. 

Edges in a graph often denote sensitive information, such as 
a patient's condition in a healthcare network or a transaction in a financial network \cite{fan2019graph, Rao_2021, graph:usage:medical:kazi2023ia, graph:usage:medical:ahmedt2021graph, graph:usage:medical:credit:wang2021temporal, graph:usage:medical:credit:sukharev2020ewsgcn,graph:usage:anomoly:xu2021towards, mao2022medgcn}.
Unveiling these connections without consent infringes upon the privacy of the entities involved. 
For instance, consider a healthcare coordination network \cite{mao2022medgcn, wu2021leveraging} maintained by a model service provider, where nodes represent patients and healthcare providers. 
Patients may establish links/edges with various providers, support groups, or other patients, where these connections indicate the types of care received, the patient’s conditions, or their support networks. Exposure of such links/edges could lead to the leakage of sensitive health information, potentially violating privacy regulations such as HIPAA \cite{HIPAA1996} in the United States.
Hence, understanding edge inference attacks is crucial to comprehending the vulnerabilities inherent within GNNs, thereby devising mechanisms to preserve privacy by safeguarding these relationships.

Previous research \cite{Inference_Attacks:link_steal:he2020stealing, Inference_Attacks:link_steal:wu2021linkteller, devil_in_disguise} highlights edge privacy vulnerabilities in GNN models but suffers from three key limitations:

     \textbullet\ Dynamic Graph Assumptions: Previous methods assume adversaries can freely perturb the graph, yet their attacks falter when the graph evolves due to external changes, such as updates from other entities. This restricts their applicability in realistic, collaborative, or rapidly evolving graph environments \cite{Inference_Attacks:link_steal:wu2021linkteller, devil_in_disguise}.

     \textbullet\ Architectural Dependency: Previous methods often show effectiveness on specific GNN architectures, but struggle to generalize across diverse architectures, limiting their robustness \cite{Inference_Attacks:link_steal:wu2021linkteller, devil_in_disguise, Inference_Attacks:link_steal:he2020stealing}.

    \textbullet\ Granularity: Previous works are limited in their focus, concentrating only on distinguishing between connected and unconnected node pairs, without addressing the more precise and practical task of inferring specific edges \cite{Inference_Attacks:link_steal:wu2021linkteller, devil_in_disguise, Inference_Attacks:link_steal:he2020stealing, he2024maui}.

\begin{figure}[h!]
    \centering  
    \includegraphics[width=\columnwidth]{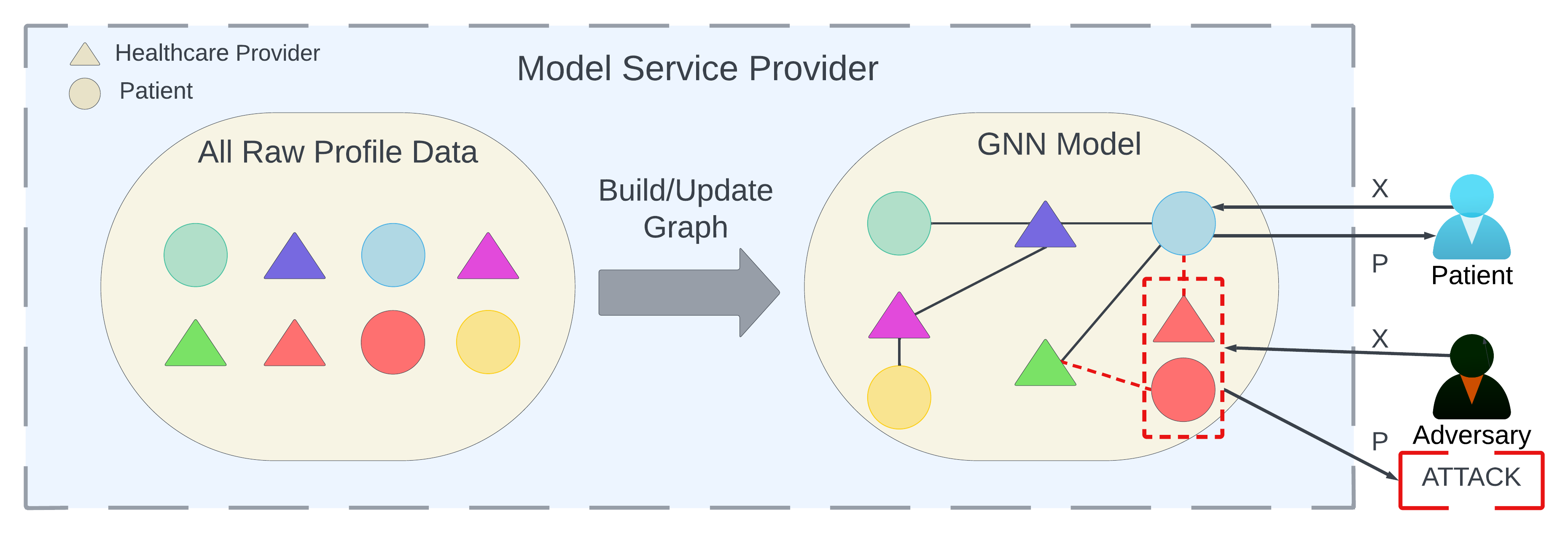}
    \caption{Our attack scenario: an adversary accesses some nodes in the graph and depends on the model's predictions for these nodes to deduce private information about other nodes to which they do not have access.}
    \label{fig:user_model_inteteraction}
\end{figure}

To address the limitations of previous works, we propose a realistic threat model to explore edge privacy attacks in dynamic graphs. Specifically, we focus on scenarios where GNNs predict outcomes on evolving graphs, such as social or healthcare networks \cite{robinson2024relbench, borisyuk2024lignn, mao2022medgcn, wu2021leveraging}. In this setup, a service provider continuously updates the graph based on user behavior (e.g., profile updates or new connections) and provides predictions while restricting users' access to sensitive data.

As depicted in Figure \ref{fig:user_model_inteteraction}, an adversary, such as a healthcare provider, might aim to infer connections between their patient and other providers to uncover sensitive information. Using their existing connections, the adversary could introduce new nodes (e.g., a new patient) to interact with the target provider, such as by scheduling an appointment. By analyzing the model's predictions for these auxiliary nodes, the adversary can deduce hidden relationships within the graph.

\textbf{Problem and Scope.} 
This paper examines privacy risks associated with the exposure of private edges in graphs during the inference phase. 
Specifically, we focus on inductive GNNs, which excel in generalizing to unseen nodes and structures. These models are particularly well-suited for real-world applications where graphs evolve dynamically, with new nodes and edges emerging over time \cite{borisyuk2024lignn, robinson2024relbench, inductive:usage:amazon}.
Our work marks the first exploration of edge inference attacks targeting inductive GNNs in dynamic graph settings, shedding light on critical privacy vulnerabilities and advancing the understanding of these risks.

\textbf{Challenges.}
Unlike prior work~\cite{Inference_Attacks:link_steal:he2020stealing, Inference_Attacks:link_steal:wu2021linkteller, wu2024link, he2024maui}, in our settings, the adversary operates under constrained circumstances, lacking the privilege to query the model regarding the targeted node pairs (adversary aims to infer the presence of edge between a targeted node pair). 
This limitation necessitates a reliance on indirect information, escalating the complexity of any potential attack. 
This complexity is further exacerbated by the diverse message aggregation methods employed by various GNN models, complicating the task of generalizing the attack across different architectures.
Moreover, the dynamic nature of graphs introduces another layer of challenges. 
This continuous evolution hinders an adversary's ability to analyze and leverage the model output.

\textbf{Key Intuition and New Attacks.} 
Our approach is rooted in the message aggregation process of GNNs, where information flows between nodes through their connections. This aggregation process inherently encodes information about the graph’s structure, as messages are passed along edges and influence predictions based on the paths connecting nodes. By observing how messages are exchanged between two nodes, an adversary can estimate the underlying path and infer structural information, such as the distance between them.
Specifically, the adversary manipulates auxiliary nodes (nodes under their control) and observes the resulting changes in the predictions of target nodes. These observations capture the flow of information along the graph's edges, enabling the adversary to infer the local topology. 
We develop two attack strategies based on this principle: Magnitude-based attacks and Direction-based attacks. Magnitude-based attacks estimate the distance between nodes by measuring the extent of change in output probabilities. Direction-based attacks focus on the patterns of change, leveraging the consistent way messages propagate to nearby nodes.

\textbf{Evaluation.} 
We conduct a comprehensive evaluation of our attacks across four widely-recognized inductive GNN models, namely, Graph Attention Network (GAT) \cite{model:gat}, Graph Convolutional Network (GCN) \cite{model:gcn}, Graph Isomorphism Network (GIN) \cite{model:gin}, and GraphSAGE \cite{model:sage}. 
Our assessment spans seven static graph datasets, including Flickr \cite{dataset:flickr}, LastFM \cite{dataset:lastfm}, and Twitch-\{DE, EN, FR, PT, RU\} \cite{dataset:twitch}, as well as two dynamic graph datasets, tgbl-flight \cite{huang2023temporal} and Dgraph-fin \cite{huang2022dgraph}. 
Our extensive testing demonstrates that our attacks outperform all previous methods in the context of static graphs, for which those methods \cite{Inference_Attacks:link_steal:he2020stealing, Inference_Attacks:link_steal:wu2021linkteller, devil_in_disguise} were originally designed. 
Furthermore, our approach maintains robustness in dynamic scenarios where previous methods fail.
Additionally, when tested against state-of-the-art differential privacy-based defenses \cite{Inference_Attacks:link_steal:wu2021linkteller}, our evaluations reveal that privacy risks remain significant when the differential privacy budget is set at a level that keeps the model usable.

\textbf{Contributions.} In this paper, our primary objective is to explore the potential risks to edge privacy posed by inductive GNNs. In summary, we make the following contributions.

  \ding{114}  We introduce a novel, practical threat model and present several query-based edge inference attacks specifically tailored for inductive GNNs. Our approach determines whether two nodes are connected without requiring direct or query access to the target nodes.

 \ding{114} We showcase consistent effectiveness across a diverse range of inductive GNN architectures with different message aggregation mechanisms.
  We analyzed the shortcomings of previous works and conducted extensive experiments on four popular inductive GNNs. Our results show that our approach consistently outperforms three state-of-the-art attack methods.

  \ding{114} We are the first to explore edge inference attacks in dynamic graphs, specifically focusing on scenarios where the graph keeps changing during the attack. We conducted experiments using two real-life dynamic graph datasets and two synthetic dynamic graphs. The findings demonstrate that our method remains effective, even when all previous attack approaches fail.

  \ding{114}  We subject our attacks to cutting-edge differential privacy-based defenses, demonstrating the intrinsic challenge of concurrently achieving both privacy and utility against our attacks.

\section{Preliminaries}

\subsection{Graph Neural Network (GNN)}
Graph Neural Networks (GNNs), a class of deep learning models, can be utilized for various tasks within the graph data domain, primarily aligning with three categories: node-level tasks (e.g., node classification \cite{model:gcn, model:sage, model:gin, model:gat}), graph-level tasks (e.g., graph classification \cite{model:gin, lee2019selfattention, ying2019hierarchical}), and edge-level tasks (e.g., link prediction \cite{zhang2018link}).
The core ideology of most GNNs revolves around the neighborhood aggregation strategy. 
This process allows them to generate representations of nodes that capture not only their features but also the features of their surrounding nodes. 
Let $G=(V, E)$ be a graph where $V$ represents the set of nodes and $E$ represents the set of edges. 
We denote the 1-hop neighbors of a node $v \in V$ as $\mathcal{N}(v)$. When generating representation for $v$, after performing $l$ iterations of message-passing and aggregation, node $v$'s representation, denoted as $h^{l}_{v}$, captures structural information within its $l$-hop  neighborhood. GNNs typically consist of multiple graph convolution layers, each of which can be described as:
$${\scriptstyle h_{v}^{l} = f (h^{(l-1)}_{v},AGG(h^{(l-1)}_{v}, h^{(l-1)}_{u}))| u \in \mathcal{N}(v)}$$
where $f(.)$ is a function that obtains a new representation $h_{v}^{l}$ for node $v$ based on its current representation $h^{l-1}_{v}$ and the aggregated features from its 1-hop neighbors, i.e., $\mathcal{N}(v)$. The function $AGG(.)$ represents the specific aggregation mechanism. Note that the number of layers in a GNN model determines how far it can reach through the network. 
 
GNNs can be trained under two distinct settings, namely, transductive \cite{model:gcn, wu2020comprehensive, zhou2020graph, zhang2020deep} and inductive \cite{model:sage, model:gat, model:gin, dataset:flickr, xu2018representation, rong2019dropedge}. In the transductive setting, the GNN learns from labeled and unlabeled nodes within a fixed graph during training, subsequently predicting the labels for the unlabeled nodes post-training. A significant limitation of this approach is the necessity for retraining the models when new nodes are added to the graph. 
Contrarily, the inductive setting has garnered more popularity due to its ability to generalize the learned GNN model to previously unseen nodes, making it well-suited for real-life graphs that continually evolve \cite{gnn:survey:xie2022self, bilot2023graph, anil2023inductive, model:sage}. 
In light of this, our research centers exclusively on inductive GNNs, exploring four widely employed architectures: GAT \cite{model:gat}, GCN \cite{model:gcn}, GIN \cite{model:gin}, and GraphSAGE \cite{model:sage}.

\textbf{Graph Convolutional Network (GCN).} 
GCN \cite{model:gcn} updates each node's features by aggregating features from its neighbors and itself. The layer-wise propagation rule for a GCN can be written as:
$${\scriptstyle h_{v}^{l} = \sigma(W^{(l-1)}\cdot MeanAgg(\{ h_{u}^{(l-1)}: u \in \mathcal{N}(v) \cup \{v\} \}))}$$
where $W$ is a learnable weight matrix, $\sigma$ is a non-linear activation function, and $MeanAgg$ calculates the mean of the feature vectors of node $v$ and its 1-hop neighbors $\mathcal{N}(v)$.

\textbf{Graph Attention Network (GAT).}
GAT \cite{model:gat} computes attention coefficients that indicate the importance of each neighbor's information. The attention-based aggregation in GAT is represented as follows where $W$ is a trainable weight matrix, $\sigma$ is an activation function, and $\alpha_{vu}$ is the attention coefficient indicating how much attention node $v$ should pay to node $u$ during aggregation:
$$ {\scriptstyle h_{v}^{l} = \sigma(\sum_{u \in \mathcal{N}(v)}\alpha^{(l-1)}_{vu} W^{(l-1)} h^{(l-1)}_{u})}$$

\textbf{Graph Sampling and Aggregation (GraphSAGE).} 
GraphSAGE \cite{model:sage} aims to generate embeddings by sampling and aggregating features from a node's neighbors. With commonly used mean aggregation as an example, the aggregated information for node $v$ is computed as follows where $W$ is the weight matrix and $\sigma$ is the activation function:
$${\scriptstyle h_{v}^{l}=\sigma (W^{l-1} \cdot CONCAT(h_{v}^{l-1}, MeanAgg({h^{l-1}_{u}:u \in \mathcal{N}(v)) )})}$$

\textbf{Graph Isomorphism Network (GIN).} 
GIN \cite{model:gin} uses a parametrized aggregation function that can capture the graph structure by considering both the node itself and its neighbors. 
The update rule is defined as follows where $\epsilon$ is a learnable parameter and
$MLP$ is a multi-layer perceptron:
$${\scriptstyle h_{v}^{l} = MLP^{(l-1)}((1 + \epsilon_{v}^{(l-1)}) \cdot h_{v}^{(l-1)} + \sum_{u \in \mathcal{N}(v)} h_{u}^{(l-1)})}$$

\section{Attack Methodology}

\begin{table*}
\scriptsize
\centering
\begin{tabular}{ccccc}
    \hline
    \multirow{2}{*}{\textbf{Attack}} & \multicolumn{2}{c}{\textbf{Attacker's Knowledge}}  & \multirow{2}{*}{\textbf{Attacker's Capability}} & \multirow{2}{*}{\textbf{Who Modifies the Graph}}\\
        & \textbf{Node Outputs} & \textbf{Node Features}  \\
    \hline
    \vspace{1pt} 
    {Link Stealing Attack ($LSA$) \cite{Inference_Attacks:link_steal:he2020stealing}} & All & All & Query GNN model on all nodes  & {None} \\

    \vspace{1pt} 
    {LinkTeller Attack ($LTA$) \cite{Inference_Attacks:link_steal:wu2021linkteller}} & All & All  & Query GNN model on all nodes, modify all nodes & {Only Adversary} \\

    {Infiltration Inference Attack ($IIA$) \cite{devil_in_disguise}} & Limited & Limited & Add/modify limited nodes, add/remove edges & {Only Adversary} \\
        
    {Our proposed attacks} & Limited & Limited & Add/modify limited nodes, add edges & {Everyone} \\
    \hline
\end{tabular}
\caption{Comparison of threat models among our and state-of-the-art edge inference attacks on GNNs.}
\label{table:threat_model_comparison}
\vspace{-0.3cm}
\end{table*}

\subsection{Threat Model}
\noindent\textbf{Graph Structure and GNN Specification.} 
In our setting, the service provider maintains a private graph, where nodes represent entities for classification and edges denote relationships or interactions. The service provider has exclusive access to the entire graph, including all nodes and connections, while users are restricted to accessing only the nodes relevant to them. The graph is dynamic, reflecting updates such as new interactions, profile changes, or the creation of new accounts, ensuring it stays aligned with real-world behavior.  
These updates inherently introduce opportunities for graph manipulation, as users, by design, can interact with the graph to influence its structure within the scope of their access. 

Formally, the graph is modeled as a series of snapshots, each representing a static state: $\mathcal{G} = {G_1, G_2, \dots, G_T}$, where $G_T$ is the most recent snapshot. The service provider determines the update frequency—whether weekly, daily, hourly, or in real time—depending on the required freshness and relevance of the data. The model operates on the most current snapshot.

This framework reflects the practical applications of GNNs across various domains, such as financial networks \cite{graph:usage:medical:credit:sukharev2020ewsgcn, graph:usage:medical:credit:wang2021temporal, graph:usage:anomoly:xu2021towards, wang2021review}, medical networks \cite{graph:usage:medical:ahmedt2021graph, graph:usage:medical:kazi2023ia}, and intrusion detection systems \cite{bilot2023graph, graph:usage:anomoly:xu2023understanding}. In these settings, graphs are constructed directly from raw data, leaving them vulnerable to manipulation. Adversaries can exploit this process by introducing new nodes, edges, or features, mimicking legitimate user interactions. For example, with the growing adoption of relational deep learning \cite{robinson2024relbench}, which transforms relational databases into graphs, adversaries can indirectly manipulate the graph by injecting new rows into the underlying database.

We conduct experiments in both \emph{static} and \emph{dynamic} graph scenarios to examine this issue thoroughly. 
Here, static and dynamic refer to whether external changes to the graph occur while the adversary is conducting an attack. Formally, it depends on whether the service provider updates $G_{T}$ to $G_{T+1}$ based solely on the adversary's behavior or includes changes made by others.
The static scenario becomes possible if the adversary can pinpoint a time when no other changes are being made to the graph, or if they can execute their attack swiftly during real-time updates, assuming that no other significant changes to the graph will happen in that interval.
On the other hand, a dynamic scenario prepares for the worst-case situation by assuming that the graph may change significantly during the attack, i.e., in between the attack steps.

Previous research has focused exclusively on static scenarios \cite{Inference_Attacks:link_steal:wu2021linkteller, Inference_Attacks:link_steal:he2020stealing, devil_in_disguise}. Table \ref{table:threat_model_comparison} shows the comparison of the threat models. 
\emph{We are the first to investigate GNN edge inference attacks within dynamic graph contexts}. 

In this work, we focus on inductive GNNs in dynamic scenarios, building on prior methodologies \cite{Inference_Attacks:link_steal:wu2021linkteller, devil_in_disguise}. Inductive GNNs are well-suited for dynamic graphs as they generalize to unseen nodes and adapt to evolving structures without retraining. This adaptability makes them a dominant choice in dynamic applications, including those at Amazon \cite{inductive:usage:amazon}, LinkedIn \cite{borisyuk2024lignn}, and Kumo.ai \cite{robinson2024relbench}.

Following the inductive approach \cite{model:sage, model:gat, model:gin}, the GNN model is first trained on a foundational graph populated with labeled nodes (the training phase). 
It is then utilized on the continually evolving graph (the inference phase).
All users have only black-box access to the GNN model, meaning they can observe the outputs but not the internal workings of the model. The model makes predictions based on the private graph and provides the users with prediction probabilities (node outputs).

\noindent\textbf{Adversary’s Goal.} The adversary aims to ascertain whether a connection exists between two nodes, a goal that holds practical relevance across numerous real-world scenarios. For example, in financial networks, stakeholders might need to uncover the connections of a specific entity to grasp the essential relationships influencing its financial activities.

\noindent\textbf{Adversary's Capability.}
The adversary acts as a user during the inference phase and possesses the following capabilities:

     \textbullet\ Access to auxiliary nodes: The adversary can introduce or control specific nodes within the graph, referred to as auxiliary nodes ($A$).
    
     \textbullet\ Establish new edges: The adversary can create edges between auxiliary nodes $A$ and other nodes in the graph, mimicking actions such as new transactions or interactions.
    
    \textbullet\ Modify node features: The adversary can alter the features of nodes in $A$, simulating profile updates or similar changes.
    
     \textbullet\ Query the model: The adversary can query the GNN model for predictions on nodes in $A$.

These capabilities align with typical user interactions in dynamic environments where inductive GNNs are deployed \cite{borisyuk2024lignn, robinson2024relbench, devil_in_disguise}, 
\added{such as social networks or healthcare networks. For instance, on LinkedIn \cite{borisyuk2024lignn}, an attacker could pose as an alumnus of a specific institution to establish connections.} 
Unlike prior works assuming static graphs or full adversarial control \cite{Inference_Attacks:link_steal:wu2021linkteller, devil_in_disguise}, our model considers more realistic conditions where benign user actions also contribute to graph evolution.

\noindent\textbf{Adversary's Extra Knowledge.}
Prior works \cite{devil_in_disguise, Inference_Attacks:link_steal:wu2021linkteller, Inference_Attacks:link_steal:he2020stealing} either assume extra knowledge, such as subgraph density or shadow dataset, for threshold selection or fail to provide explicit methods for setting thresholds. To address this gap, we propose a dual approach: leveraging extra knowledge when available or deriving thresholds directly from the adversary’s capabilities. This ensures a practical and flexible solution, accommodating both knowledge-rich and knowledge-limited scenarios, as detailed in Section \ref{sec:threshold}.



\begin{algorithm}
\scriptsize
    \caption{General Pipeline of Our Attacks}
    \label{alg:general}
    \begin{algorithmic}[1] 
    \Require
        GNN model $GNN(\cdot)$, Target node $t$, Candidate node set $C$, estimated degree of target node $\hat{d}$,  Attack method $AttackMethod()$
    \Ensure
        {A boolean vector for nodes in the candidate set, indicating the presence of an edge between $t$ and each $c \in C$}
        \For{each node $c \in C$}
            \State $LPS[c] \leftarrow AttackMethod(GNN, t, c)$ 
        \EndFor
        \State $threshold \leftarrow (\textit{$\hat{d} + 1)$-th largest value in LPS}$
        \State $Res \leftarrow [\textbf{True if } LPS[c]] > threshold \textbf{ else False for } c \in C]$  
        \State \Return $Res$
    \end{algorithmic}
\end{algorithm}

\subsection{Methodological Limitations of Existing Works and Key Tasks to Solve}
\label{sec:existing_limitation}
Prior works face several challenges that limit their applicability and reliability. Our methodology is designed to address these limitations by solving the following key tasks:

    \textbullet\ \textbf{Limitation 1 (L1):}
    Existing attacks operate at a low granularity, focusing on distinguishing connected node pairs from unconnected ones rather than directly determining if a specific pair is connected. This approach addresses a less granular version of the problem and bypasses the true challenge: differentiating directly connected nodes from closely related but unconnected ones, which is critical for real-world applicability. 
    This issue is compounded by flawed experimental setups that rely on random sampling for negative samples. Since most node pairs in a graph are far apart, even simple attacks can easily classify them as unconnected. The true challenge lies in distinguishing directly connected pairs from those closely related ones, but such closely related pairs are underrepresented in random sampling. This leads to inflated performance metrics, even when attacks fail in these cases. To address this limitation, attacks should target the harder task of distinguishing directly connected pairs from close but unconnected ones to ensure meaningful connection inference.
    

    \textbf{Task 1 (T1):} Reframe the problem to focus on inferring the existence of an edge between 2 nodes rather than simply distinguishing connected from unconnected pairs, ensuring that the methodology and evaluation are robust to node pair selection.

    \textbullet\ \textbf{Limitation 2 (L2):} Poor Generalization Across GNN Architectures.
    Existing methods often perform well on specific GNN architectures (e.g., GCNs) but fail on others, such as GATs, limiting their adaptability. This lack of generalization reduces their applicability in diverse real-world scenarios where a variety of GNNs are deployed.
    
    \textbf{Task 2 (T2):} Ensure that the proposed method performs consistently across a wide range of GNN models.

    \textbullet\ \textbf{Limitation 3 (L3):} Limited Robustness Against Dynamic Graphs.
        Prior works often assume static graphs during attacks, even when allowing adversarial modifications. These approaches fail to account for real-world conditions where graphs evolve dynamically due to benign changes made by other users or systems.
        
    \textbf{Task 3 (T3):} Develop a robust attack methodology capable of adapting to evolving graph structures, reflecting realistic deployment scenarios.

\subsection{Proposed Attacks}

\noindent\textbf{Key Intuitions.}
Our attack leverages the message aggregation mechanism inherent in GNNs, where the flow of information is a two-way street: nodes absorb information from their neighbors and, in turn, impart their data back into the network. This bidirectional flow serves dual purposes in our strategy. 
In a scenario where two nodes, $v$ and $u$, are interconnected. Node $v$ assimilates information from $u$ through message aggregation, enabling us to learn insights about $u$'s private attributes by analyzing $v$'s output. This helps us solve the lack of direct access to the target nodes.
Moreover, the messages exchanged between $v$ and $u$ navigate along specific paths within the graph. As a message progresses along a path, it is transformed by each node it encounters, which means that the final message received by each node is influenced by the specific paths it has traversed. Therefore, even if the same initial message is sent to different nodes, the final messages they receive will differ due to variations in their respective paths.
Consequently, the message received by node $u$ from node $v$ will contain information reflective of the path between them.
This observation provides a way to infer information about the path between nodes.
\added{We provide a formal justification about our intuition in Appendix~\ref{app:justification:intuition}. 
}

\noindent\textbf{Overall Attack Design.}
We organize our attack into two phases:

    \textbullet\ \emph{Information Extraction (P1):} Extract information from the target model to assign a score to each node pair, where a higher score reflects closer proximity between nodes. This phase addresses T2 and T3 by ensuring adaptability across GNN architectures and dynamic graphs.
    
    \textbullet\ \emph{Threshold Determination (P2):} Establish a threshold based on the scores to identify whether a node pair is connected. This phase addresses T1 by reframing the problem to focus on direct edge inference.

\subsection{Attack Phase 1: Extracting Statistics}
In Phase 1, we identify a statistic to distinguish connected from unconnected node pairs, forming the basis for Phase 2. The adversary adopts a node-centric approach, introducing auxiliary nodes (A) and connecting them to the target node (t) and candidate nodes ($c \in C$, nodes the adversary tests for connection to the target node). By querying the GNN and analyzing prediction probability vectors from auxiliary nodes, the adversary computes a Link Possibility Score (LPS) for each target-candidate pair. 
This LPS serves as a measure of proximity. Our key intuitions lead to the design of multiple attacks, all adhering to the same pipeline depicted in Algorithm \ref{alg:general}.

\begin{figure}[ht]
    \centering
    \begin{subfigure}[b]{0.4\linewidth}
        \includegraphics[width=\linewidth]{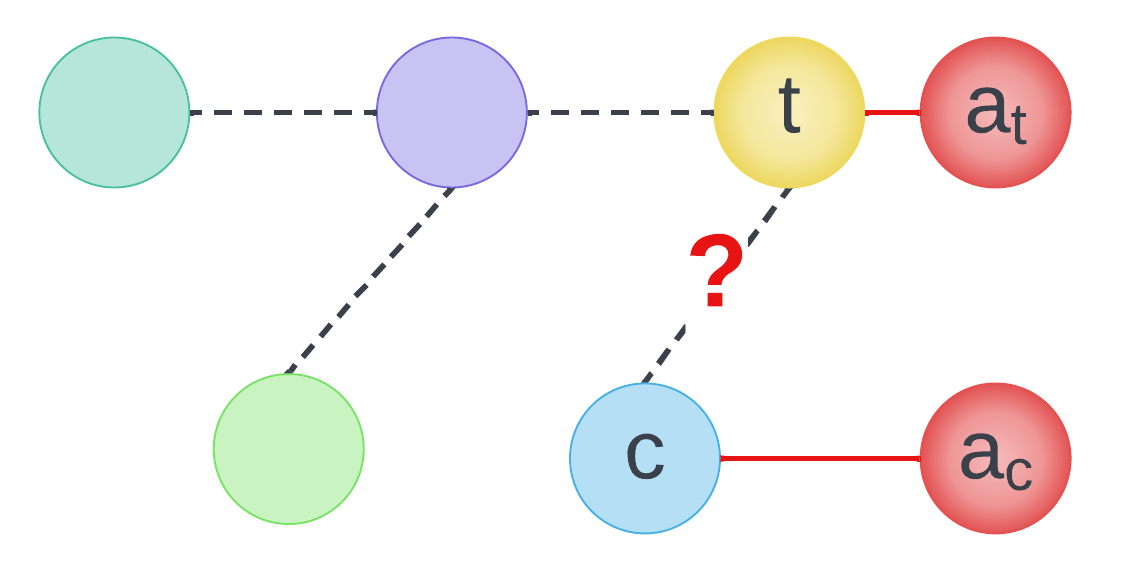}
        \caption{{\footnotesize (a) SIM and INF}}
        \label{fig:attack:methodology:node_insertion:a}
    \end{subfigure}
\hfill
    \begin{subfigure}[b]{0.4\linewidth}
        \centering
        \includegraphics[width=\linewidth]{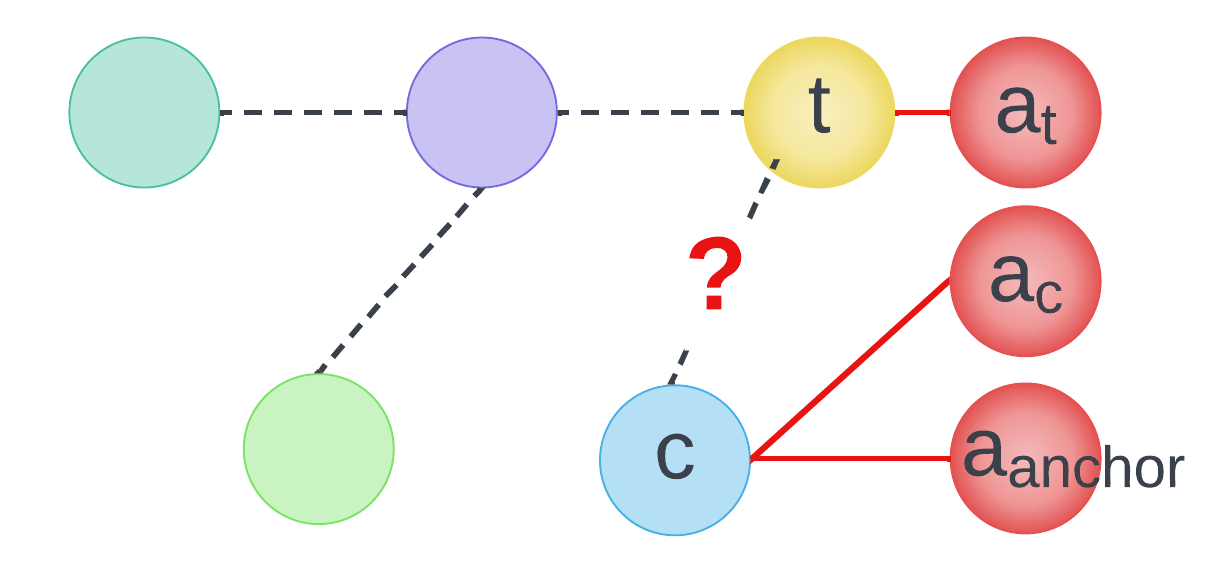}
        \caption{{\footnotesize (b) INF\textit{--}DIR and INF\textit{--}MAG}}
        \label{fig:attack:methodology:node_insertion:b}
    \end{subfigure}
    \caption{GNNBleed attacks: $t$, $C$, and $a_n$ represent the target node, the candidate set, and the auxiliary nodes, respectively. Black dashed lines represent existing edges within the graph, concealed from the adversary's knowledge, and red lines indicate edges introduced by the adversary.}
    \label{fig:attack:methodology:node_insertion}
\end{figure}

\noindent\textbf{Attack Strategies.}
Our goal is to infer information about the paths connecting two nodes by analyzing the exchange of messages between them. The key idea is that the influence one node exerts on others provides a means to observe how messages propagate along these paths. Specifically, we track how changes to a node’s features affect the predictions of its neighbors, using this influence to estimate the flow of information between nodes.
To illustrate this concept, consider two nodes, $u$ and $v$. If an adversary modifies the features of node $u$ by a perturbation factor of $\alpha$, making it $(1+\alpha)u$, this will impact the output probabilities of $v$ if $v$ can receive information from $u$. By observing $v$'s predicted probability before ($p_{v}$) and after ($p_{v}'$) this change, the adversary can determine the resulting difference in $v$'s output probability, quantified as $p_{v} - p_{v}'$. This difference can be used to infer the properties of the message path between $u$ and $v$.
In this work, influence is generated through such perturbations across all methods. 
\added{In this work, we treat the perturbation factor as a simple yet effective measure of adversarial influence. Rather than focusing on how to generate these perturbations, our primary goal is to analyze how model outputs respond to adversarial changes in node features. This perspective is also adopted by \cite{Inference_Attacks:link_steal:wu2021linkteller}, where similar perturbation-based approaches are used.
Different methods—such as modifying categorical node features or altering the graph structure—can be employed to generate influence depending on the scenario.
}
For further details on influence generation, see Appendix~\ref{app:evaluation:auxiliary_node_features}.

Prior research \cite{devil_in_disguise, Inference_Attacks:link_steal:wu2021linkteller} focuses on the magnitude of influence between nodes, typically measuring the L2 norm of changes in output probabilities and correlating these with the distance between nodes. These methods identify nodes with high-magnitude influence on each other as connected. However, they face two key challenges:

    1. Aggregation Mechanism Dependence: The magnitude of influence in GNNs does not always correlate with node distance. Message aggregation mechanisms, such as GCN's \emph{mean} function, average messages along paths, causing influence to diminish over distance. This explains the success of prior methods on GCNs. However, their effectiveness drops against models like GAT, which use attention-based mechanisms to selectively weigh nodes. For example, close nodes with low attention weights may exhibit minimal influence, contradicting assumptions based on proximity.

    2. Dynamic Graph Scenarios:
    In real-world dynamic graphs, relying solely on influence magnitude is insufficient. When multiple agents modify the graph, high influence magnitudes can result from any nearby influence source, not necessarily the adversary. This makes it harder to isolate adversarial effects in evolving graph environments.

To solve these problems, we employ two distinct methodologies.
The first attack, named the \emph{Magnitude-based Influence Attack}, focuses on the degree of change in the output probabilities. This attack is tailored to be effective across various GNNs with different message aggregation methods, making it well-suited for scenarios where the graph is relatively static.
The second attack, the \emph{Direction-based Influence Attack}, centered on analyzing the patterns of output probability changes, demonstrating its resilience in dynamic graph scenarios.

\noindent\textbf{Magnitude-Based Influence Attack ($INF\textit{--}MAG$)}
This attack strategy relies on the correlation between the distance between two nodes and the magnitude of influence one node exerts on the other. 
\added{Appendix~\ref{app:intuition} illustrates this relationship, showing that as distance increases, the magnitude of change decreases.
}
To improve accuracy across different GNN architectures, we propose a method that accounts for variations in message aggregation mechanisms, ensuring the magnitude of influence becomes a more reliable indicator of node proximity. As in prior works, we quantify influence using the L2 norm of changes in output probabilities. 
The challenge lies in mitigating the variability introduced by GNN-specific aggregation weights, which can distort the magnitude of influence. To address this, we introduce an auxiliary node termed the {anchor node}, $a_{anchor}$, which is used to help estimate the weight.
The anchor node shares identical features with $a_t$ and $a_c$ and is directly connected to the candidate node $c$ (depicted in Figure \ref{fig:attack:methodology:node_insertion:b}). 
Because $a_c$ and $a_{anchor}$ are connected to the $c$, their distance would always be 2. However, the response of $a_{anchor}$ to changes in $a_c$'s features can differ based on the varying weights associated with $a_c$, even though their distance remains constant.
A strong influence of $a_c$ on $a_{anchor}$ ($I(a_c,a_{anchor})$) suggests a great weight of $a_c$, thus $a_c$ would have a strong influence on all the other nodes aggregate information from it.  
This setup allows us to calibrate the influence of $a_c$ on $a_t$ relative to the change observed in $a_{anchor}$'s output. 
We calculate the LPS as $LPS = \frac{I(a_c,a_t)}{I(a_c,a_{anchor})}$. This metric adjusts the perceived influence of $a_c$ on $a_t$ based on the relative influence on $a_{anchor}$, thereby accounting for the variations in weight and providing a more accurate measure of connectivity strength.

\begin{algorithm}
\scriptsize
    \caption{$INF\textit{--}MAG$}
    \label{alg:attack_function:inf_3}
    \begin{algorithmic}[1]
        \Require{
            GNN API $GNN(\cdot)$, target node $t$, candidate node $c$
        } 
        \Ensure{
            LPS, which quantifies the likelihood that $t$ and $c$ are connected
        } 
        \State $G' \leftarrow G$ update the API internal graph with modifications as follows:
        \State -- Add nodes $a_t$, $a_c$, $a_{anchor}$.
        \State -- Establish edges: $(t, a_t)$, $(c, a_c)$ and $(c, a_{anchor})$ .
        \State $p_{a_t}, p_{a_{anchor}} \leftarrow GNN(a_t), GNN(a_{anchor})$
        \State $G'' \leftarrow G'$ perturb the feature of $a_c$
        \State $p'_{a_t}, p'_{a_{anchor}} \leftarrow GNN(a_t), GNN(a_{anchor})$
        \State $LPS \leftarrow I(a_c,a_t) / I(a_c, a_{anchor})$
        \State \Return  $LPS$
    \end{algorithmic}
\end{algorithm}

\textbf{Implementation.} To get the $LPS$ for a node pair $(t, c)$, we initially insert three auxiliary nodes $\{a_t,a_c, a_{anchor}\}$ as depicted in Figure \ref{fig:attack:methodology:node_insertion:b}. 
We then obtain their initial outputs, denoted $p_{a_t}$ and $p_{a_{anchor}}$. Following this, we alter the feature of $a_c$ using $feature_{a_c}' = (1+\alpha) feature_{a_c}$ and subsequently acquire the updated outputs $p_{a_t}'$ and $p_{a_{anchor}}'$.
Next, we calculate the LPS score as $LPS = \frac{I(a_c,a_t)}{I(a_c, a_{anchor})}$, where $I(a_c,a_t) = \|p_{a_t}' - p_{a_t}\|_2$. Higher $LPS$ values indicate a stronger likelihood of an edge, and lower values suggest otherwise. Algorithm \ref{alg:attack_function:inf_3} demonstrates the attack steps.

\noindent\textbf{Direction-Based Influence Attack ($INF\textit{--}DIR$)}
Direction-Based Influence Attack is based on the principle that nodes nearby typically exhibit similar reactions to external influences due to their shared topology. 
This common context causes these nodes to process information from the influence source similarly, resulting in comparable updates to their output probabilities.
Consequently, when two nodes exhibit similar patterns of output changes, it suggests they are near each other in the graph. 
This principle allows the method to remain effective in dynamic scenarios, even with multiple sources of influence. 
When two nodes are close to each other, they are likely to receive information from a similar set of sources, thereby exhibiting similar changes in their output probabilities.
\begin{addedenv}
Appendix~\ref{app:intuition} illustrates this relationship, showing that as distance increases, the similarity of change decreases.
\end{addedenv}

\begin{algorithm}
\scriptsize
    \caption{$INF\textit{--}DIR$}
    \label{alg:attack_function:inf_2}
    \begin{algorithmic}[1]
        \Require{
            GNN API $GNN(\cdot)$, target node $t$, candidate node $c$
        }
        \Ensure{
            LPS, which quantifies the likelihood that $t$ and $c$ are connected
        }
        \State $G' \leftarrow G$ update the API internal graph with modifications as follows:
        \State -- Add nodes $a_t$, $a_c$, and $a_{anchor}$.
        \State -- Establish edges: $(t, a_t)$, $(c, a_c)$, and $(c, a_{anchor})$.
        \State $p_{a_t}, p_{a_{anchor}} \leftarrow GNN(a_t), GNN(a_{anchor})$
        \State $G'' \leftarrow G'$ perturb the feature of $a_c$
        \State $p'_{a_t}, p'_{a_{anchor}} \leftarrow GNN(a_t), GNN(a_{anchor})$
        \State $LPS \leftarrow 1-distance\_metrics(p'_{a_t} - p_{a_t}, p'_{a_{anchor}} - p_{a_{anchor}})$
        \State \Return  $LPS$
        \end{algorithmic}
\end{algorithm}

\noindent\textbf{Implementation Details.} 
This attack involves inserting three auxiliary nodes: $a_t$, connected to the target node $t$, and the other two, $a_{anchor}$ and $a_c$, connected to the candidate node $c$ (depicted in Figure \ref{fig:attack:methodology:node_insertion:b}). 
Here, $a_{anchor}$ is a reference node, indicating how nodes two hops away from the influence source $a_{c}$ respond to changes. 
Nodes exhibiting reactions similar to $a_{anchor}$ are likely in a similar context, suggesting proximity to both $a_{anchor}$ and $a_{c}$.
We record the output probabilities both before ($p_{a_t}$, $p_{a_{anchor}}$) and after ($p_{a_t}'$, $p_{a_{anchor}}'$) the feature changes. 
The LPS is then computed as $LPS = 1 - distance\_metrics(p_{a_t}' - p_{a_t}, p_{a_{anchor}}' - p_{a_{anchor}})$. 
A smaller distance and a higher LPS value imply a likely edge between $t$ and $c$. 
Our experiment utilizes eight different distance metrics, including Cosine, Euclidean, Correlation, Chebyshev, Bray-Curtis, Canberra, Manhattan, and Square-Euclidean distances, to identify the most effective for this analysis. The detailed execution of this attack is described in Algorithm \ref{alg:attack_function:inf_2}.

\noindent\textbf{Further Adapted Direction-Based Influence Attack ($INF\text{--}DIR^*$)}
Graphs that evolve unpredictably introduce noise between queries, complicating analysis. To address this, we extend the $INF\text{--}DIR$ approach by performing multiple queries and analyzing trends in output changes over time. This longitudinal analysis mitigates transient noise from random graph fluctuations while preserving consistent adversarial influence.

\noindent\textbf{Implementation Details.} 
This multi-query attack collects data by repeating the influence generation process after node injection, recording output changes as lists: $PCL_a$ for the anchor node and $PCL_t$ for the target node. These changes are then preprocessed using one of four techniques:

    \textbullet\ Concatenation: Combine all components in each $PCL$.
    
    \textbullet\ Mean/Median: Calculate the mean or median for each dimension across the vectors in $PCL$.
    
    \textbullet\ PCA: Reduce dimensionality by projecting concatenated vectors onto a lower-dimensional space.

The preprocessed vectors, $Processed_{a_t}$ and $Processed_{a_{anchor}}$, are then compared using the formula: 
$$LPS = 1 - distance\_metrics(Processed_{a_t}, Processed_{a_{anchor}})$$
A smaller distance implies a higher LPS, indicating a likely edge between the target and anchor nodes.


\subsection{Attack Phase 2: Threshold Selection.}
\label{sec:threshold}
Building on Phase 1, where we distinguish between connected and unconnected node pairs, Phase 2 transforms this into the task of determining whether two specific nodes are connected. To achieve this, we adopt a node-centric approach tailored to each node’s unique influence distribution.

\noindent\textbf{What makes a good threshold?}
\deleted{
A good threshold should account for the variability in influence distributions across different nodes. Factors such as node degree, neighborhood structure, and aggregation mechanisms contribute to diverse influence patterns. For example, high-degree nodes aggregate information from many neighbors, leading to weaker individual signals, while low-degree nodes are more sensitive to changes in individual neighbors. Similarly, aggregation mechanisms like attention further amplify these differences by selectively weighing messages, resulting in unique influence distributions for each node.
Thus, a global threshold assuming uniform distributions across nodes would fail to account for these variations, leading to potential misclassification. Instead, the threshold should be tailored to each target node's specific influence distribution for accurate and adaptive connectivity determination. 
}
\noindent \added{A good threshold should account for variability in influence distributions across nodes, shaped by degree, neighborhood structure, and aggregation. High-degree nodes aggregate weaker individual signals, while low-degree nodes are more sensitive to changes. Attention further amplifies these differences.

\textbf{Figure~\ref{fig:threshold}} shows that low-degree nodes experience larger magnitude changes, while high-degree nodes exhibit weaker signals. Notably, 2-hop neighbors of low-degree nodes resemble 1-hop neighbors of high-degree nodes, making a global threshold ineffective. It would misclassify weakly connected nodes around low-degree targets as “connected” while missing real connections around high-degree targets.

Thus, threshold selection should be node-specific. A formal justification is in Appendix~\ref{app:justification:threshold}.

\begin{figure}[htbp]
  \centering
  \includegraphics[width=0.7\columnwidth]{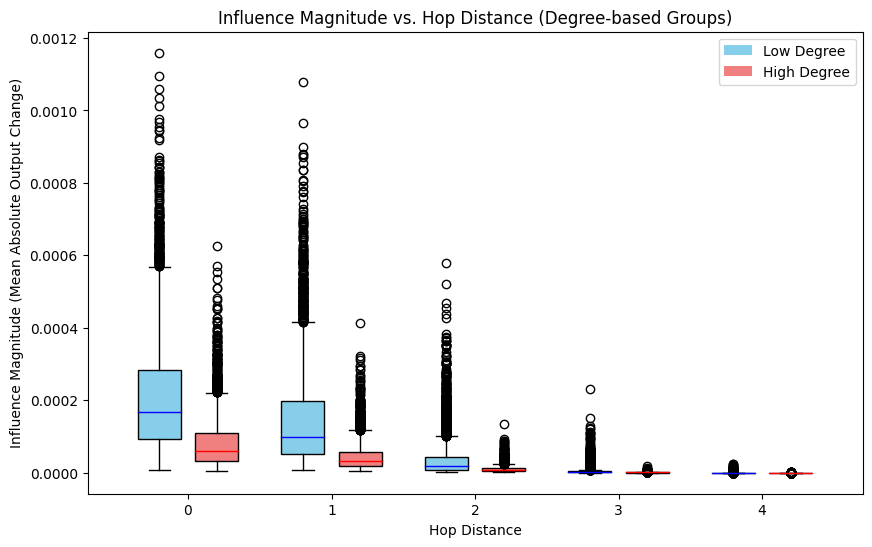}
  \caption{Magnitude of influence for nodes with different degree distributions in the LastFM dataset using a GCN model.}
  \label{fig:threshold}
\end{figure}
}

\noindent\textbf{How to derive a node-centric threshold?}
To determine a threshold for a target node, the adversary requires insight into the node’s influence distribution. 
Specifically, they need a reference point to understand what threshold value would effectively distinguish connected from unconnected nodes for that particular target. 
Using the statistics obtained from Phase 1, the adversary can set a threshold by identifying what scores indicate true connections for the target node. 
This information can be derived from a shadow dataset or directly inferred from the target graph. 

Based on the GNN model's implementation and the adversary's interaction capabilities, we propose two approaches for threshold selection:

    \textbullet\ In-Graph Thresholding: If the adversary has knowledge of some connections to the target node, they can use the scores of these known connected nodes to construct a score distribution and set an accurate threshold. In the worst-case scenario, where no such knowledge exists, the adversary can establish a direct connection to the target node via auxiliary nodes. By observing the scores of these auxiliary nodes, the adversary can infer a threshold based on the observed influence. This approach leverages the adversary’s ability to interact directly with the target graph.
    
    \textbullet\ Extra Knowledge-Based Thresholding: The key idea is to leverage additional knowledge to approximate the threshold distribution for different nodes. If the adversary has access to shadow data or partial properties of the target node (e.g., degree or feature attributes), this information can be used to estimate the score distribution for connected nodes. By migrating the threshold distribution from nodes with similar properties or from shadow datasets, the adversary can define more accurate thresholds. This approach aligns with existing   methods that utilize extra knowledge for threshold determination \cite{Inference_Attacks:link_steal:wu2021linkteller}.

\deleted{
In this work, we use the lowest score learned from these methods as the threshold. More discussions can be found in Appendix~\ref{app:threshold}. }
\begin{addedenv}
{\noindent\textbf{Threshold Selection in This Work.}
In our experiments, we take a simple heuristic approach by using the lowest score among those derived via the above methods as the threshold. This choice stems from the intuition that higher scores indicate genuine connections, and thus setting a lower bound is a conservative way to label edges. We emphasize that we are demonstrating possible ways of thresholding rather than prescribing a single optimal solution. Further discussion on threshold design and rationale can be found in Appendix~\ref{app:threshold}. }    
\end{addedenv}

\noindent\textbf{Mapping Threshold Selection to Pairwise Connectivity.}
Given a node pair, we can address the question of whether they are connected by selecting one as the target node and the other as the candidate node. By comparing the candidate's influence against the threshold tailored to the target node’s unique distribution, we can determine if a connection exists.

\section{Evaluation}
Our experiments span two scenarios: static and dynamic, each requiring unique designs and dataset selections. We first outline the common setups for both scenarios, followed by scenario-specific details and datasets.

\subsection{Common Experiment Setup}
\label{sec:experiment}
\noindent \textbf{Models.}  
We evaluate the attacks on four types of GNN models: GAT \cite{model:gat}, GCN \cite{model:gcn}, GIN \cite{model:gin}, and GraphSAGE \cite{model:sage}. Unless otherwise specified, results in the main paper are based on a 4-layer architecture. The rationale for selecting a 4-layer setup, along with results from attacks on 3-layer models, is provided in Appendix~\ref{app:4_layer_choice}.

\noindent \textbf{Metrics.} 
We evaluate the attacks using precision (the ratio of connected nodes among the ones recognized as true), recall (the ratio of identified connected nodes to all connected nodes), F1-score (the harmonic mean of precision and recall), and ROC-AUC (a threshold-independent metric evaluating the discriminatory power of classification tasks). Due to space constraints in the main paper, we primarily present results for the F1-score and ROC-AUC.

\noindent \textbf{Auxiliary Node Feature.} The features of the auxiliary nodes are randomly sampled from the target graph to simulate a scenario where the adversary joins the graph as a regular user.  

\noindent\textbf{Baseline Attacks.} We compare our attacks with \emph{three} state-of-the-art attacks, as described below. While these attacks operate under different threat models, we include them in our evaluation to provide a comprehensive performance comparison:

    \noindent\underline{\emph{Link Stealing Attack (LSA)~\cite{Inference_Attacks:link_steal:he2020stealing}}}: For our baseline, we use LSA-0, as it most closely aligns with our threat model. This approach utilizes the correlation distance between nodes' output probabilities to infer the existence of edges.
    
    \noindent\underline{\emph{LinkTeller Attack (LTA) \cite{Inference_Attacks:link_steal:wu2021linkteller}}}: LTA discerns links by analyzing the magnitude of influence one node exerts on others. The influence is generated by changing a node's feature.
    
    \noindent\underline{\emph{Infiltration Inference Attack (IIA) \cite{devil_in_disguise}}}: IIA examines the influence exerted by a newly added node with a zero feature vector. However, IIA is ineffective against models like GIN that use \emph{sum} for message aggregation. Since a node with a zero feature vector does not alter the overall sum of aggregated features, it fails to exert any influence, rendering the attack ineffective.
    
We also introduce two baseline attacks that adapt LSA and LTA to our threat model (depicted in Figure \ref{fig:attack:methodology:node_insertion:a}):

    \noindent\underline{\emph{Similarity Attack (SIM)}}: This attack determines the existence of edges by computing the correlation distance between the output probabilities of auxiliary nodes.
    
    \noindent\underline{\emph{Naive Influence Attack (INF)}}: This attack infers connections by examining the magnitude of the influence exerted by one auxiliary node on others.

\noindent\textbf{Threshold Selection.} 
We evaluate our attack using two threshold selection methods: one for fair comparison with existing works and another for assessing performance under different threshold selection methods. 

\noindent\textbf{Uniform Threshold for Fair Comparison.}
    To ensure a fair comparison with prior works, which lack robust thresholding methods \cite{Inference_Attacks:link_steal:he2020stealing, devil_in_disguise}, we use a standardized approach. Following \cite{Inference_Attacks:link_steal:wu2021linkteller}, we adopt a top-$k$ selection method, where $k$ is set to the ground truth number of connected node pairs. This approach evaluates the distinguishability of the generated scores.

\noindent\textbf{Threshold Selection for Determining Connections.}
    This threshold selection evaluates the attack's effectiveness in directly determining whether two nodes are connected. We compare the methods proposed in Section \ref{sec:threshold}, emphasizing their feasibility and performance across different scenarios.

\noindent\textbf{Test Set Construction.} 
To evaluate the performance of link inference attacks and ensure practical applicability, we adopt a node-centric test set design:

    \textbullet\ We randomly sample 100 target nodes, $T$.
    
        \textbullet\ For each target node $t \in T$, we construct node pairs $(t, c)$, where $c \in C_t$, with the candidate set $C_t$ defined as:
        
        \emph{Positive Samples} — Directly connected neighbors of $t$.
        
        \emph{Negative Samples} — Nodes within a two-hop distance from $t$, but not directly connected. 
\begin{addedenv}
{Specifically, for each target node, we include all its direct connections as positive samples, while negative samples consist of all nodes within two hops that are not direct neighbors. This results in a test set whose size corresponds to the number of two-hop neighbors per target node.}

\end{addedenv}

\emph{Problem with Random Sampling:}
Existing works \cite{devil_in_disguise, Inference_Attacks:link_steal:wu2021linkteller, Inference_Attacks:link_steal:he2020stealing} primarily use random sampling for negative samples, often resulting in distant nodes outside the GNN’s message-passing range. Such nodes show no influence changes, making them trivially separable from connected nodes. This inflates performance metrics by simplifying the task to identifying outliers rather than detecting meaningful connections.
By relying on simplistic setups, these evaluations fail to test attack robustness in distinguishing directly connected nodes from nearby unconnected nodes—arguably the core challenge in link inference. This oversight leads to flawed experimental designs that poorly reflect real-world scenarios.

\noindent\textbf{Theoretical Justification for Using 2-Hop Nodes as Negative Samples.}
We use 2-hop nodes as negative samples to better evaluate attack robustness, as they are the most challenging to distinguish from directly connected neighbors. This difficulty arises from two factors:

\textbullet\ Proximity-Based Scoring: GNNs propagate stronger influence to closer nodes, leading to higher scores for 2-hop nodes that are spatially close to the target.

\textbullet\ Shared Structure: 2-hop nodes often share neighbors with the target, causing their scores to closely resemble those of directly connected nodes.

This setup tests an attack’s ability to differentiate direct connections from close-but-unconnected nodes, demonstrating robustness in more realistic scenarios. Success in such stringent conditions validates the method’s generalizability to less restrictive setups, where negative samples are easier to distinguish due to weaker influence and reduced structural similarity. 
\begin{addedenv}
{Detailed theoretical justification can be found in Appendix~\ref{app:2hopselection}.}
    
\end{addedenv}

\subsection{Static Graph Scenario}
\label{sec:static}

\noindent \textbf{Datasets.}
We evaluate our attacks on seven datasets: Flickr \cite{dataset:flickr}, LastFM \cite{dataset:lastfm}, and Twitch-{DE, EN, FR, PT, RU} \cite{dataset:twitch}. Flickr is an image relationship dataset, while LastFM and Twitch are social network datasets.
Appendix~\ref{app:evaluation:data_statistics} presents them in detail. 

\noindent\textbf{Target Node Selection.} 
We randomly chose 100 nodes for each dataset as target nodes, and all results in the later sections represent the average across these nodes.

\noindent\textbf{Perturbation Factor.} We set the perturbation factor $\alpha=0.1$ for LTA and our attacks. In static scenarios, the exact value of $\alpha$ is less critical as long as it introduces noticeable changes in the model output since the adversary is the sole source of perturbations.

\definecolor{ForestGreen}{RGB}{34, 139, 34}  
\definecolor{softgreen}{RGB}{139, 69, 19}     
\definecolor{1}{RGB}{25, 105, 25}        
\definecolor{softblue}{RGB}{70, 129, 207}       
\definecolor{softred}{RGB}{165, 52, 52}         
\definecolor{softcyan}{RGB}{44, 174, 158}       
\definecolor{softorange}{RGB}{205, 133, 0}      
\definecolor{softpurple}{RGB}{96, 0, 96}        
\definecolor{darkbrown}{RGB}{109, 54, 14}       
\definecolor{softpink}{RGB}{205, 85, 145}       
\definecolor{softyellow}{RGB}{205, 205, 0}      
\definecolor{darklime}{RGB}{40, 165, 40}        
\definecolor{darkgray}{RGB}{139, 139, 139}      
\definecolor{darkolive}{RGB}{85, 114, 28}       
\definecolor{darkpeach}{RGB}{205, 175, 149}     
\definecolor{darkcoral}{RGB}{205, 102, 64}      
\definecolor{darkaquamarine}{RGB}{102, 205, 170}
\begin{figure*}[!ht]
    \centering
    \begin{tikzpicture}
        \begin{axis}[hide axis, xmin=0, xmax=1, ymin=0, ymax=1,
                     legend columns=-1, 
                     legend entries={$SIM$, $INF$, $INF\text{--}DIR$, $INF\text{--}MAG$, $LTA$, $LSA$, $IIA$},
                    legend style={          
                        text=black,        
                        draw=gray, fill=gray!10,
                        legend columns=1,
                        cells={anchor=west},
                        font=\footnotesize\itshape,
                        /tikz/every even column/.append style={column sep=0.5cm}
                    },
                     legend to name=static:f1:legend] 
            \addlegendimage{softred,fill=softred, line width=3pt}
            \addlegendimage{softblue, fill=softblue, line width=3pt}
            \addlegendimage{softgreen,fill=softgreen, line width=3pt}         \addlegendimage{softyellow,fill=softyellow, line width=3pt}
            \addlegendimage{softcyan,fill=softcyan, line width=3pt}
            \addlegendimage{softpurple,fill=softpurple, line width=3pt}
            \addlegendimage{darkpeach,fill=darkpeach, line width=3pt}
        \end{axis}
    \end{tikzpicture}
    \ref{static:f1:legend} 
    \hfill

    \raisebox{2.2cm}{\rotatebox{90}{F1-score}}
    \begin{subfigure}{0.24\linewidth}
        \centering
        \begin{tikzpicture}
            \begin{axis}
            [ybar=0.01cm, 
            bar width=0.04cm,
            width=0.85\textwidth,  
            enlarge x limits=0.15, 
            enlarge y limits=false,
            ylabel near ticks,
            xlabel near ticks,
            scale only axis,
            ymin=0,
            ymax=1,
            yticklabel style={font=\tiny},
            xtick=data,
            xticklabels={LastFM, Flickr, DE, RU, PT, FR, EN},
            xticklabel style={xshift=1ex, rotate=45, anchor=east, font=\tiny},
            xlabel={GAT},
            x label style={at={(axis description cs:0.5,-0.2)},anchor=north},
            y label style={font=\small, at={(axis description cs:-0.1,0.5)}},
            major x tick style={opacity=0},   
            minor x tick num=1,
            minor tick length=1 ex,
            ymajorgrids= true,
            xmajorgrids= true,
            grid style={white},
            axis background/.style={fill=gray!15},
            axis line style={white, line width=0pt},
            xtick style={draw=none},
            ytick style={draw=none},
            ]
            \addplot [style={softred, fill = softred}] coordinates {
            (1,0.31) 
            (2,0.34) 
            (3,0.28) 
            (4,0.21)
            (5,0.32)
            (6,0.28)
            (7, 0.14)
            };
            
            \addplot [style={softblue, fill=softblue}] coordinates {
                (1,0.73) 
                (2,0.86) 
                (3,0.56) 
                (4,0.45)
                (5,0.65)
                (6, 0.51)
                (7, 0.80)
            };
            \addplot [style={softgreen, fill=softgreen}] coordinates {
                (1,0.73) 
                (2,0.99) 
                (3,0.81) 
                (4,0.82)
                (5,0.75)
                (6, 0.83)
                (7, 0.75)
            };
            \addplot [style={softyellow, fill=softyellow}] coordinates {
                (1,0.84) 
                (2,0.99) 
                (3,0.88) 
                (4,0.91)
                (5, 0.82)
                (6, 0.96)
                (7, 0.95)
            };
            \addplot [style={softcyan, fill=softcyan}] coordinates {
                (1, 0.53) 
                (2,0.79)   
                (3,0.57) 
                (4,0.51)
                (5,0.51)
                (6, 0.47)
                (7,0.62)
            };
            \addplot [style={softpurple, fill=softpurple}] coordinates {
                (1,0.42) 
                (2,0.29) 
                (3,0.28) 
                (4,0.17)
                (5,0.25)
                (6,0.28)
                (7, 0.15)
            };
            \addplot [style={darkpeach, fill = darkpeach}] coordinates {
                (1,0.67) 
                (2,0.77) 
                (3,0.65) 
                (4,0.55)
                (5,0.76)
                (6, 0.62)
                (7, 0.76)
            };
            \end{axis}
        \end{tikzpicture}
        \label{fig:F1_GAT}
    \end{subfigure}
    \hfill
    \begin{subfigure}{0.24\textwidth}
        \centering
        \begin{tikzpicture}
            \begin{axis}
            [ybar=0.01cm, 
            bar width=0.04cm,
            width=0.85\textwidth,  
            enlarge x limits=0.15, 
            enlarge y limits=false,
            ylabel near ticks,
            xlabel near ticks,
            scale only axis,
            ymin=0,
            ymax=1,
            yticklabel style={font=\tiny},
            xtick=data,
            xticklabels={LastFM, Flickr, DE, RU, PT, FR, EN},
            xticklabel style={xshift=1ex, rotate=45, anchor=east, font=\tiny},
            xlabel={GCN},
            x label style={at={(axis description cs:0.5,-0.2)},anchor=north},
            y label style={font=\tiny, at={(axis description cs:-0.1,0.5)}},
            major x tick style={opacity=0},
            minor x tick num=1,
            minor tick length=1 ex,
            ymajorgrids= true,
            xmajorgrids= true,
            grid style={white},
            axis background/.style={fill=gray!15},
            axis line style={white, line width=0pt},
            xtick style={draw=none},
            ytick style={draw=none}
            ]
            \addplot [style={softred, fill = softred}] coordinates {
            (1,0.36) 
            (2,0.32) 
            (3,0.26) 
            (4, 0.22)
            (5,0.32)
            (6,0.25)
            (7, 0.14)
            };
            
            \addplot [style={softblue, fill=softblue}] coordinates {
                (1,0.77) 
                (2,0.91) 
                (3,0.58) 
                (4,0.68)
                (5,0.69)
                (6, 0.62)
                (7, 0.89)
            };
            \addplot [style={softgreen, fill=softgreen}] coordinates {
    
                (1,0.80) 
                (2,0.99) 
                (3,0.98) 
                (4,0.90)
                (5,0.91)
                (6, 0.88)
                (7, 0.97)
            };
            
            \addplot [style={softyellow, fill=softyellow}] coordinates {
                (1,0.87) 
                (2,0.99) 
                (3,1.0) 
                (4,0.95)
                (5,0.99)
                (6, 0.99)
                (7, 0.97)
            };
            
            \addplot [style={softcyan, fill=softcyan}] coordinates {
                (1, 0.76) 
                (2,0.95) 
                (3,0.86) 
                (4,0.80)
                (5,0.79)
                (6, 0.88)
                (7, 0.86)
            };
            
            \addplot [style={softpurple, fill=softpurple}] coordinates {
                (1,0.49) 
                (2,0.33) 
                (3,0.27) 
                (4,0.19)
                (5,0.29)
                (6,0.24)
                (7, 0.14)
            };
            \addplot [style={darkpeach, fill = darkpeach}] coordinates {
                (1,0.82) 
                (2,0.95) 
                (3,0.82) 
                (4,0.85)
                (5,0.85)
                (6, 0.75)
                (7, 0.92)
            };
            \end{axis}
            \end{tikzpicture}
            \label{fig:F1_GCN}
    \end{subfigure}
    \hfill
    \begin{subfigure}{0.24\textwidth}
        \centering
        \begin{tikzpicture}
            \begin{axis}
            [ybar=0.01cm, 
            bar width=0.04cm,
            width=0.85\textwidth,  
            enlarge x limits=0.15, 
            enlarge y limits=false,
            ylabel near ticks,
            xlabel near ticks,
            scale only axis,
            ymin=0,
            ymax=1,
            yticklabel style={font=\tiny},
            xtick=data,
            xticklabels={LastFM, Flickr, DE, RU, PT, FR, EN},
            xticklabel style={xshift=1ex, rotate=45, anchor=east, font=\tiny},
            xlabel={GIN},
            x label style={at={(axis description cs:0.5,-0.2)},anchor=north},
            y label style={font=\tiny, at={(axis description cs:-0.1,0.5)}},
            major x tick style={opacity=0},
            minor x tick num=1,
            minor tick length=1 ex,
            ymajorgrids= true,
            xmajorgrids= true,
            grid style={white},
            axis background/.style={fill=gray!15},
            axis line style={white, line width=0pt},
            xtick style={draw=none},
            ytick style={draw=none}
            ]
                
            \addplot [style={softred, fill = softred}] coordinates {
                (1, 0.31) 
                (2,0.23) 
                (3,0.31) 
                (4,0.24)
                (5,0.35)
                (6, 0.25)
                (7, 0.19)
            };
   
            \addplot [style={softblue, fill=softblue}] coordinates {
                (1,0.75) 
                (2,0.86) 
                (3, 0.41) 
                (4,0.41)
                (5, 0.44)
                (6, 0.44)
                (7, 0.90)
            };
            \addplot [style={softgreen, fill=softgreen}] coordinates {
                (1,0.97) 
                (2, 1.0) 
                (3, 0.86) 
                (4, 0.93)
                (5, 0.91)
                (6, 0.92)
                (7, 0.99)
            };
            
            \addplot [style={softyellow, fill=softyellow}] coordinates {
                (1,0.97) 
                (2, 1.0) 
                (3, 0.99) 
                (4, 0.99)
                (5, 0.99)
                (6, 0.99)
                (7, 0.99)
            };
            \addplot [style={softcyan, fill=softcyan}] coordinates {
                (1, 0.86) 
                (2, 0.67) 
                (3, 0.78) 
                (4, 0.88)
                (5, 0.80)
                (6, 0.75)
                (7, 0.90)
            };
            \addplot [style={softpurple, fill=softpurple}] coordinates {
                (1, 0.50) 
                (2, 0.28) 
                (3, 0.23) 
                (4, 0.28)
                (5, 0.30)
                (6, 0.23)
                (7, 0.19)
            };
            \addplot [style={darkpeach, fill = darkpeach}] coordinates {
                (1,0.) 
                (2,0.) 
                (3,0.) 
                (4,0.)
                (5,0.)
                (6, 0.)
                (7, 0.)
            };
            \end{axis}
            \end{tikzpicture}
            \centering 
            \label{fig:F1_GIN}
    \end{subfigure}%
    \hfill
    \begin{subfigure}{0.24\textwidth}
        \centering 
        \begin{tikzpicture}
            \begin{axis}
            [ybar=0.01cm, 
            bar width=0.04cm,
            width=0.85\textwidth,  
            enlarge x limits=0.15, 
            enlarge y limits=false,
            ylabel near ticks,
            xlabel near ticks,
            scale only axis,
            ymin=0,
            ymax=1,
            yticklabel style={font=\tiny},
            xtick=data,
            xticklabels={LastFM, Flickr, DE, RU, PT, FR, EN},
            xticklabel style={xshift=1ex, rotate=45, anchor=east, font=\tiny},
            xlabel={SAGE},
            x label style={at={(axis description cs:0.5,-0.2)},anchor=north},
            y label style={font=\tiny, at={(axis description cs:-0.1,0.5)}},
            major x tick style={opacity=0},
            minor x tick num=1,
            minor tick length=1 ex,
            ymajorgrids= true,
            xmajorgrids= true,
            grid style={white},
            axis background/.style={fill=gray!15},
            axis line style={white, line width=0pt},
            xtick style={draw=none},
            ytick style={draw=none}
            ]
            \addplot [style={softred, fill = softred}] coordinates {
                (1, 0.25) 
                (2,0.19) 
                (3,0.27) 
                (4,0.20)
                (5,0.29)
                (6, 0.26)
                (7, 0.14)
            };
    
            \addplot [style={softblue, fill=softblue}] coordinates {
                (1,0.75) 
                (2,0.93) 
                (3, 0.41) 
                (4, 0.59)
                (5, 0.50)
                (6, 0.54)
                (7, 0.60)
            };
            \addplot [style={softgreen, fill=softgreen}] coordinates {
                (1,0.79) 
                (2,0.83) 
                (3,0.64) 
                (4,0.84)
                (5, 0.59)
                (6, 0.66)
                (7, 0.88)
            };
            \addplot [style={softyellow, fill=softyellow}] coordinates {
                (1,0.82) 
                (2,0.99) 
                (3, 0.86) 
                (4, 0.94)
                (5, 0.93)
                (6, 0.94)
                (7, 0.96)
            };
            \addplot [style={softcyan, fill=softcyan}] coordinates {
                (1, 0.72) 
                (2,0.92) 
                (3,0.80) 
                (4,0.77)
                (5, 0.77)
                (6, 0.79)
                (7, 0.83)
            };
            \addplot [style={softpurple, fill=softpurple}] coordinates {
                (1,0.35) 
                (2,0.21) 
                (3,0.26) 
                (4,0.20)
                (5, 0.26)
                (6, 0.25)
                (7, 0.15)
            };
            \addplot [style={darkpeach, fill = darkpeach}] coordinates {
                (1,0.79) 
                (2,0.94) 
                (3, 0.77) 
                (4, 0.82)
                (5, 0.77)
                (6, 0.73)
                (7, 0.80)
            };
            \end{axis}
            \end{tikzpicture}
            \label{fig:f1_SAGE}
    \end{subfigure}
    \caption{{Comparative analysis of attack performances: the graph illustrates average F1-scores from all attacks, computed across seven datasets and four types of GNN models. Each score is averaged from attacks on 100 target nodes. 
    }}
    \label{fig:attack_comparasion_f1}
\end{figure*}

\begin{table*}[! h]
    \centering
    \tiny
    \caption{{AUC metrics for various attacks: rows signify datasets, while columns illustrate attacks against different models.}}
    \label{tab:attack_performance:auc}
    \setlength\tabcolsep{6pt} 
    \begin{tabularx}{\textwidth}{l *{4}{c} *{4}{c} *{4}{c} *{4}{c} *{4}{c}} 
        \toprule
        \multirow{2}{*}{\textbf{Dataset}}  &  \multicolumn{4}{c}{\textbf{INF}} &  \multicolumn{4}{c}{\textbf{INF--DIR}} &  \multicolumn{4}{c}{\textbf{INF--MAG}}  &  \multicolumn{4}{c}{\textbf{LTA}}  &  \multicolumn{4}{c}{\textbf{IIA}}  \\
        \cmidrule(lr){2-5} \cmidrule(lr){6-9} \cmidrule(lr){10-13} \cmidrule(lr){14-17} \cmidrule(lr){18-21}
        & \textbf{GCN} & \textbf{SAGE} & \textbf{GAT} & \textbf{GIN} 
        & \textbf{GCN} & \textbf{SAGE} & \textbf{GAT} & \textbf{GIN} 
        & \textbf{GCN} & \textbf{SAGE} & \textbf{GAT} & \textbf{GIN} 
        & \textbf{GCN} & \textbf{SAGE} & \textbf{GAT} & \textbf{GIN} 
        & \textbf{GCN} & \textbf{SAGE} & \textbf{GAT} & \textbf{GIN} \\
        \midrule
        \texttt{LastFM}  
        &           0.81       & 0.76    & 0.79  & 0.77
        &           0.98       & 0.98    & 0.98  & \textbf{0.97} 
        &         \textbf{0.99}       & \textbf{0.99}   & \textbf{0.99}  & \textbf{0.97} 
        &           0.98       & 0.98    & 0.92  & 0.86
        &           0.98       & 0.98    & 0.91  & - \\
        \texttt{Flickr}  
        &           0.98       & 0.97    &   0.97   & 0.95
        &            \textbf{0.99}     &\textbf{0.99}     &   0.95   & 0.99
        &        \textbf{0.99}        & \textbf{0.99}    &  \textbf{0.99}  & \textbf{0.99} 
        &         \textbf{0.99}         & \textbf{0.99}     &   0.94   & 0.94
        &             0.98     & \textbf{0.99}    &   0.93   & - \\
       \texttt{Twitch-EN}  
       &          0.98     & 0.97    &   0.97    &  0.90
       &            \textbf{0.99}  & \textbf{0.99}     & 0.95      &  \textbf{0.99} 
       &       \textbf{0.99}       & \textbf{0.99}    & \textbf{0.99}      &  \textbf{0.99}
       &       \textbf{0.99}       & \textbf{0.99}   & 0.94      &  0.90
       &        \textbf{0.99}       & 0.98    & 0.92      &  - \\
        \texttt{Twitch-DE}  
        &            0.91   & 0.64  & 0.82 &  0.84
        &        \textbf{0.99}       & 0.93  & 0.94 &  \textbf{0.99} 
        &        \textbf{0.99}       & \textbf{0.96 } & \textbf{0.97} &  0.98
        &              0.93 & 0.93  & 0.76 &  0.91
        &              0.95 & 0.95  & 0.88 &  - \\
       \texttt{Twitch-FR}  
       &         0.98    & 0.97  &   0.97 &  0.70
       &         \textbf{0.99}     & \textbf{0.99}   & 0.95   &  \textbf{0.99} 
       &        \textbf{0.99}     & \textbf{0.99}   & \textbf{0.99}   &  \textbf{0.99} 
       &        \textbf{0.99}    &\textbf{0.99}   & 0.94   &  0.89
       &         \textbf{0.99}   & \textbf{0.99}   & 0.95   &  - \\
       \texttt{Twitch-PT}  
       &           0.98     & 0.97 &   0.97 & 0.71
       &              \textbf{0.99}  & \textbf{0.99}  & 0.95   & \textbf{0.99} 
       &        \textbf{0.99}  & \textbf{0.99}  & \textbf{0.99}   & \textbf{0.99} 
       &        \textbf{0.99}     & \textbf{0.99}     & 0.94  & 0.95
       &         \textbf{0.99}     &\textbf{0.99}    & 0.94  & - \\
        \texttt{Twitch-RU}  
        &      0.91     & 0.77    &  0.79   & 0.71
        &    \textbf{0.99}     & 0.96    &  0.95   & \textbf{0.99} 
        &   \textbf{0.99}       &\textbf{0.99}  &\textbf{0.99}   &\textbf{0.99} 
        &         0.91  & 0.96    &  0.79   & 0.95
        &         0.96  & 0.96    &  0.85   & - \\
        \bottomrule
    \end{tabularx}
\end{table*}

\noindent\textbf{Results.} 
The F1-scores for all attack methods are illustrated in Figure \ref{fig:attack_comparasion_f1}, and the AUCs are reported in Table \ref{tab:attack_performance:auc}. 
Due to space constraints, only the AUCs for influence-based attacks are included in the main paper.
Additional results are detailed in Appendix~\ref{app:similarity:auc}. 
For SIM and $INF\text{--}DIR$, in which we require using distance metrics, we present their results only with the best-performing metrics, correlation distance for SIM and Bray–Curtis distance for $INF\text{--}DIR$, due to space limitations. Refer to Appendix~\ref{app:evaluation:sim:full} for full results.

Reviewing the results, similarity-based attacks like LSA and SIM consistently perform poorly, with F1-scores around 0.2. This is due to their reliance on the assumption that nodes in close proximity share similar features and, therefore, similar output probabilities—an assumption that often fails across diverse datasets.
While LTA shows good results with GCN (F1-score surpassing 0.8), it demonstrates diminished efficacy against GAT (F1-score drops to $\sim$0.5). 
The results are anticipated because LTA does not account for variations in message aggregation methods. While GCNs use \emph{mean} as the aggregation function, which aligns with the premises of this work, this approach is not supported by GATs, which aggregate messages differently.
Similarly, IIA performs well against GCN but fails against GIN and shows suboptimal results against GAT. 
These outcomes suggest that both attacks struggle to adapt to models employing diverse message aggregation mechanisms. 
INF, though use the same methodology as LTA, is less effective due to its lack of direct access to target node pairs. This shows that simply extending LTA's method to our more challenging threat model is inadequate.

Our $INF\text{--}DIR$ attack demonstrates greater consistency and superior performance than LTA and IIA. 
It performs well when evaluated against GCN and GIN (F1-score surpassing 0.8 and AUC surpassing 0.95) and maintains its effectiveness when tested with GAT (F1-score around 0.8 and AUC surpassing 0.95). These results highlight the potential of investigating the pattern of output changes.

The $INF\text{--}MAG$ attack consistently surpasses competing strategies in terms of performance, demonstrating superiority across various model architectures and datasets. It reliably secures F1-scores exceeding 0.8 and frequently surpasses 0.9 in several tests. Moreover, it attains AUC scores as high as 0.99 in most evaluations. This robust performance is primarily credited to its strategic neutralization of weight impacts. By utilizing an anchor node, the $INF\text{--}MAG$ attack effectively minimizes the influence of weights, thereby accentuating distance as the key determinant in the influence's magnitude.

\added{Existing methods occasionally match our performance on certain model–dataset combinations, particularly with GCNs. As discussed in Limitation 2 (Section~\ref{sec:existing_limitation}), these methods do not account for the per-node weights during aggregation, making them less general. However, in settings where graph edges are evenly distributed and the model is a GCN using mean aggregation, each node’s weight is also relatively uniform. Under these conditions, methods that ignore weights can still yield strong results. More broadly, this illustrates that specific model–dataset pairings may be especially vulnerable to attack, allowing even simple techniques to accurately infer edges.
} 

\added{We provide attack results averaged over 500 target nodes in Appendix~\ref{app:more_nodes}, which show a similar pattern to those presented in the main paper.
}

\definecolor{softgrey}{RGB}{211, 211, 211} 
\subsection{Dynamic Graph Scenario}
\label{sec:evo}

\noindent\textbf{Datasets.}
We use two real-world dynamic graph datasets—tgbl-flight \cite{huang2023temporal} and Dgraph-fin \cite{huang2022dgraph}—converted into weekly static graph snapshots. tgbl-flight represents flight data, with nodes as airports and edges as airline connections, while Dgraph-fin is a financial dataset with nodes as users and edges representing emergency financial connections. These datasets primarily capture structural changes, such as the addition of edges. Further details are provided in Appendix~\ref{app:evaluation:data_statistics}.

To analyze dynamic scenarios, we create synthetic datasets by introducing artificial changes into static graphs (LastFM and Twitch-EN). This setup includes node- and structure-level dynamics, enabling controlled experiments. These synthetic graphs provide deeper insights into how adversarial modifications interact with natural graph evolution and test attack robustness under diverse conditions.

\noindent\textbf{Synthetic Dynamics Design.} 
The evolution rate ($e$) determines the extent of changes, adjusting node features within [$x(1-e)$, $x(1+e)$] and proportionally adding nodes and edges within the target node’s neighborhood. In the experiment, we apply global node feature changes, meaning all nodes in the graph experience random feature modifications. Structural changes are restricted to the 2-hop neighborhood to ensure they occur near the target node, directly impacting its local structure.

\noindent\textbf{Target Node Selection.} 
For the synthetic datasets, we randomly select 100 target nodes. For real-world datasets, we sample 100 nodes near structural changes, ensuring the tests reflect realistic dynamic conditions rather than static graph scenarios. 

\noindent\textbf{Baseline Attacks.}
We benchmark $INF-DIR^*$ against influence-based attacks, including LTA, IIA, $INF-MAG$, and $EVO$, which follows $INF-DIR^*$'s methodology but excludes adversarial influence generation. Similarity-based attacks are excluded due to their ineffectiveness in static settings and limited relevance in dynamic environments.
To evaluate query frequency, we tested $INF-DIR^*$ and $EVO$ with up to 15 queries, as discussed in Section \ref{dis:number_queries}. The main paper presents the best performance metrics of $INF-DIR^*$ using Concatenation preprocessing, with full results and variations available in Appendix~\ref{app:dynamic:distance_metrics_selection}.
For fairness, LTA, IIA, and $INF-MAG$ were adapted to use 15 queries, summing scores across queries, and are denoted as LTA$^*$, IIA$^*$, and $INF-MAG^*$.
All queries were conducted with natural graph updates, aligning adversarial manipulations with real-time graph evolution. More queries also introduce additional noise from the graph's natural changes.

\noindent\textbf{Evalution Method: Signal-to-Noise Ratio.}
We evaluate attack performance in dynamic scenarios using a Signal-to-Noise Ratio (SNR) framework, which quantifies the effectiveness of adversarial perturbations (signal) against natural graph evolution (noise). Natural changes in dynamic graphs can obscure or mimic adversarial modifications, complicating the isolation of attack impacts. By treating the perturbation rate as the signal and the graph change rate as noise, the SNR framework systematically analyzes the robustness of attack methods under varying dynamics.
Experiments varied three key parameters:

    \textbullet\ Perturbation Rate: With the feature change rate fixed at 0.02 and the local structural change rate at 0.2, the perturbation rate was adjusted (e.g., 0.1–1.1). For feature perturbation attacks, this defines the magnitude of changes to the target node’s features. For edge addition attacks (e.g., IIA), it corresponds to the fraction of new edges added relative to the target node’s degree.

    \textbullet\ Feature Change Rate: With the perturbation rate fixed at 1 and the local structural change rate at 0.2, the feature change rate varied (e.g., 0.00001–0.05).

     \textbullet\ Structural Change Rate: With the perturbation rate fixed at 1 and the feature change rate at 0.02, the local structural change rate varied (e.g., 0.1–0.6).

\begin{figure*}[ht]
    \centering

        \caption{
            \parbox[t]{1.2\linewidth}{\centering A. Varying Perturbation Rate: \\ Higher Perturbation Rate → Higher SNR.}
        }
    \end{subfigure}
    \hspace{15mm}%
    \begin{subfigure}[t]{0.25\textwidth}
    \centering 
        %
            \caption{
            \parbox[t]{1.2\linewidth}{\centering B. Varying Feature Change Rate: \\ Higher Change Rate → Lower SNR.}
        }
    \end{subfigure}
    \hspace{15mm}%
    \begin{subfigure}[t]{0.25\textwidth}
        \centering 

        %
        \caption{
            \parbox[t]{1.2\linewidth}{\centering  C. Varying Local Structure Change Rate: \\ Higher Change Rate → Lower SNR.}
        }
    \end{subfigure}
    \caption{Signal-to-Noise Ratio analysis on GCN with the LastFM Dataset.}
    \label{fig:snr}
\end{figure*}

\begin{figure*}[!ht]
    \centering
    %
        \caption{\hspace{5pt}GAT}
    \end{subfigure}%
    \hfill
    \begin{subfigure}{0.24\textwidth}
        \centering
        %
        \caption{\hspace{5pt}GCN}
    \end{subfigure}%
    \hfill
    \begin{subfigure}{0.24\textwidth}
        \centering
        %
            \centering
        \caption{\hspace{5pt}GIN}
        \label{fig:F1_GIN_f1}
    \end{subfigure}%
    \hfill
    \begin{subfigure}{0.24\textwidth}
        \centering 
        %
        \caption{\hspace{5pt}SAGE}
        \label{fig:f1_SAGE_f1}
    \end{subfigure}
    \caption{F1-score of attacks in dynamic scenarios.}
    \label{fig:attack:dynamic:neighbor_change:f1:new}
\end{figure*}

\begin{figure}[!h]
    \centering
    %
    \caption{F1-score of attacks in dynamic scenarios with negative samples constrained to nodes undergoing structural changes on GCN.}
    \label{fig:dynamic:real:changed_nodes}
    \vspace{-0.5cm}
\end{figure}

\noindent\textbf{Signal-to-Noise Ratio Analysis.}
Figure \ref{fig:snr} presents the SNR analysis results on GCN using the LastFM dataset (full results in Appendix~\ref{app:snr}), with similar trends observed across other models. 
Magnitude-based methods show potential for effectiveness at very high SNR levels, where adversarial signals dominate natural noise. However, achieving such a high SNR requires excessive perturbations, which compromise stealth and destabilize the graph, rendering these methods impractical in real-world scenarios.
In contrast, direction-based attacks consistently demonstrate robustness across all SNR levels. By leveraging stable patterns in output changes, they remain effective even in noisy and dynamic environments, proving their reliability and practicality for evolving graph scenarios.

    Figure \ref{fig:snr} A (Varying Perturbation Rate): Magnitude-based methods perform poorly at all perturbation rates, as higher perturbations amplify noise and risk detection without improving performance. Direction-based attacks consistently outperform magnitude-based ones, requiring fewer perturbations to remain effective.
    
    Figure \ref{fig:snr} B (Varying Feature Change Rate): Direction-based attacks generally outperform magnitude-based methods. Magnitude-based methods show good performance at very low feature change rates, consistent with their strong performance in static scenarios. This is because structural changes are sparse (rate of 0.2), affecting only a subset of nodes, while feature changes are global, impacting all nodes. Limited structural noise enables magnitude-based attacks to perform well initially, but as feature change rates increase, noise overshadows adversarial signals, causing degradation, while direction-based attacks remain robust. 
    
    Figure \ref{fig:snr} C (Varying Structural Change Rate): At high structural change rates, all attacks degrade significantly as excessive structural changes overshadow adversarial influence. However, direction-based methods still outperform magnitude-based ones, maintaining higher F1-scores (down to 0.4) despite the noise from widespread structural changes.

\noindent\textbf{Performance Evaluation on Real-World and Synthetic Datasets.}
Building on the insights from the SNR analysis, we extended our evaluation to a broader set of real-world and synthetic datasets to assess the generalizability of our findings. Here we only show the result for $INF-MAG^*$, LTA$^*$, and IIA$^*$ for space limitation and for they outperform the original version. 
For synthetic graphs, we used a feature change rate of 0.02, a structural change rate of 0.2 and fixed the perturbation rate at 1. These values align with the controlled conditions in the SNR analysis, providing a consistent baseline for evaluating attack performance. Figure \ref{fig:attack:dynamic:neighbor_change:f1:new} shows the F1-scores. Other metrics can be found in Appendix~\ref{app:dynamic:results}. 

In real-world datasets, where only structural changes occur, $INF-MAG^*$, LTA$^*$, and IIA$^*$ appear to perform well, consistent with the SNR analysis where static node features support strong attack performance. However, this performance is misleading as it lacks stability. To investigate further, we evaluated attack performance exclusively on nodes undergoing structural changes. 
The results, shown in Figure \ref{fig:dynamic:real:changed_nodes}, reveal poor performance for magnitude-based attacks. In contrast, $INF-DIR^*$ remains robust, effectively handling structural noise. The full result can be found in Appendix~\ref{app:dynamic:changed_nodes}.
For synthetic graphs, $INF-DIR^*$ continues to perform robustly across all models, demonstrating adaptability and effectiveness even in highly dynamic environments. Magnitude-based attacks, however, fail under these conditions, which is consistent with the SNR analysis.

In conclusion, the effectiveness of magnitude-based attacks heavily depends on whether target nodes are affected by natural graph changes. Nodes unaffected by changes maintain performance comparable to static scenarios. However, for nodes undergoing changes, magnitude-based attacks fail, highlighting their lack of robustness. Adversaries would need prior knowledge of target node stability to achieve consistent results. In contrast, direction-based attacks remain reliable and effective, showcasing resilience under evolving graph conditions.

\begin{table}[!h]
    \centering
    \caption{Performance of $INF\text{--}MAG$ (static scenario) and $INF\text{--}DIR^*$ (dynamic scenario) with different threshold selections on the LastFM dataset.}
    \label{tab:dynamic:different_threshold}
    \scriptsize
    \setlength{\tabcolsep}{6pt}  
    \begin{tabular}{llSSSS}  
        \toprule
        \multirow{2}{*}{\textbf{Threshold}} & \multirow{2}{*}{\textbf{Model}} 
        & \multicolumn{2}{c}{\textbf{Static}} & \multicolumn{2}{c}{\textbf{Dynamic}} \\ 
        \cmidrule(lr){3-4} \cmidrule(lr){5-6}
        & & {\textbf{Recall}} & {\textbf{Precision}} & {\textbf{Recall}} & {\textbf{Precision}} \\
        \midrule
        \multirow{4}{*}{In-graph}   
        & GCN  & 0.65 & 0.99 & 0.67 & 0.63  \\  
        & SAGE & 0.68 & 0.99 & 0.66 & 0.54   \\ 
        & GAT  & 0.68 & 0.94 & 0.67 & 0.54  \\ 
        & GIN  & 0.70 & 0.99 & 0.65 & 0.80    \\ 
        \midrule
        \multirow{4}{*}{Knowledge}     
        & GCN  & 0.76 & 0.96 &  0.67 & 0.88  \\  
        & SAGE & 0.96 & 0.90 & 0.66 & 0.82 \\ 
        & GAT  & 0.92 & 0.88 & 0.57 & 0.74 \\ 
        & GIN  & 0.86 & 0.92 & 0.66 & 0.92  \\ 
        \bottomrule
    \end{tabular}
\end{table}

\vspace{-0.3cm}
\subsection{Threshold Selection Methods}

\deleted{Table \ref{tab:dynamic:different_threshold} compares attack performance using in-graph and knowledge-based thresholds. Results for $INF-MAG$ in static and $INF-DIR^*$ in dynamic scenarios are shown on the LastFM dataset. 
}

\deleted{
As expected, static scenarios show better metrics, consistent with findings that they are more vulnerable to attacks.
As expected, static scenarios yield better metrics, which aligns with previous findings that static graphs are generally more susceptible to attacks. For in-graph thresholds, we assume that the adversary knows 20\% of the ground-truth scores for connected nodes. In contrast, knowledge-based thresholds rely on selecting the optimal threshold derived from the model’s output similarity to the target node. The rationale is that GNNs encode structural information within their embeddings, so nodes with similar outputs often share comparable local structures—and therefore require similar thresholds.
Overall, Knowledge-based thresholds achieve better performance. The difference between these two methods reflects the adversary's knowledge level about the target's score distribution—partial (20\%) for in-graph thresholds and a more comprehensive distribution for knowledge-based thresholds derived from similar nodes. 
Both approaches achieve acceptable precision and recall, highlighting the feasibility of using score distribution exploration to set thresholds under different threat models.
Additional results and analyses, including evaluations of the most constrained scenario where no extra knowledge is used with in-graph thresholding, are provided in Appendix~\ref{app:threshold}. Refining threshold estimation methods for varying threat models remains an avenue for future work.}

\added{Table~\ref{tab:dynamic:different_threshold} presents attack performance using in‐graph and knowledge‐based thresholds for both the static ($INF$--$MAG$) and dynamic ($INF$--$DIR$*) scenarios on the LastFM dataset.

We stress that these two thresholding approaches rely on different assumptions about the adversary’s knowledge, so the goal is not to rank or directly compare them, but rather to demonstrate that each can yield a usable threshold for an attack.
For in‐graph thresholds, we assume the adversary knows scores for 20\% of the target’s connected nodes. By contrast, knowledge‐based thresholds select an optimal threshold derived from model‐output similarity to the target node, leveraging the idea that GNN embeddings encode structural properties, so nodes with similar outputs often share comparable local structures—and thus require similar thresholds.

As expected, static scenarios yield stronger metrics, aligning with previous findings that static graphs are generally more vulnerable to attacks. In dynamic scenarios, however, the attack shows lower performance due to the inherent challenges in handling evolving graph structures. These observations underscore that an effective attack must not only reliably extract information from the model but also employ a robust method for threshold selection.

Additional experiments and analyses—covering an even more constrained scenario with no extra knowledge for in‐graph thresholding and results on the Twitch/EN dataset—are provided in Appendix~\ref{app:threshold}. Future work includes refining threshold selection techniques to handle a wider range of adversarial assumptions and threat models.
}




\definecolor{mygreen}{RGB}{34, 139, 34} 
\definecolor{myblue}{RGB}{100, 149, 237} 
\definecolor{myred}{RGB}{205, 92, 92}    
\definecolor{mycyan}{RGB}{64, 224, 208}  
\begin{figure*}[t]
    \centering

    \ref{lg:dp:defense}
    \hfill
    
    \raisebox{1.2cm}{\rotatebox{90}{\small F1 score}}%
    \begin{minipage}[b]{0.46\textwidth}
    \centering
    \captionsetup[subfigure]{labelformat=empty, font=small}
    \captionsetup[subfigure]{labelformat=empty, font=tiny} 
    \begin{subfigure}{0.22\textwidth}
        \centering
        \caption{\hspace{20pt}GAT}
        %
    \end{subfigure} 
    \begin{subfigure}{0.22\textwidth}
        \centering
        \caption{\hspace{20pt}GCN}
        %

    \end{subfigure} 
\begin{subfigure}{0.22\textwidth}
        \centering
        \caption{\hspace{20pt}GIN}
        %
    \end{subfigure}
 \begin{subfigure}{0.22\textwidth}
        \centering
        \caption{\hspace{20pt}GraphSage}
        %
    \end{subfigure}
    \vspace{-0.2cm}
    \caption*{\vspace{-0.2cm} EdgeRand}
    \end{minipage}%
    \begin{minipage}[b]{0.46\textwidth}
        \centering
        \captionsetup[subfigure]{labelformat=empty, font=small}
        \captionsetup[subfigure]{labelformat=empty, font=tiny} 
    \begin{subfigure}{0.22\textwidth}
    \centering
    \caption{\hspace{20pt}GAT}
    %

    \end{subfigure} 
\begin{subfigure}{0.22\textwidth}
        \centering
        \caption{\hspace{20pt}GCN}
        %

    \end{subfigure} 
\begin{subfigure}{0.22\textwidth}
    \centering
    \caption{\hspace{20pt}GIN}
    %
    \phantomsubcaption
    \end{subfigure}
    \begin{subfigure}{0.22\textwidth}
    \centering
    \caption{\hspace{20pt}GraphSage}
    %
    \end{subfigure}
    \vspace{-0.2cm}
    \caption*{\vspace{-0.2cm} LapGraph}
    \end{minipage}
    \caption{Evaluating $INF\text{--}MAG$ attack performance on LastFM dataset using EdgeRand and LapGraph defenses.}
    \label{fig:def:edgerand:lastfm}
\end{figure*}

 \begin{figure}[ht]
     \centering
    %
            \vspace{-0.2cm}
            \caption{\vspace{-0.2cm} \hspace{15pt} \footnotesize EdgeRand}
        \end{subfigure}
        \begin{subfigure}{0.20\textwidth}
        \centering
            %
            \vspace{-0.2cm}
            \caption{\vspace{-0.2cm}\hspace{15pt} \footnotesize LapGraph}
        \end{subfigure}
     \caption{F1 score of $INF\textit{--}DIR^*$ against DP on LastFM.}
     \label{fig:dynamic:defense:dp}
 \end{figure}

\section{Evaluation Against Defenses}
To assess the resilience of our attacks, we implement two cutting-edge defense approaches based on Differential Privacy (DP), EdgeRand \cite{Inference_Attacks:link_steal:wu2021linkteller} and LapGraph \cite{Inference_Attacks:link_steal:wu2021linkteller}, which introduce noise into the graph to obscure its actual structure. These methods are well-suited for general GNN architectures, unlike defenses such as GAP and LPGNet \cite{sajadmanesh2023gap, def:kolluri2022lpgnet}, which require specific model designs, or Grid \cite{lou2024grid}, which targets the training phase rather than inference-time attacks like ours. 
Specifically, EdgeRand disrupts the adjacency matrix by randomly flipping its entries based on a Bernoulli distribution, while LapGraph employs the Laplacian mechanism to modify the entire adjacency matrix, ensuring that the graph's overall density is preserved.
For LapGraph, we varied the privacy budget $\epsilon$ between 2 and 10. For EdgeRand, the budget ranged from 5 to 10, as using a budget below 5 resulted in out-of-memory issues. We trained various models on these perturbed graphs and evaluated the effectiveness of our attack against them. 
Note that particular attention should be given to the attack performance when the privacy budget is \emph{8} or higher, as models in this range outperform an MLP that relies solely on node features for predictions.
The model utility and more discussion about the defense can be found in Appendix~\ref{app:dp:model_utility}. 

\noindent\textbf{Static Graph Scenario.} 
Figure \ref{fig:def:edgerand:lastfm} presents a comparative evaluation of our $INF-MAG$ attack and LTA, for a range of privacy budgets on the LastFM Dataset. 
The results highlight the difficulty in balancing model utility and privacy. When the privacy budget is low ($\epsilon < 6$), attack performance is poor. However, at higher privacy budgets ($\epsilon \geq 8$), the models' utility improves, but so does the effectiveness of attacks, indicating a weak defense. 
\added{
\noindent LTA and our attacks operate under different threat models. Note that the defense adds noise to the message passing in the graph, which affects each attack differently. Our method relies on indirect analysis with longer node distances, leading to more accumulated noise but still demonstrating robust performance. In contrast, LTA’s analysis is typically two hops shorter, so it experiences less accumulated noise. When the model is still sufficiently accurate (i.e., the defense uses a higher privacy budget), both methods exhibit similar performance, underlining the robustness of our approach.
}

\noindent\textbf{Dynamic Graph Scenario.}
In this setup, we apply DP to the latest snapshot. The impact of these DP-based defenses on the LastFM dataset is detailed in Figure \ref{fig:dynamic:defense:dp}. The result indicates that our $INF-DIR^*$ attack remains effective in scenarios where the model is operational, especially when the privacy budget exceeds 8.

\noindent\textbf{Possible Defense.}
We propose integrating defenses into the message aggregation process to counter adversarial manipulations. As shown in Section~\ref{app:node_sampling}, a simple node sampling mechanism can disable magnitude-based attacks and significantly reduce the effectiveness of direction-based attacks by ignoring adversarial nodes. Industry practices, such as selective message aggregation \cite{borisyuk2024lignn}, demonstrate that full aggregation is unnecessary, enabling security enhancements without sacrificing utility.
The approach dynamically adjusts aggregation using adaptive filtering and anomaly detection to mitigate adversarial influence. 
\added{Further details and defense strategies, including a discussion on combining graph manipulation and node sampling for defense, are provided in Appendix~\ref{app:new_defense}. }

\section{Discussion}
\label{sec:discussion}

\added{
\noindent\textbf{Scalability}
The scalability of our approach can be analyzed in two dimensions:
    
    \emph{Graph Expansion Scalability:} Our method introduces only three auxiliary nodes, which are reusable across different attacks. The number of added edges scales as $O(3n)$, where $n$ represents the number of target node pairs. 

    \emph{Computational Scalability:} The attack requires $O(n)$ queries to obtain model outputs, followed by statistical analysis.
}

\added{
\noindent\textbf{Perturbation's Impact on Model Utility.}
The perturbations introduced in the graph influence the messages propagated to neighboring nodes, which can degrade the model’s performance. 
Figure~\ref{fig:utility} illustrates the effect of perturbations on model utility. 
As expected, 1-hop neighbors suffer the greatest decline in accuracy, as they directly absorb the modified information. 2-hop nodes experience a moderate degradation in accuracy, while 3-hop and 4-hop nodes remain largely unaffected. These findings align with our theoretical expectations—nodes closer to the perturbation source are more vulnerable to its effects. Higher influence rates worsen performance as randomly added auxiliary nodes disrupt the graph structure.



\begin{figure}[htbp]
    \centering
        \centering
        \includegraphics[width=0.7\columnwidth]{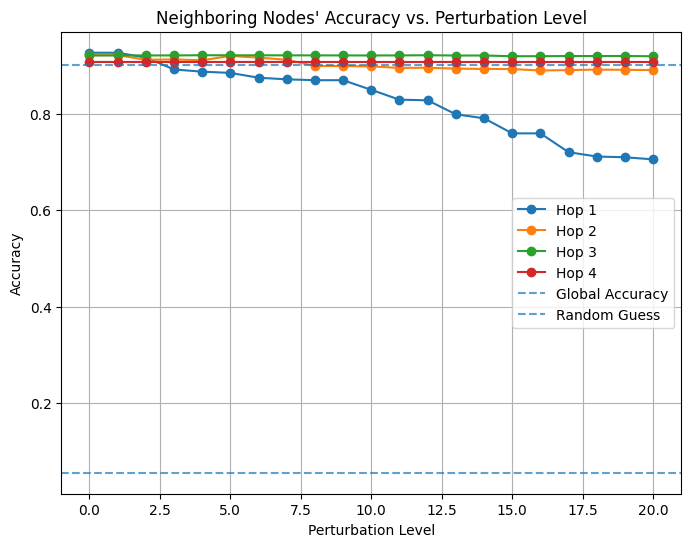}
    \caption{Impact of perturbation on neighboring nodes' performance. Accuracy degradation is highest for 1-hop nodes. Results are averaged over all nodes in the LastFM dataset using GCN.}
    \label{fig:utility}
\end{figure}


}

\noindent\textbf{Imperceptibility of Adversary's Actions.} 
Our attack involves perturbing the graph during the inference phase, making stealth a critical aspect. We evaluate imperceptibility across two dimensions: changes to nodes/edges and alterations to node features. Full results and detailed discussion are in Appendix~\ref{app:imperceptibility}.

\noindent\textbf{Node/Edge Imperceptibility.} 
Traditional graph attack methods rely on fixed perturbation budgets to maintain stealth, which is insufficient in dynamic graphs. Using model explanation techniques like GNNExplainer \cite{ying2019gnnexplainer} and PGExplainer \cite{luo2020parameterized}, we find that adversarially inserted nodes rarely rank among the top influencers for target nodes, demonstrating their stealth. 

\noindent\textbf{Feature Change Imperceptibility.} 
Our experiments reveal that effective attacks often require perturbations exceeding natural graph evolution rates. Distributing changes across multiple nodes could improve concealment, a direction we reserve for future work.

\noindent\textbf{Advanced Influence Generation Method.} One of our future directions is to develop methods for generating distinct patterns that achieve strong performance at lower SNRs. Preliminary experiments reveal that in GAT models, auxiliary nodes with features similar to the target node receive higher weights, enhancing their influence (see Appendix~\ref{app:evaluation:auxiliary_node_features}).


\noindent\textbf{Impact of Node Sampling on Attacks.}
\label{dis:node_sampling}
Node sampling is implemented in industrial GNN applications \cite{borisyuk2024lignn, robinson2024relbench} to improve scalability and efficiency. Thus we explore how node sampling impacts our attacks. Magnitude-based attacks fail when the adversary's nodes are not consistently sampled, as their influence cannot propagate effectively. Direction-based attacks show greater robustness across scenarios but degrade at very low sampling rates due to limited shared information. Introducing unpredictable sampling or excluding adversarial nodes can reduce attack effectiveness, though systems remain vulnerable if adversaries manipulate the sampling process. The result can be found in Appendix~\ref{app:node_sampling}

\noindent\textbf{Number of Queries.}
\label{dis:number_queries}
We evaluated our adapted attacks ($INF-MAG^*$ and $INF-DIR^*$) by varying the number of queries from 1 to 15. For magnitude-based attacks, increasing the number of queries does not improve performance; instead, it amplifies noise, further degrading results.
In contrast, $INF-DIR^*$ outperforms magnitude-based methods even with one query, with performance improving as queries increase, leveraging richer informational patterns. Full results are in Appendix~\ref{app:number_queries}.

\noindent\textbf{Hard-label scenario.}
Our attack methodology can be extended to the hard-label scenario by analyzing the magnitude or pattern of perturbations required to decisively change a node’s output. 

\noindent\textbf{Limitation.}
Our threat model is not universally applicable to all scenarios; however, it represents a realistic and common setting where participants in a graph can act imperceptibly as benign users while performing attacks \cite{borisyuk2024lignn}. This aligns with practical cases in dynamic graphs, where adversaries can exploit the system's natural interactions to manipulate connections or features. By focusing on this scenario, our work highlights vulnerabilities that can arise in widely used graph-based systems, ensuring the threat model remains both practical and relevant. 

\section{Related Work}
GNN attacks are broadly categorized into adversarial \cite{bojchevski2019adversarial, dai2018adversarial, ma2020towards, zhang2021backdoor, xi2021graph, mu2021hard, wang2023turning} and privacy attacks \cite{devil_in_disguise, Inference_Attacks:link_steal:wu2021linkteller, Inference_Attacks:link_steal:he2020stealing, Inference_Attacks:graph_attribute, Inference_Attacks:node_level:he2021, Inference_Attacks:graph_reconstruction:Duddu2020, attack:inference:group_property2023}. 
Adversarial attacks are designed to undermine the accuracy of GNN classifications, typically through poisoning a portion of the training data or executing evasion attacks that alter the graph during the inference phase. 
Privacy attacks, however, focus on revealing confidential graph information. These attacks vary in scope, including graph-level attacks which seek to infer properties or even reconstruct entire graphs \cite{Inference_Attacks:graph_reconstruction:Duddu2020, attack:inference:group_property2023, Inference_Attacks:graph_attribute, Inference_Attacks:graph_reconstruction:zhang2021}, node/edge level attacks which target the disclosure of specific nodes or edges \cite{Inference_Attacks:link_steal:wu2021linkteller,Inference_Attacks:link_steal:he2020stealing, devil_in_disguise, Inference_Attacks:node_level:he2021, he2024maui, wu2024link}, and model level attacks that focus on the underlying model's characteristics \cite{shen2022model, zhuangunveiling}. 
Our research is focused explicitly on edge inference attacks, reviewing relevant literature within this area.

\noindent\textbf{Edge Inference Attacks.}
He \emph{et al.} \cite{Inference_Attacks:link_steal:he2020stealing} developed an attack to deduce connections in training graphs of transductive GNNs based on the premise that connected nodes exhibit similar features and output probabilities. However, their approach lacks robustness across various datasets due to specific assumptions about the dataset characteristics.
Meanwhile, Wu \emph{et al.} \cite{Inference_Attacks:link_steal:wu2021linkteller} introduced the LinkTeller Attack (LTA), which explores how changes to a node's features impact its neighbors' output probabilities to infer all the edges in a subgraph. LTA fails to generalize to GNNs with different message aggregation mechanisms, showing strong performance against GCNs but fails against GATs. 
Moreover, LTA operates under a scenario where the adversary has complete control over all nodes in the graph, including access to all predictions and the ability to modify node features, which differs from our setup. 
Meng et al. \cite{devil_in_disguise} introduced the Infiltration Inference Attack (IIA), which shares a similar intuition as LTA. However, while it proposes a threat model that accommodates a dynamic graph, allowing the adversary to change the graph, the attack's effectiveness diminishes in scenarios where the graph is actively modified by others, highlighting its limitations in truly dynamic environments. 
Recent works \cite{wu2024link, he2024maui} build on these approaches but inherit similar limitations.

\noindent\textbf{Privacy Preservation in GNN Models}.
Various strategies have been introduced to protect the privacy of GNN models, among which Differential Privacy (DP) is notably prevalent. 
Originating from the foundational work of Dwork \emph{et al.} \cite{dwork2006differential}, DP provides a systematic method to uphold privacy, particularly pertinent to machine learning models. It asserts that the outputs of a model when trained on adjacent datasets (those diverging by a single data point at most), should be nearly identical. 
Typically, DP is achieved by adding calibrated noise to the result of a function computed on the data. Various forms of DP have been proposed to extend it to a graph context. 
Two principal methods to embed DP within GNN for edge privacy are identified: Firstly, incorporating DP noise into the graph \cite{Inference_Attacks:link_steal:wu2021linkteller, lou2024grid}, and secondly, introducing DP noise to internal layers of the GNN model \cite{sajadmanesh2023gap, def:kolluri2022lpgnet}.


\section{Conclusion}
This paper introduces a novel threat model and attacks for edge inference in GNNs, showing that an adversary with black-box access can infer a node’s neighbors without direct graph access. 
Our approach addresses two major limitations of prior work: lack of generalization across GNN architectures and ineffectiveness in dynamic graph scenarios. 
By leveraging directional patterns in message aggregation, our method remains robust against noise and graph evolution. 
Additionally, we reformulate the problem at a finer granularity, directly determining connections between nodes. 
Extensive evaluations on nine datasets and four GNN architectures validate its effectiveness, highlighting critical GNN vulnerabilities in evolving environments.


\begin{acks}
This research received no specific grant from any funding agency in the public, commercial, or not-for-profit sectors.
\end{acks}

\bibliographystyle{ACM-Reference-Format}
\bibliography{main}

\appendix
\section*{Appendix}

\section{Additional Information}

\subsection{Node Feature and Influence Generation}
\label{app:evaluation:auxiliary_node_features}
Auxiliary nodes play two critical roles in our experiments:
1. They determine the type of influence generated since influence is created by reweighting node features, with different initial features representing various kinds of influence.
2. We analyze the model's output for the auxiliary node; variations in node features can lead to distinct reactions to the same influence.

We consider five distinct strategies to set the features of auxiliary nodes:

\ding{70}  \textit{Random:} The features are picked randomly.

\ding{70}  \textit{Target Node Duplication:} Same as the target node.

\ding{70}  \textit{Mean:} Average of all nodes' features.

\ding{70}  \textit{Typical:} The features of the node having the shortest Euclidean distance to the mean are adopted.

\ding{70}  \textit{Median:} Median of all nodes' features.

In the static scenario, the results of $INF\textit{-}MAG$ with various auxiliary node feature selections are presented in Table \ref{tab:attack:different_node_feature:full}. Due to space constraints, we only display the AUC score for the GAT model, as the outcomes across different models are consistent. The results indicate that different node feature selection strategies do not significantly affect the attack's performance.

In the dynamic scenario, which is more challenging, we further explored the different combinations of auxiliary node features to explore possible ways to improve the attack. 
We conducted experiments assuming the adversary knows the features of the target and candidate nodes. 
We tested three different strategies for setting the features of the auxiliary nodes: 1. \emph{SAME}, where all inserted nodes share identical features; 2. \emph{DIFF1}, where the anchor node and the node linked to the target share features; and 3. \emph{DIFF2}, where the anchor node and the node linked to the candidate share features.
The results indicate that the $INF\textit{--}DIR^*$ performs consistently across these different configurations. We have not included detailed results due to space constraints.

\textbf{Influence Generation.} 
\label{dis:inf_gen}
We tested various influence generation methods as mentioned in the main paper. The results indicate that while influence generation has a minimal impact in static scenarios, it shows potential in dynamic environments. 
This outcome aligns with our methodological focus on exploring how influence traverses different candidate nodes. 
Essentially, for our analysis, we only require an influence that can lead to changes in node output probabilities. The specific method of generating this influence and its characteristics are not crucial. Whether the magnitude of the source influence is 1 or 10,000 does not affect the relative ranking of LPS calculated for each candidate node as all the candidate nodes are under the same influence, and thus does not impact the inference results. 
In dynamic scenarios, where external noise complicates the detection of adversarial influence, the key advantage of a specific influence generation method lies in its ability to distinguish the adversary's influence from random noise. Crafting a strategy that produces detectable output changes, which are noticeable to the adversary but remain undetected by the model owner, poses a significant challenge. This challenge is vital for improving the effectiveness of attacks in environments where random fluctuations may mask malicious activities. Exploring more effective ways to generate influence will be the focus of our future work.

\begin{addedenv}
    
\subsection{Intuition}
\label{app:intuition}
Figure~\ref{fig:attack:intuition:mag} illustrates as distance increases, the similarity of change decreases. Figure~\ref{fig:attack:intuition:dir} illustrates that as distance increases, the similarity of change decreases.

\begin{figure}[ht]
    \centering
    \begin{subfigure}[b]{0.45\linewidth}
        \includegraphics[height=2.5cm]{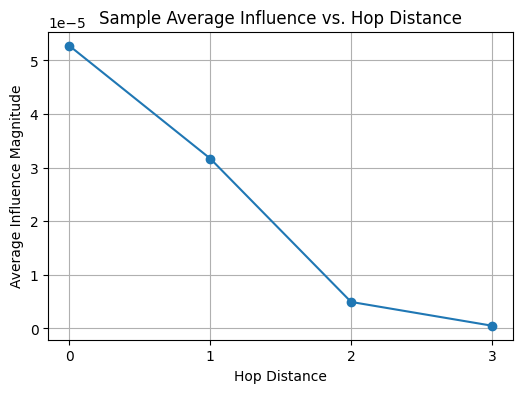}
        \caption{(a) Magnitude of change decreases with distance.}
        \label{fig:attack:intuition:mag}
    \end{subfigure}
    \hspace{0.2cm} 
    \begin{subfigure}[b]{0.45\linewidth}
        \centering
        \includegraphics[height=2.5cm]{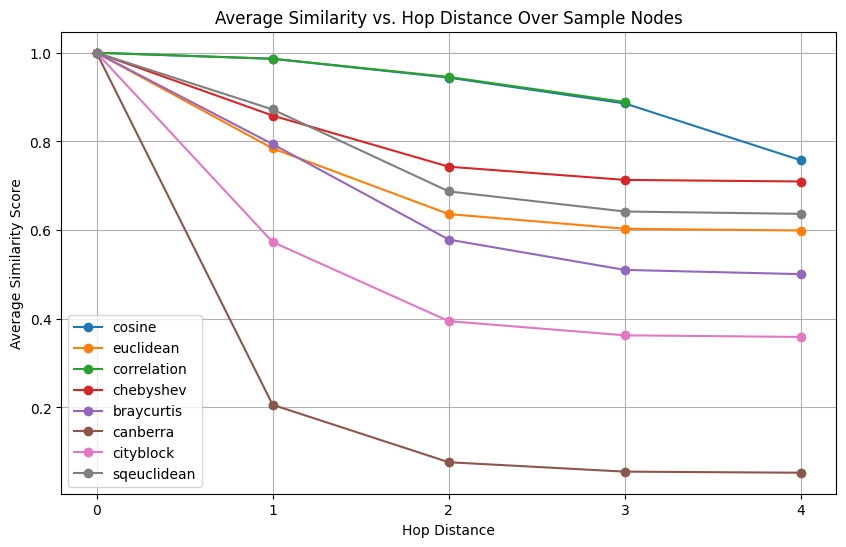}
        \caption{(b) Closer nodes exhibit higher similarity in change.}
        \label{fig:attack:intuition:dir}
    \end{subfigure}
    \caption{Impact of distance on perturbation effects in GCN (LastFM dataset). We perturb each node in the graph and measure the resulting changes in its neighbors at different hop distances. The results are averaged over all nodes, using each as an influence source. The findings indicate that both the similarity and magnitude of change can be used to infer the distance between nodes.}
    \label{fig:attack:intuition}
\end{figure}
\end{addedenv}

\subsection{Choice of 4-layer GNN architecture}
\label{app:4_layer_choice}
Our choice for a 4-layer architecture stems from the pivotal role the number of layers, represented as $\emph{l}$, plays in GNNs. 
The parameter $\emph{l}$ sets the cap for how many layers of deep information can traverse within the GNN. 
Consequently, any alterations to the feature of a node will resonate across its $\emph{l}$-hop neighbors. 
To illustrate, consider two nodes, $v$ and $u$, with an edge in between. 
Our attacks introduce two new nodes, $a_t$ and $a_c$, linked to $v$ and $u$, respectively, forming edges $(a_t,v)$ and $(a_c,u)$, resulting in $a_t$ and $a_c$'s distance to be 3, as illustrated in Figure \ref{fig:attack:methodology:node_insertion:b}. 
This suggests that our influence attacks would be most effective when the number of layers is 3, as in this case, the message change would happen between $a_t$ and $a_c$ only when the candidate node is connected to the target node. When the number of layers of the model increases, our attack would show degradation in performance as the message would change between $a_t$ and $a_c$ when the target node pair is not connected.
To maintain a balanced evaluation, we have utilized GNNs with 4 layers. 
Additionally, using four-layer models complicates the attack in dynamic scenarios, as each node can aggregate information from more nodes. This increased connectivity introduces more noise, further challenging the attack's effectiveness.
Previous attacks \cite{Inference_Attacks:link_steal:he2020stealing, devil_in_disguise, Inference_Attacks:link_steal:wu2021linkteller}, primarily using 2-layer and 3-layer GNNs to evaluate their attack. 
However, this choice potentially oversimplifies their attack scenario. 
In the case of \cite{devil_in_disguise} which uses 2-layer GNNs for the experiment, the nodes analyzed for influence are spaced at least three hops apart, which mirrors the setup in our work. 
However, utilizing a 2-layer model restricts the flow of information, making it impossible for data to propagate from one node to another if they are three hops away. Consequently, the influence score between any two such nodes would always be zero. 
We lack access to their code, but from a theoretical standpoint, this approach appears problematic.

We also have the result of our attack's performance on 3-layer GNN in Figure \ref{fig:app:dynamic:3layer}. Overall attacks show similar performance on 3-layer and 4-layer GNNs.

\begin{figure*}[!ht]
    \centering
    \begin{tikzpicture}
        \begin{axis}[hide axis, xmin=0, xmax=1, ymin=0, ymax=1,
                     legend columns=-1, 
                     legend entries={$INF-DIR^*$, $INF-MAG^*$, $LTA^*$, $IIA^*$, $EVO$},
                    legend style={          
                        text=black,        
                        draw=gray, fill=gray!10,
                        legend columns=1,
                        cells={anchor=west},
                        font=\footnotesize\itshape,
                        /tikz/every even column/.append style={column sep=0.5cm}
                    },
                     legend to name=figappdynamiccombinedLegend] 
            \addlegendimage{softred,fill=softred, line width=3pt}
            \addlegendimage{softblue, fill=softblue, line width=3pt}
            \addlegendimage{softorange, fill=softorange, line width=3pt}
            \addlegendimage{ForestGreen,fill=ForestGreen, line width=3pt}
            \addlegendimage{softpurple,fill=softpurple, line width=3pt}

        \end{axis}
    \end{tikzpicture}
    \ref{figappdynamiccombinedLegend} 
    \hfill

    \raisebox{1.8cm}{\rotatebox{90}{F1-score}}
    \begin{subfigure}{0.24\linewidth}
        \centering
        \begin{tikzpicture}
            \begin{axis}
        [ybar=0.01cm, 
        bar width=0.08cm,
        width=0.85\textwidth,  
        height=0.65\textwidth,  
        enlarge x limits={abs=0.35},  
        enlarge y limits=false,
        ylabel near ticks,
        xlabel near ticks,
        scale only axis,
        ymin=0,
        ymax=1,
        yticklabel style={font=\tiny},
        xtick=data,
        xticklabels={LastFM, EN, Flight, Dgraph},
        xticklabel style={rotate=0, xshift=0pt, yshift=4pt, anchor=north , font=\tiny},
        x label style={at={(axis description cs:0.5,-0.1)}, anchor=north},
        y label style={font=\small, at={(axis description cs:-0.1,0.5)}},
        major x tick style={opacity=0},
        minor x tick num=1,
        minor tick length=1 ex,
        ymajorgrids=true,
        xmajorgrids=true,
        grid style={white},
        axis background/.style={fill=gray!15},
        axis line style={white, line width=0pt},
        xtick style={draw=none},
        ytick style={draw=none},
        ]
            \addplot [style={softred, fill=softred}] coordinates {
                (1,0.65) 
                (2,0.54) 
                (3, 0.71)
                (4, 0.72)
            };
            \addplot [style={softblue, fill=softblue}] coordinates {
                (1,0.43) 
                (2,0.34) 
                (3, 0.46)
                (4, 0.51)
            };
            \addplot [style={softorange, fill=softorange}] coordinates {
                (1,0.23) 
                (2,0.25) 
                (3,0.48) 
                (4,0.41) 
            };
            \addplot [style={ForestGreen, fill=ForestGreen}] coordinates {
                (1,0.22) 
                (2,0.21) 
                (3,0.44) 
                (4,0.56) 
            };
            \addplot [style={softpurple, fill=softpurple}] coordinates {
                (1,0.31) 
                (2,0.32) 
                (3,0.23) 
                (4,0.26) 
            };
            \end{axis}
        \end{tikzpicture}
        \caption{\hspace{5pt}GAT}
    \end{subfigure}%
    \hfill
    \begin{subfigure}{0.24\textwidth}
        \centering
        \begin{tikzpicture}
            \begin{axis}
        [ybar=0.01cm, 
        bar width=0.08cm,
        width=0.85\textwidth,  
        height=0.65\textwidth,  
        enlarge x limits={abs=0.35},  
        enlarge y limits=false,
        ylabel near ticks,
        xlabel near ticks,
        scale only axis,
        ymin=0,
        ymax=1,
        yticklabel style={font=\tiny},
        xtick=data,
        xticklabels={LastFM, EN, Flight, Dgraph},
        xticklabel style={rotate=0, xshift=0pt, yshift=4pt, anchor=north , font=\tiny},
        x label style={at={(axis description cs:0.5,-0.1)}, anchor=north},
        y label style={font=\small, at={(axis description cs:-0.1,0.5)}},
        major x tick style={opacity=0},
        minor x tick num=1,
        minor tick length=1 ex,
        ymajorgrids=true,
        xmajorgrids=true,
        grid style={white},
        axis background/.style={fill=gray!15},
        axis line style={white, line width=0pt},
        xtick style={draw=none},
        ytick style={draw=none},
        ]
            \addplot [style={softred, fill=softred}] coordinates {
                (1,0.69) 
                (2,0.72) 
                (3, 0.91)
                (4, 0.84)
            };
            \addplot [style={softblue, fill=softblue}] coordinates {
                (1,0.44) 
                (2,0.18) 
                (3,0.70) 
                (4,0.71)
            };
            \addplot [style={softorange, fill=softorange}] coordinates {
                (1,0.16) 
                (2,0.28)
                (3,0.54) 
                (4,0.56) 
            };
            \addplot [style={ForestGreen, fill=ForestGreen}] coordinates {
                (1,0.28) 
                (2,0.21) 
                (3,0.72) 
                (4,0.71) 
            };
            \addplot [style={softpurple, fill=softpurple}] coordinates {
                (1,0.29) 
                (2,0.34) 
                (3,0.33) 
                (4,0.30) 
            };
            \end{axis}
            \end{tikzpicture}
        \caption{\hspace{5pt}GCN}
    \end{subfigure}%
    \hfill
    \begin{subfigure}{0.24\textwidth}
        \centering
        \begin{tikzpicture}
            \begin{axis}
        [ybar=0.01cm, 
        bar width=0.08cm,
        width=0.85\textwidth,  
        height=0.65\textwidth,  
        enlarge x limits={abs=0.35},  
        enlarge y limits=false,
        ylabel near ticks,
        xlabel near ticks,
        scale only axis,
        ymin=0,
        ymax=1,
        yticklabel style={font=\tiny},
        xtick=data,
        xticklabels={LastFM, EN, Flight, Dgraph},
        xticklabel style={rotate=0, xshift=0pt, yshift=4pt, anchor=north , font=\tiny},
        x label style={at={(axis description cs:0.5,-0.1)}, anchor=north},
        y label style={font=\small, at={(axis description cs:-0.1,0.5)}},
        major x tick style={opacity=0},
        minor x tick num=1,
        minor tick length=1 ex,
        ymajorgrids=true,
        xmajorgrids=true,
        grid style={white},
        axis background/.style={fill=gray!15},
        axis line style={white, line width=0pt},
        xtick style={draw=none},
        ytick style={draw=none},
        ]
            \addplot [style={softred, fill=softred}] coordinates {
                (1,0.86) 
                (2,0.84) 
                (3,0.80) 
                (4,0.92) 
            };
            \addplot [style={softblue, fill=softblue}] coordinates {
                (1,0.34) 
                (2,0.55) 
                (3,0.72) 
                (4,0.70) 
            };
            \addplot [style={softorange, fill=softorange}] coordinates {
                (1, 0.19) 
                (2, 0.20) 
                (3,0.70) 
                (4,0.59) 
            };
            \addplot [style={ForestGreen, fill=ForestGreen}] coordinates {
                (1, 0) 
                (2, 0) 
                (3,0) 
                (4,0) 
            };
            \addplot [style={softpurple, fill=softpurple}] coordinates {
                (1,0.38) 
                (2,0.33)
                (3,0.43) 
                (4,0.32) 
            };
            \end{axis}
            \end{tikzpicture}
            \centering
        \caption{\hspace{5pt}GIN}
    \end{subfigure}%
    \hfill
    \begin{subfigure}{0.24\textwidth}
        \centering 
        \begin{tikzpicture}
            \begin{axis}
        [ybar=0.01cm, 
        bar width=0.08cm,
        width=0.85\textwidth,  
        height=0.65\textwidth,  
        enlarge x limits={abs=0.35},  
        enlarge y limits=false,
        ylabel near ticks,
        xlabel near ticks,
        scale only axis,
        ymin=0,
        ymax=1,
        yticklabel style={font=\tiny},
        xtick=data,
        xticklabels={LastFM, EN, Flight, Dgraph},
        xticklabel style={rotate=0, xshift=0pt, yshift=4pt, anchor=north , font=\tiny},
        x label style={at={(axis description cs:0.5,-0.1)}, anchor=north},
        y label style={font=\small, at={(axis description cs:-0.1,0.5)}},
        major x tick style={opacity=0},
        minor x tick num=1,
        minor tick length=1 ex,
        ymajorgrids=true,
        xmajorgrids=true,
        grid style={white},
        axis background/.style={fill=gray!15},
        axis line style={white, line width=0pt},
        xtick style={draw=none},
        ytick style={draw=none},
        ]
            \addplot [style={softred, fill=softred}] coordinates {
                (1,0.76) 
                (2,0.72) 
                (3,0.83) 
                (4,0.90) 
            };
            \addplot [style={softblue, fill=softblue}] coordinates {
                (1,0.26) 
                (2,0.20)
                (3,0.62) 
                (4,0.75) 
            };
            \addplot [style={softorange, fill=softorange}] coordinates {
                (1, 0.22) 
                (2, 0.13) 
                (3,0.63) 
                (4,0.67) 
            };
            \addplot [style={ForestGreen, fill=ForestGreen}] coordinates {
                (1,0.19) 
                (2,0.22) 
                (3,0.52) 
                (4,0. 76) 
            };
            \addplot [style={softpurple, fill=softpurple}] coordinates {
                (1,0.42) 
                (2,0.32) 
                (3,0.33) 
                (4,0.31) 
            };
            \end{axis}
            \end{tikzpicture}
        \caption{\hspace{5pt}SAGE}
    \end{subfigure}
    \caption{F1-score of attacks in dynamic scenarios on 3-layer GNNs.}
    \label{fig:app:dynamic:3layer}
\end{figure*}

\subsection{Test Set Construction}
\label{app:2hopselection}
In this work, we primarily evaluate attack performance using 2-hop node pairs as negative samples. This choice is deliberate: success in this challenging setup can extend to more relaxed scenarios while including distant negative samples (beyond the GNN's message-passing range) risks inflating performance metrics artificially. In static scenarios, distant negative samples show no observable influence changes, making them trivially easy to distinguish. As a result, random sampling hides the true difficulty of the task. For example, if the test set includes both randomly sampled negatives and 2-hop pairs, an attack labeling all 2-hop pairs as connected could still achieve high metrics because random samples contribute little to the challenge.

\noindent\textbf{Why Success in This Setup Extends to More Relaxed Setups.} Success in distinguishing directly connected nodes from 2-hop unconnected nodes demonstrates attack robustness under challenging conditions. This generalizes to more relaxed setups because:

    \textbullet\ Weaker Influence for Distant Nodes: In relaxed setups, negative samples farther from the target node receive weaker influence, creating a clearer distinction between connected and unconnected nodes.

    \textbullet\ Simpler Structural Patterns: Distant negative samples often lack shared neighbors or structural overlap with the target node, further reducing the similarity in influence scores and simplifying separation.

    \textbullet\ Real-World Relevance: Many real-world datasets inherently include distant nodes as negatives. Success in distinguishing close-but-unconnected nodes directly validates the attack’s ability to handle practical scenarios and less challenging alternatives.

By focusing on 2-hop negatives, our evaluation rigorously addresses the core challenge of link inference: distinguishing meaningful connections from topologically similar alternatives. This approach ensures that demonstrated robustness in this setup translates to confidence in the attack’s generalizability to broader and less stringent scenarios.

\subsection{Dataset} 
\label{app:evaluation:data_statistics}
We experiment with the following datasets:

\noindent\underline{\emph{Twitch}} contains six separate graphs, each reflecting data from a distinct country. Users represent nodes, while edges represent follower connections. Node features correspond to the embeddings of games played by users. The dimensions of the features remain the same across different graphs. The task is binary classification to predict whether a user streams mature content.  \\ 
\noindent\underline{\emph{Flickr}} is an image relationship dataset where each node represents an image. The features of each node are characterized by bag-of-words models derived from their descriptions, while the edges between them indicate common attributes shared by the images. The task is to classify the images into one of the 7 classes.\\
\noindent\underline{\emph{LastFM}} is a social network graph where nodes symbolize LastFM users from Asia, and the edges represent friendships between these users. The node features are extracted based on the artists liked by the users. The task in this dataset is to predict the users' nationality. \\
\noindent\underline{\emph{tgbl-flight}} This dataset is a crowd-sourced international flight network from 2019 to 2022. The airports are modeled as nodes, while the edges are flights between airports on a given day. The task is to predict whether a flight will happen between two specific airports on a future date. \\ 
\noindent\underline{\emph{Dgraph-fin}} Dgraph-fin dataset source from Finvolution users. The node feature derived from the basic personal profile is a vector with 17 dimensions. Each dimension of the node attribute corresponds to a distinct element of the personal profile, such as age and gender. The edges between nodes represent emergency contact records. The task of this dataset is to find malicious users.

For the Twitch dataset, we train the GNN model on the Twitch-ES dataset and then apply it to the graphs of the other five countries.
For both the Flickr and LastFM datasets, we randomly partition the data into training, validation, and test sets using ratios of 70\%, 15\%, and 15\%, respectively. The model accesses the subgraph composed of only the training nodes during training. During testing and validation, predictions are made based on the entirety of the graph.
For the tgbl-flight and Dgraph-fin datasets, we first preprocess the data from its original format into a series of weekly snapshots. We then select the 50th snapshot to serve as the base graph and apply the same settings used for static graphs to train the model.
We generated labels based on the traffic volume for the tgbl-flight dataset, which is designed for link prediction tasks. Nodes with traffic in the top 50\% are labeled as 'busy', while the remaining nodes are labeled as 'not busy'.
We provide the dataset statistics in Table \ref{tab:exp:dataset_statistics}.

\begin{table}[!h]
\small
\centering
\begin{tabular}{ccccc}
    \hline
    \textbf{Dataset} & \textbf{Nodes} & \textbf{Edges} & \textbf{Features} & \textbf{Classes} \\
    \hline
    Twitch-DE & 9,498 & 315,774 & 128 & 2 \\ 
    Twitch-EN & 7,126 & 77,774 & 128 & 2  \\ 
    Twitch-ES & 4,648 & 123,412 & 128 & 2 \\
    Twitch-FR & 6,551 & 231,883 & 128 & 2 \\
    Twitch-PT & 1,912 & 64,510 & 128 & 2 \\ 
    Twitch-RU & 4,385 & 78,993 & 128 & 2 \\ 
    Flickr & 89,250  & 899,756  & 500 &  7 \\ 
    LastFM & 7,624 & 55,612 & 128 & 18 \\ 
    tgbl-flight & 18,143 & 67,169,570 & 20 & 2 \\ 
    Dgraph-fin & 3,700,550 & 4,300,999 & 17 & 2 \\ 

    \hline
\end{tabular}
\caption{Dataset statistics}
\label{tab:exp:dataset_statistics}
\end{table}

\subsection{Impact of Node Sampling on Attacks.}
\label{app:node_sampling}
To improve both accuracy and efficiency, many modern GNNs employ node sampling techniques, enabling each node to aggregate information from only a subset of its neighbors, rather than all of them. In industrial applications, methods such as Personalized PageRank \cite{borisyuk2024lignn} and time-based sampling \cite{robinson2024relbench} are commonly used.
This sampling introduces a new dynamic for adversaries: their nodes may or may not be sampled. In this paper, we explore the impact of node sampling on attack performance. We investigate three key scenarios: (1) when the adversary's nodes are always included in the sample, (2) when they are excluded, and (3) when the adversary's nodes may or may not be included.
We vary the sampling rate, defined as the proportion of neighbors sampled, between 10\% and 50\%. By examining these cases, our analysis reveals how different sampling strategies can either enhance or mitigate adversarial influence in GNNs, providing insights into the vulnerabilities and robustness of GNNs under varying node sampling conditions. 

Here we evaluate the attacks in dynamic scenarios, as in static scenarios the adversary is the sole agent of influence, thus if their nodes are not selected none of the attacks would work as the model output would not change. For simplicity, we use the single version of the attack for evaluation.
In Figure \ref{fig:res:node_sampling} we show the result of attacks on GraphSage with the LastFM dataset. The result demonstrates that all magnitude-based attacks fail when the adversary’s node is not sampled for message aggregation. This outcome is expected, as these attacks rely on the adversary’s influence to alter the model’s inference process. When this influence is absent, the attack naturally becomes ineffective.
Conversely, when the adversary's node is included in the sample, we observe improved attack performance at lower sampling rates. This is because lower sampling rates reduce the inclusion of external noise, allowing the adversary's influence to become more pronounced. 
However, in scenario (3), where nodes are sampled randomly, the attack remains ineffective. This is because the success of the attack hinges on the adversary's node being sampled consistently across two consecutive rounds—something the adversary cannot predict or control. 

In summary, for magnitude-based attacks to succeed, the adversary’s node must be consistently sampled during message aggregation, and if the sampling mechanism is explored by the adversary, the system would be more vulnerable.

In contrast, direction-based attacks demonstrate greater robustness across all scenarios. 
When the nodes are sampled, the attack performs better at lower sampling rates due to reduced external noise.
In scenarios (2) and (3), where the adversary’s node is either excluded from sampling or sampled randomly, the attack performance remains strong. This is because direction-based attacks can capture the natural evolution and these scenarios are close to our $EVO$ attack where no adversary influence is involved. 
However, as the sampling rate decreases further in scenarios (2) and (3), attack performance degrades. With lower sampling rates, each node aggregates information from only a small number of neighbors, reducing the likelihood that the target node pairs receive information from the same set of sources, leading to divergence in their model outputs.

In summary, direction-based attacks show promise when node sampling occurs, but excluding the adversary's node from message aggregation can still degrade their effectiveness. 
If adversaries can manipulate the sampling process, they can enhance their attack performance, particularly when their nodes are consistently sampled.

\begin{figure*}[t]
    \centering
    \begin{tikzpicture}
        \begin{axis}[hide axis, xmin=0, xmax=1, ymin=0, ymax=1,
                     legend columns=-1, 
                     legend entries={{$INF-MAG$, $INF-DIR$, $LTA$, $IIA$}},
                    legend style={          
                        text=black,        
                        draw=gray, fill=gray!10,
                        legend columns=1,
                        cells={anchor=west},
                        legend pos=north west,
                        font=\footnotesize\itshape,
                        /tikz/every even column/.append style={column sep=0.5cm}
                    },
                    legend to name=dp:defense] 
                     
            \addlegendimage{mygreen,fill=mygreen, line width=3pt}
            \addlegendimage{myblue, fill=myblue, line width=3pt}
            \addlegendimage{myred, fill=myred, line width=3pt}
            \addlegendimage{mycyan, fill=mycyan, line width=3pt}
        \end{axis}
    \end{tikzpicture}
    \ref{dp:defense}
    \hfill
    
    \raisebox{2.3cm}{\rotatebox{90}{F1-score}}%
    \centering
    \begin{subfigure}{0.32\textwidth}
        \centering
        \caption{\hspace{20pt} Adversarial Node Sampled}
        \begin{tikzpicture}
            \begin{axis}
         [ width=0.85\textwidth,  
            height=0.6\textwidth,
            enlarge x limits=0.15, 
            enlarge y limits=false,
            ylabel near ticks,
            xlabel near ticks,
            scale only axis,
            ymin=0,
            ymax=1,
            yticklabel style={},
            xtick=data,
            xticklabels={10\%, 20\%, 30\%, 40\%, 50\%},
            xticklabel style={yshift=0ex, rotate=0},
            xlabel={Sample Rate},
            x label style={at={(axis description cs:0.5,-0.2)},anchor=north, font=\footnotesize},
            y label style={at={(axis description cs:-0.1,0.5)}},
            major x tick style={opacity=0},
            minor x tick num=1,
            minor tick length=1 ex,
            ymajorgrids= true,
            xmajorgrids= true,
            grid style={white},
            axis background/.style={fill=gray!15},
            axis line style={white, line width=0pt},
            xtick style={draw=none},
            ytick style={draw=none},
            ]
        \addplot [mygreen, mark=*, line width=1pt] coordinates {
            (0,0.80) (1,0.60) (2,0.57) (3,0.54) (4, 0.44)
        };
        \addplot [myblue, mark=*, line width=1pt] coordinates {
            (0,0.50) (1,0.23) (2,0.26) (3,0.23) (4, 0.26)

        };
        \addplot [myred, mark=*, line width=1pt] coordinates {
             (0,0.23) (1,0.19) (2,0.21) (3,0.22) (4, 0.23)

        };
        \addplot [mycyan, mark=*, line width=1pt] coordinates {
            (0,0.2) (1,0.21) (2,0.23) (3,0.23) (4, 0.22)

        };

        \end{axis}
        \end{tikzpicture}
    \end{subfigure} 
    \begin{subfigure}{0.32\textwidth}
        \centering
        \caption{\hspace{20pt} Adversarial Node Not Sampled}
        \begin{tikzpicture}
            \begin{axis}
         [ width=0.85\textwidth,  
            height=0.6\textwidth,
            enlarge x limits=0.15, 
            enlarge y limits=false,
            ylabel near ticks,
            xlabel near ticks,
            scale only axis,
            ymin=0,
            ymax=1,
            yticklabel style={},
            xtick=data,
            xticklabels={10\%, 20\%, 30\%, 40\%, 50\%},
            xticklabel style={yshift=0ex, rotate=0},
            xlabel={Sample Rate},
            x label style={at={(axis description cs:0.5,-0.2)},anchor=north, font=\footnotesize},
            y label style={at={(axis description cs:-0.1,0.5)}},
            major x tick style={opacity=0},
            minor x tick num=1,
            minor tick length=1 ex,
            ymajorgrids= true,
            xmajorgrids= true,
            grid style={white},
            axis background/.style={fill=gray!15},
            axis line style={white, line width=0pt},
            xtick style={draw=none},
            ytick style={draw=none},
            ]
        \addplot [mygreen, mark=*, line width=1pt] coordinates {
            (0,0.22) (1,0.35) (2,0.33) (3,0.34) (4, 0.37)
        };
        \addplot [myblue, mark=*, line width=1pt] coordinates {
            (0,0.22) (1,0.21) (2,0.20) (3,0.21) (4, 0.20)

        };
        \addplot [myred, mark=*, line width=1pt] coordinates {
             (0,0.22) (1,0.23) (2,0.22) (3,0.20) (4, 0.20)

        };
        \addplot [mycyan, mark=*, line width=1pt] coordinates {
            (0,0.22) (1,0.21) (2,0.21) (3,0.22) (4, 0.23)

        };
        \end{axis}
        \end{tikzpicture}

    \end{subfigure} 
    \begin{subfigure}{0.32\textwidth}
        \centering
        \caption{\hspace{20pt} Random Sampling}
        \begin{tikzpicture}
            \begin{axis}
         [ width=0.85\textwidth,  
            height=0.6\textwidth,
            enlarge x limits=0.15, 
            enlarge y limits=false,
            ylabel near ticks,
            xlabel near ticks,
            scale only axis,
            ymin=0,
            ymax=1,
            yticklabel style={},
            xtick=data,
            xticklabels={10\%, 20\%, 30\%, 40\%, 50\%},
            xticklabel style={yshift=0ex, rotate=0},
            xlabel={Sample Rate},
            x label style={at={(axis description cs:0.5,-0.2)},anchor=north, font=\footnotesize},
            y label style={at={(axis description cs:-0.1,0.5)}},
            major x tick style={opacity=0},
            minor x tick num=1,
            minor tick length=1 ex,
            ymajorgrids= true,
            xmajorgrids= true,
            grid style={white},
            axis background/.style={fill=gray!15},
            axis line style={white, line width=0pt},
            xtick style={draw=none},
            ytick style={draw=none},
            ]
        \addplot [mygreen, mark=*, line width=1pt] coordinates {
            (0,0.22) (1,0.30) (2,0.36) (3,0.44) (4, 0.46)
        };
        \addplot [myblue, mark=*, line width=1pt] coordinates {
            (0,0.22) (1,0.21) (2,0.20) (3,0.21) (4, 0.20)

        };
        \addplot [myred, mark=*, line width=1pt] coordinates {
             (0,0.22) (1,0.23) (2,0.22) (3,0.20) (4, 0.20)

        };
        \addplot [mycyan, mark=*, line width=1pt] coordinates {
            (0,0.22) (1,0.21) (2,0.21) (3,0.22) (4, 0.23)

        };
        \end{axis}
        \end{tikzpicture}
    \end{subfigure}
    \caption{Attack performance when node sampling method is used in GraphSAGE on LastFM dataset.}
    \label{fig:res:node_sampling}
\end{figure*}

\subsection{Possible Defense}
\label{app:new_defense}

\begin{addedenv}
    
{
\subsubsection{Combine node sampling and graph perturbation as defense}
\label{app:defense:combined}

\paragraph{Why Perturbations and Stochastic Sampling Share the Same Intuition}

\textbf{Graph perturbations} (e.g., adding/removing edges, modifying node features) and \textbf{stochastic node sampling} (randomly selecting neighbors during training) both revolve around the same principle:

\begin{quote}
\textit{They alter or randomize the information flow that each node receives, preventing over-reliance on specific neighbors or edges.}
\end{quote}

\paragraph{Graph Perturbations}

\begin{itemize}
    \item \textbf{Concept}: Small random (or structured) changes to the graph, such as EdgeDrop, prevent dependence on a fixed structure.
    \item \textbf{Effect}: The GNN encounters multiple “versions” of the graph, reducing overfitting and limiting an attacker's ability to exploit specific edges.
\end{itemize}

\paragraph{Stochastic Node Sampling}

\begin{itemize}
    \item \textbf{Concept}: Instead of aggregating from \textit{all} neighbors, randomly sample a subset (e.g., GraphSAGE).
    \item \textbf{Effect}: Random neighbor selection injects variability, preventing the model from fixating on specific adjacency structures.
\end{itemize}

Thus, \textbf{both} techniques introduce \textbf{controlled randomness} into the aggregation process, making the GNN more resistant to targeted attacks.

\paragraph{Combining Perturbations and Stochastic Sampling}

Since both methods share a common foundation—\textbf{introducing randomness and reducing fixed adjacency reliance}—they can be \textbf{combined}:

\paragraph{Approach: Perturbation at Data Level + Sampling at Model Level}

\begin{itemize}
    \item \textbf{Data-Level Perturbation}: Periodically randomize adjacency or features (e.g., DropEdge, edge flipping).
    \item \textbf{Model-Level Sampling}: Apply stochastic neighbor sampling (GraphSAGE-style) in the forward pass.
\end{itemize}

\paragraph{Advantages}

\begin{itemize}
    \item \textbf{Double Layer of Defense}: Even if an attacker influences edges, random sampling may ignore them.
    \item \textbf{Distributed Risk}: Structural noise and sampling diffuse attacks, requiring widespread manipulation to be effective.
\end{itemize}

\paragraph{Tradeoffs}

\begin{itemize}
    \item \textbf{Computational Overhead}: Frequent perturbations can be costly.
    \item \textbf{Hyperparameter Tuning}: Noise levels (edge modifications) and sampling rates must be tuned per dataset.
\end{itemize}

\paragraph{Conclusion}

In summary, \textbf{graph perturbations} and \textbf{stochastic node sampling} both introduce \textbf{controlled randomness} into GNN aggregation, \textbf{reducing reliance on fixed structures} and improving robustness. By combining these strategies, GNNs can achieve stronger \textbf{defenses against adversarial attacks}, provided the computational costs remain manageable.

}
\end{addedenv}

\subsubsection{More potential defenses}
To effectively defend against influence-based attacks on GNNs, it is essential to enhance the robustness of the message aggregation process. The core idea is to modify the message aggregation process in GNNs to dynamically identify and mitigate the influence of potentially adversarial nodes or edges. This can be achieved by integrating adaptive filtering mechanisms and anomaly detection into the aggregation step, ensuring that only reliable and consistent information is propagated through the network. 
Here are some outlines of different types of possible defenses:
\begin{enumerate}
    \item Dynamic Weighting: Implement an adaptive weighting scheme where messages from neighbors are weighted based on their reliability. Reliable neighbors contribute more to the aggregation, while suspicious or inconsistent ones are down-weighted or ignored.

    \item Attention-Based Filtering: Enhance existing attention mechanisms to incorporate robustness. For instance, use robust attention scores that can identify and reduce the influence of outlier nodes exhibiting abnormal behavior.

    \item Anomaly Detection in Message Passing: Integrate statistical methods to monitor the distribution of incoming messages. Messages that significantly deviate from the norm can be flagged as anomalies and excluded from aggregation.

    \item Consensus-Based Aggregation with Multiple Aggregators: Utilize multiple aggregation functions (e.g., mean, median, max) and require consensus among them for updating node representations. 
\end{enumerate}

\section{Additional Results}
\subsection{Result with more target node}
\label{app:more_nodes}
Figure~\ref{app:more_nodes:lastfm} presents the attack results on LastFM, averaged over 500 nodes, while Figure~\ref{app:more_nodes:en} shows the results for Twitch-EN.

\begin{figure}[!h]
    \centering
    \begin{tikzpicture}
        \begin{axis}[
            ybar=0.01cm, 
            bar width=0.1cm,  
            width=0.7\linewidth,  
            height=0.3\linewidth,  
            enlarge x limits={abs=0.2},  
            enlarge y limits=false,
            ylabel near ticks,
            xlabel near ticks,
            scale only axis,
            ymin=0,
            ymax=1,
            yticklabel style={font=\scriptsize},
            xtick=data,
            xticklabels={GCN, GAT , GIN, SAGE },
            xticklabel style={rotate=0, xshift=0pt, yshift=3pt, anchor=north , font=\scriptsize},
            x label style={at={(axis description cs:0.5,-0.1)}, anchor=north, font=\scriptsize},
            y label style={at={(axis description cs:-0.1,0.5)}, font=\scriptsize},
            major x tick style={opacity=0},
            minor x tick num=1,
            minor tick length=1 ex,
            ymajorgrids=true,
            xmajorgrids=true,
            grid style={white},
            axis background/.style={fill=gray!15},
            axis line style={white, line width=0pt},
            xtick style={draw=none},
            ytick style={draw=none},
             legend style={          
                 at={(1,1.24)}, 
                text=black,        
                draw=gray, fill=gray!10,
                legend columns=5,
                cells={anchor=west},
                font=\tiny\itshape,
                /tikz/every even column/.append style={column sep=0.5cm}
            },
        ]   
            \addplot [style={softblue, fill=softblue}] coordinates {
                (1,0.69) 
                (2,0.65) 
                (3,0.68) 
                (4,0.62) 
            };
            \addlegendentry{$INF$}

            \addplot [style={softgreen, fill=softgreen}] coordinates {
                (1,0.94) 
                (2,0.92) 
                (3,0.98) 
                (4,0.99) 
            };
            \addlegendentry{$INF\text{--}DIR$}
            \addplot [style={softyellow, fill=softyellow}] coordinates {
                (1,0.99) 
                (2,0.93) 
                (3,0.99) 
                (4,0.99) 
            };
            \addlegendentry{$INF\text{--}MAG$}

            \addplot [style={softcyan, fill=softcyan}] coordinates {
                (1,0.61) 
                (2,0.47) 
                (3,0.36) 
                (4,0.58) 
            };
            \addlegendentry{$LTA$}

            \addplot [style={darkpeach, fill=darkpeach}] coordinates {
                (1,0.82) 
                (2,0.80) 
                (3,0.0) 
                (4,0.93) 

            };
            \addlegendentry{$IIA$}

        \end{axis}
    \end{tikzpicture}
    \caption{F1-score of attacks on LastFM averaged over 500 target nodes.}
    \label{app:more_nodes:lastfm}
\end{figure}

\begin{figure}[!h]
    \centering
    \begin{tikzpicture}
        \begin{axis}[
            ybar=0.01cm, 
            bar width=0.1cm,  
            width=0.7\linewidth,  
            height=0.3\linewidth,  
            enlarge x limits={abs=0.2},  
            enlarge y limits=false,
            ylabel near ticks,
            xlabel near ticks,
            scale only axis,
            ymin=0,
            ymax=1,
            yticklabel style={font=\scriptsize},
            xtick=data,
            xticklabels={GCN, GAT , GIN, SAGE },
            xticklabel style={rotate=0, xshift=0pt, yshift=3pt, anchor=north , font=\scriptsize},
            x label style={at={(axis description cs:0.5,-0.1)}, anchor=north, font=\scriptsize},
            y label style={at={(axis description cs:-0.1,0.5)}, font=\scriptsize},
            major x tick style={opacity=0},
            minor x tick num=1,
            minor tick length=1 ex,
            ymajorgrids=true,
            xmajorgrids=true,
            grid style={white},
            axis background/.style={fill=gray!15},
            axis line style={white, line width=0pt},
            xtick style={draw=none},
            ytick style={draw=none},
             legend style={          
                 at={(1,1.24)}, 
                text=black,        
                draw=gray, fill=gray!10,
                legend columns=5,
                cells={anchor=west},
                font=\tiny\itshape,
                /tikz/every even column/.append style={column sep=0.5cm}
            },
        ]   
            \addplot [style={softblue, fill=softblue}] coordinates {
                (1,0.57) 
                (2,0.44) 
                (3,0.35) 
                (4,0.43) 
            };
            \addlegendentry{$INF$}

            \addplot [style={softgreen, fill=softgreen}] coordinates {
                (1,0.90) 
                (2,0.79) 
                (3,0.99) 
                (4,0.80) 
            };
            \addlegendentry{$INF\text{--}DIR$}
            \addplot [style={softyellow, fill=softyellow}] coordinates {
                (1,0.99) 
                (2,0.90) 
                (3,0.99) 
                (4,0.97) 
            };
            \addlegendentry{$INF\text{--}MAG$}

            \addplot [style={softcyan, fill=softcyan}] coordinates {
                (1,0.65) 
                (2,0.47) 
                (3,0.32) 
                (4,0.63) 
            };
            \addlegendentry{$LTA$}

            \addplot [style={darkpeach, fill=darkpeach}] coordinates {
                (1,0.78) 
                (2,0.63) 
                (3,0.0) 
                (4,0.87) 

            };
            \addlegendentry{$IIA$}

        \end{axis}
    \end{tikzpicture}
    \caption{F1-score of attacks on Twitch/EN averaged over 500 target nodes.}
    \label{app:more_nodes:en}
\end{figure}

\subsection{ROC-AUC of similarity-based attacks in static graph}
\label{app:similarity:auc}
AUC of Similarity-based attacks are shown in Table \ref{tab:attack_performance:auc:simi:2}.
\begin{table}[!htbp]
    \centering
    \tiny
    \caption{{AUC metrics for Similarity-based attacks.}}
    \label{tab:attack_performance:auc:simi:2}
    \setlength\tabcolsep{4.5pt} 
    \begin{tabular}{l *{8}{c}} 
        \toprule
        \multirow{2}{*}{\textbf{Dataset}}  &  \multicolumn{4}{c}{\textbf{\emph{SIM}}} &  \multicolumn{4}{c}{\textbf{\emph{LSA}}}  \\
        \cmidrule(lr){2-5} \cmidrule(lr){6-9} 
        & \textbf{GCN} & \textbf{SAGE} & \textbf{GAT} & \textbf{GIN} & \textbf{GCN} & \textbf{SAGE} & \textbf{GAT} & \textbf{GIN} \\ 
        \midrule
        \texttt{LastFM}  
        &           0.81       & 0.76    & 0.79  & 0.72
        &           0.91       & 0.91    & 0.91  & 0.83 \\
        \texttt{Flickr}  
        &           0.73       & 0.64    &   0.74   & 0.67
        &             0.76     & 0.64    &   0.74   & 0.68  \\
       \texttt{Twitch-EN}  
       &    0.73           & 0.64    &    0.74   &  0.59
       &             0.76  & 0.64    & 0.74      &  0.58 \\
        \texttt{Twitch-DE}  
        &    0.58           &  0.58 & 0.58 &  0.59
        &             0.57  & 0.59  & 0.57 &  0.69 \\
       \texttt{Twitch-FR}  
       &         0.73    & 0.64  &  0.74  &  0.65
       &         0.76    & 0.64  & 0.74   &  0.65 \\
       \texttt{Twitch-PT}  
       &    0.73       & 0.64    &    0.74  & 0.57
       &              0.76  & 0.64 & 0.74   & 0.66 \\
        \texttt{Twitch-RU}  
        &    0.58       & 0.59    &  0.59   & 0.57
        &         0.59  & 0.61    &  0.59   & 0.66 \\
        \bottomrule
    \end{tabular}
\end{table}

\subsection{Result for SIM and INF-DIR Using Different Distance Metrics in Static Graph Scenario}
\label{app:evaluation:sim:full}
The results are available in Table \ref{tab:attack_performance:sim:full}. 

\begin{table}[!ht]
    \centering
    \tiny
    \caption{{F1-score of $SIM$ and $INF-DIR$ attack with different distance metrics in static scenario}}
    \label{tab:attack_performance:sim:full}
    \setlength\tabcolsep{3pt} 
    \begin{tabular}{ll *{8}{c}} 
        \toprule
        \multirow{2}{*}{\textbf{Dataset}} & \multirow{2}{*}{\textbf{Metric}} &  \multicolumn{4}{c}{\textbf{\emph{SIM}}} &  \multicolumn{4}{c}{\textbf{\emph{INF-DIR}}}  \\
        \cmidrule(lr){3-6} \cmidrule(lr){7-10} 
        & & \textbf{GCN} & \textbf{SAGE} & \textbf{GAT} & \textbf{GIN} & \textbf{GCN} & \textbf{SAGE} & \textbf{GAT} & \textbf{GIN} \\ 
        \midrule
        
        \multirow{8}{*}{LastFM}   
        & Cosine          & 0.55 & 0.39 & 0.45 & 0.37 
        & 0.73 & 0.67 & 0.67 & 0.97 \\ 
        & Euclidean       & 0.44 & 0.33 & 0.39 & 0.34 
        & 0.64 & 0.60 & 0.59 & 0.97 \\ 
        & Correlation     & 0.55 & 0.40 & 0.47 & 0.37
        & 0.96 & 0.83 & 0.74 & 0.97 \\ 
        & Cheybyshev      & 0.51 & 0.35 & 0.43 & 0.35
        & 0.62 & 0.60 & 0.61 & 0.97 \\ 
        & Bray-Curtis     & 0.42 & 0.30 & 0.36 & 0.31
        & 0.93 & 0.89 & 0.84 & 0.97\\ 
        & Canberra        & 0.44 & 0.30 & 0.35 & 0.31
        & 0.96 & 0.95 & 0.89 & 0.97 \\ 
        & Manhattan       & 0.42 & 0.30 & 0.38 & 0.33
        & 0.64 & 0.60 & 0.59 & 0.97 \\ 
        & Square-Euclidean& 0.44 & 0.30 & 0.39 & 0.34
        & 0.64 & 0.60 & 0.59 & 0.97 \\ 
        \midrule
        \multirow{8}{*}{Flickr}   
        & Cosine          & 0.31 & 0.19 & 0.34 & 0.24 
        & 0.83 & 0.76 & 0.84 & 1.0 \\ 
        & Euclidean       & 0.29 & 0.18 & 0.35 & 0.23 
        & 0.70 & 0.70 & 0.67 & 1.0 \\ 
        & Correlation     & 0.32 & 0.19 & 0.33 & 0.23
        & 0.84 & 0.76 & 0.83 & 1.0  \\ 
        & Cheybyshev      & 0.29 & 0.18 & 0.35 & 0.24
        & 0.70 & 0.69 & 0.67 & 1.0  \\ 
        & Bray-Curtis     & 0.28 & 0.20 & 0.34 & 0.22
        & 0.99 & 0.83 & 0.99 & 1.0 \\ 
        & Canberra        & 0.28 & 0.18 & 0.37 & 0.23
        & 0.99 & 0.95 & 0.99 & 1.0  \\ 
        & Manhattan       & 0.28 & 0.18 & 0.34 & 0.22
        & 0.70 & 0.70 & 0.67 & 1.0  \\ 
        & Square-Euclidean& 0.29 & 0.18 & 0.35 & 0.23
        & 0.70 & 0.70 & 0.67 & 1.0 \\ 
        \bottomrule
    \end{tabular}
\end{table}

\begin{table*}[htbp]
    \centering
    \small
    \caption{{AUC of the $INF\textit{--}MAG$ with varied inserted node features.}}
    \label{tab:attack:different_node_feature:full}
    \begin{tabular}{llrrrrrrr}
        \hline
        \textbf{Model} & \textbf{Node Feature} & \textbf{LastFM} & \textbf{Flickr} & \textbf{Twitch-DE} & \textbf{Twitch-RU} & \textbf{Twitch-PT} & \textbf{Twitch-FR} & \textbf{Twitch-EN} \\

        \multirow{5}{*}{GAT}  
                              & Random  & 0.938 & 1.0 & 0.899 & 0.961 & 0.996 & 0.959 & 0.976\\ 
                              & Typical & 0.966 & 1.0 & 0.981 & 0.986 & 0.996 & 0.987 & 0.976\\ 
                              & Duplication    & 0.969 & 1.0 & 0.944 & 0.953 & 0.996 & 0.954 & 0.976\\ 
                              & Mean    & 0.963 & 0.999 & 0.960 & 0.988 & 0.996 & 0.959 & 0.940\\ 
                              & Median  & 0.963 & 0.999 & 0.979 & 0.961 & 0.996 & 0.968 & 0.994\\ 
        \hline
    \end{tabular}
\end{table*}

\subsection{Different version of $INF-DIR^*$ }
\label{app:dynamic:distance_metrics_selection}
The result with Different Distance Metrics is illustrated in Table \ref{tab:attack_performance:infdir:distance:full}. 

The results of $INF\textit{--}DIR^*$ with different data preprocessing methods are presented in Table \ref{tab:dynamic:f1:different_preprocess}. Concatenation delivered the best performance among these techniques, while PCA was the least effective. This outcome underscores the value of high-dimensional data in preserving rich information, which is crucial when assessing the similarity of vectors for our attack. 

\begin{table}[!h]
    \centering
    \tiny
    \caption{{F1-score for $INF\textit{--}DIR^*$ with different data preprocessing method.}}
    \label{tab:dynamic:f1:different_preprocess}
    \setlength\tabcolsep{1pt} 

        \caption{\hspace{14pt}GAT}
    \end{subfigure}
    \begin{subfigure}{0.2\textwidth}
    \centering
    %
        \caption{\hspace{14pt}GCN}
    \end{subfigure}   
    \begin{subfigure}{0.2\textwidth}
    \centering
    %
        \caption{\hspace{14pt}GIN}
    \end{subfigure}   
\begin{subfigure}{0.2\textwidth}
    \centering
    %
        \caption{\hspace{14pt}GraphSage}
    \end{subfigure}   
    \caption{Model performance on LastFM dataset: The x-axis illustrates privacy budgets, and the y-axis shows accuracy.}
    \label{fig:def:model_performance}
    \end{figure*}
\begin{table}[!ht]
    \centering
    \tiny
    \caption{{F1-score of $INF\textit{--}DIR^*$ attack with different distance metrics}}
    \label{tab:attack_performance:infdir:distance:full}
    \setlength\tabcolsep{3pt} 
    \begin{tabular}{ll *{4}{c}} 
        \toprule
        \textbf{Dataset} & \textbf{Metric} & \textbf{GCN} & \textbf{SAGE} & \textbf{GAT} & \textbf{GIN} \\
        \midrule
        \multirow{8}{*}{LastFM}   
        & Cosine          & \emph{0.85} & \emph{0.65} & 0.46 & 0.74  \\ 
        & Euclidean       & 0.49 & 0.31 & 0.35 & 0.67 \\ 
        & Correlation     & 0.85 & 0.65 & 0.46 & 0.74 \\ 
        & Cheybyshev      & 0.37 & 0.23 & 0.32 & 0.66 \\ 
        & Bray-Curtis     & 0.71 & 0.54 & \emph{0.54} & \emph{0.78} \\ 
        & Canberra        & 0.68 & 0.47 & 0.43 & 0.77 \\ 
        & Manhattan       & 0.53 & 0.33 & 0.32 & 0.66 \\ 
        & Square-Euclidean& 0.49 & 0.31 & 0.35 & 0.34 \\ 
        \midrule
        \multirow{8}{*}{Twitch-EN}   
        & Cosine          & \emph{0.68} & \emph{0.67} & 0.53 & 0.66  \\ 
        & Euclidean       & 0.26 & 0.19 & 0.35 & 0.71 \\ 
        & Correlation     & 0.66 & 0.59 & 0.53 & 0.66 \\ 
        & Cheybyshev      & 0.22 & 0.21 & 0.33 & 0.70 \\ 
        & Bray-Curtis     & 0.50 & 0.39 & \emph{0.55} & \emph{0.80} \\ 
        & Canberra        & 0.54 & 0.28 & 0.47 & 0.79 \\ 
        & Manhattan       & 0.36 & 0.18 & 0.36 & 0.70 \\ 
        & Square-Euclidean& 0.23 & 0.17 & 0.35 & 0.71 \\ 
        \bottomrule
    \end{tabular}
\end{table}

\subsection{More result for Signal-to-Noise Rate analysis.}
\label{app:snr}
The result can be found at GAT (Figure \ref{fig:snr:gat}), GIN (Figure \ref{fig:snr:gin}), and GraphSage (Figure \ref{fig:snr:sage}).

\begin{figure*}[ht]
    \centering
    %
        \caption{
            \parbox[t]{1.2\linewidth}{\centering A. Varying Perturbation Rate: \\ Higher Perturbation Rate → Higher SNR.}
        }
    \end{subfigure}
    \hspace{15mm}%
    \begin{subfigure}[t]{0.25\textwidth}
    \centering 
        %
            \caption{
            \parbox[t]{1.2\linewidth}{\centering B. Varying Feature Change Rate: \\ Higher Change Rate → Lower SNR.}
        }
    \end{subfigure}
    \hspace{15mm}%
    \begin{subfigure}[t]{0.25\textwidth}
        \centering 

        %
        \caption{
            \parbox[t]{1.2\linewidth}{\centering  C. Varying Local Structure Change Rate: \\ Higher Change Rate → Lower SNR.}
        }
    \end{subfigure}
    \caption{Signal-to-Noise Ratio analysis on GAT with the LastFM Dataset.}
    \label{fig:snr:gat}
\end{figure*}

\begin{figure*}[ht]
    \centering
    %
        \caption{
            \parbox[t]{1.2\linewidth}{\centering A. Varying Perturbation Rate: \\ Higher Perturbation Rate → Higher SNR.}
        }
    \end{subfigure}
    \hspace{15mm}%
    \begin{subfigure}[t]{0.25\textwidth}
    \centering 
        %
            \caption{
            \parbox[t]{1.2\linewidth}{\centering B. Varying Feature Change Rate: \\ Higher Change Rate → Lower SNR.}
        }
    \end{subfigure}
    \hspace{15mm}%
    \begin{subfigure}[t]{0.25\textwidth}
        \centering 

        %
        \caption{
            \parbox[t]{1.2\linewidth}{\centering  C. Varying Local Structure Change Rate: \\ Higher Change Rate → Lower SNR.}
        }
    \end{subfigure}
    \caption{Signal-to-Noise Ratio analysis on GIN with the LastFM Dataset.}
    \label{fig:snr:gin}
\end{figure*}

\begin{figure*}[ht]
    \centering
    %
        \caption{
            \parbox[t]{1.2\linewidth}{\centering A. Varying Perturbation Rate: \\ Higher Perturbation Rate → Higher SNR.}
        }
    \end{subfigure}
    \hspace{15mm}%
    \begin{subfigure}[t]{0.25\textwidth}
    \centering 
        %
            \caption{
            \parbox[t]{1.2\linewidth}{\centering B. Varying Feature Change Rate: \\ Higher Change Rate → Lower SNR.}
        }
    \end{subfigure}
    \hspace{15mm}%
    \begin{subfigure}[t]{0.25\textwidth}
        \centering 

        %
        \caption{
            \parbox[t]{1.2\linewidth}{\centering  C. Varying Local Structure Change Rate: \\ Higher Change Rate → Lower SNR.}
        }
    \end{subfigure}
    \caption{Signal-to-Noise Ratio analysis on GraphSage with the LastFM Dataset.}
    \label{fig:snr:sage}
\end{figure*}

\subsection{Result of Dynamic Scenarios}
\label{app:dynamic:results}
ROC-AUC of the attack in dynamic scenarios on 4-layer GNNs can be found in Table \ref{tab:dynamic:auc:all_n:4layer}.
\begin{table}[!ht]
    \centering
    \tiny    
    \setlength\tabcolsep{4pt} 
    \begin{tabular}{ll *{4}{c}} 
        \toprule
        \textbf{Attacks} & & LastFM & EN & Flight & Dgraph \\ 
        \midrule
        \multirow{4}{*}{$INF-DIR^*$}   
        & GCN          & 0.98 & 0.92  & 0.97 & 0.98 \\ 
        & SAGE       & 0.91 & 0.91 & 0.95 & 0.96 \\ 
        & GAT     & 0.88 & 0.86 & 0.91 & 0.95 \\ 
        & GIN      & 0.92 & 0.95 & 0.95 & 0.98 \\ 
        \midrule
        \multirow{4}{*}{$INF-MAG^*$}   
        & GCN          & 0.67 & 0.64  & 0.88 & 0.91 \\ 
        & SAGE       & 0.60 & 0.62  & 0.87 & 0.89 \\ 
        & GAT     & 0.66 & 0.59 & 0.85 & 0.82 \\ 
        & GIN      & 0.61 & 0.61  & 0.86 & 0.84 \\ 
        \midrule
        \multirow{4}{*}{$LTA^*$}   
        & GCN          & 0.63 & 0.61 & 0.88 & 0.85 \\ 
        & SAGE       & 0.59 & 0.61 & 0.89 & 0.85 \\ 
        & GAT     & 0.62 & 0.58 & 0.79 & 0.82 \\ 
        & GIN      & 0.61 & 0.57 & 0.73 & 0.85\\ 
        \midrule
        \multirow{4}{*}{$IIA^*$}   
        & GCN          & 0.65 & 0.61  & 0.88 & 0.85 \\ 
        & SAGE       & 0.62 & 0.59 & 0.86 & 0.85\\ 
        & GAT     & 0.67 & 0.59 & 0.82 & 0.82\\ 
        & GIN      & - & -  & - & - \\             
            \bottomrule  
        \end{tabular}
    \caption{{AUC for attacks against evolving graph on 4-layer GNNs.}}
    \label{tab:dynamic:auc:all_n:4layer}
\end{table}

\subsection{Attack Performance on Structural Changed Nodes}
\label{app:dynamic:changed_nodes}
Figure \ref{fig:dynamic:changed_nodes:full} shows the full result if attack performance against nodes undergoes structural change. 
\begin{figure*}[!ht]
    \centering
    \begin{tikzpicture}
        \begin{axis}[hide axis, xmin=0, xmax=1, ymin=0, ymax=1,
                     legend columns=-1, 
                     legend entries={$INF-DIR^*$, $INF-MAG^*$, $LTA^*$, $IIA^*$, $EVO$},
                    legend style={          
                        text=black,        
                        draw=gray, fill=gray!10,
                        legend columns=1,
                        cells={anchor=west},
                        font=\footnotesize\itshape,
                        /tikz/every even column/.append style={column sep=0.5cm}
                    },
                     legend to name=figdynamicchangednodesfullcombinedLegend] 
            \addlegendimage{softred,fill=softred, line width=3pt}
            \addlegendimage{softblue, fill=softblue, line width=3pt}
            \addlegendimage{softorange, fill=softorange, line width=3pt}
            \addlegendimage{ForestGreen,fill=ForestGreen, line width=3pt}
            \addlegendimage{softpurple,fill=softpurple, line width=3pt}

        \end{axis}
    \end{tikzpicture}
    \ref{figdynamicchangednodesfullcombinedLegend} 
    \hfill

    \raisebox{1.8cm}{\rotatebox{90}{F1-score}}
    \begin{subfigure}{0.24\linewidth}
        \centering
        \begin{tikzpicture}
            \begin{axis}
        [ybar=0.01cm, 
        bar width=0.08cm,
        width=0.85\textwidth,  
        height=0.65\textwidth,  
        enlarge x limits={abs=0.35},  
        enlarge y limits=false,
        ylabel near ticks,
        xlabel near ticks,
        scale only axis,
        ymin=0,
        ymax=1,
        yticklabel style={font=\tiny},
        xtick=data,
        xticklabels={LastFM, EN, Flight, Dgraph},
        xticklabel style={rotate=0, xshift=0pt, yshift=4pt, anchor=north , font=\tiny},
        x label style={at={(axis description cs:0.5,-0.1)}, anchor=north},
        y label style={font=\small, at={(axis description cs:-0.1,0.5)}},
        major x tick style={opacity=0},
        minor x tick num=1,
        minor tick length=1 ex,
        ymajorgrids=true,
        xmajorgrids=true,
        grid style={white},
        axis background/.style={fill=gray!15},
        axis line style={white, line width=0pt},
        xtick style={draw=none},
        ytick style={draw=none},
        ]
            \addplot [style={softred, fill=softred}] coordinates {
                (1,0.60) 
                (2,0.62) 
                (3, 0.56)
                (4, 0.60)
            };
            \addplot [style={softblue, fill=softblue}] coordinates {
                (1,0.42) 
                (2,0.29) 
                (3, 0.20)
                (4, 0.32)
            };
            \addplot [style={softorange, fill=softorange}] coordinates {
                (1,0.22) 
                (2,0.21) 
                (3,0.20) 
                (4,0.20) 
            };
            \addplot [style={ForestGreen, fill=ForestGreen}] coordinates {
                (1,0.22) 
                (2,0.21) 
                (3,0.21) 
                (4,0.19) 
            };
            \addplot [style={softpurple, fill=softpurple}] coordinates {
                (1,0.30) 
                (2,0.30) 
                (3,0.25) 
                (4,0.24) 
            };
            \end{axis}
        \end{tikzpicture}
        \caption{\hspace{5pt}GAT}
    \end{subfigure}%
    \hfill
    \begin{subfigure}{0.24\textwidth}
        \centering
        \begin{tikzpicture}
            \begin{axis}
        [ybar=0.01cm, 
        bar width=0.08cm,
        width=0.85\textwidth,  
        height=0.65\textwidth,  
        enlarge x limits={abs=0.35},  
        enlarge y limits=false,
        ylabel near ticks,
        xlabel near ticks,
        scale only axis,
        ymin=0,
        ymax=1,
        yticklabel style={font=\tiny},
        xtick=data,
        xticklabels={LastFM, EN, Flight, Dgraph},
        xticklabel style={rotate=0, xshift=0pt, yshift=4pt, anchor=north , font=\tiny},
        x label style={at={(axis description cs:0.5,-0.1)}, anchor=north},
        y label style={font=\small, at={(axis description cs:-0.1,0.5)}},
        major x tick style={opacity=0},
        minor x tick num=1,
        minor tick length=1 ex,
        ymajorgrids=true,
        xmajorgrids=true,
        grid style={white},
        axis background/.style={fill=gray!15},
        axis line style={white, line width=0pt},
        xtick style={draw=none},
        ytick style={draw=none},
        ]
            \addplot [style={softred, fill=softred}] coordinates {
                (1,0.80) 
                (2,0.70) 
                (3, 0.85)
                (4, 0.81)
            };
            \addplot [style={softblue, fill=softblue}] coordinates {
                (1,0.26) 
                (2,0.18) 
                (3,0.20) 
                (4,0.23)
            };
            \addplot [style={softorange, fill=softorange}] coordinates {
                (1,0.16) 
                (2,0.18)
                (3,0.21) 
                (4,0.18) 
            };
            \addplot [style={ForestGreen, fill=ForestGreen}] coordinates {
                (1,0.18) 
                (2,0.21) 
                (3,0.22) 
                (4,0.20) 
            };
            \addplot [style={softpurple, fill=softpurple}] coordinates {
                (1,0.45) 
                (2,0.43) 
                (3,0.56) 
                (4,0.52) 
            };
            \end{axis}
            \end{tikzpicture}
        \caption{\hspace{5pt}GCN}
    \end{subfigure}%
    \hfill
    \begin{subfigure}{0.24\textwidth}
        \centering
        \begin{tikzpicture}
            \begin{axis}
        [ybar=0.01cm, 
        bar width=0.08cm,
        width=0.85\textwidth,  
        height=0.65\textwidth,  
        enlarge x limits={abs=0.35},  
        enlarge y limits=false,
        ylabel near ticks,
        xlabel near ticks,
        scale only axis,
        ymin=0,
        ymax=1,
        yticklabel style={font=\tiny},
        xtick=data,
        xticklabels={LastFM, EN, Flight, Dgraph},
        xticklabel style={rotate=0, xshift=0pt, yshift=4pt, anchor=north , font=\tiny},
        x label style={at={(axis description cs:0.5,-0.1)}, anchor=north},
        y label style={font=\small, at={(axis description cs:-0.1,0.5)}},
        major x tick style={opacity=0},
        minor x tick num=1,
        minor tick length=1 ex,
        ymajorgrids=true,
        xmajorgrids=true,
        grid style={white},
        axis background/.style={fill=gray!15},
        axis line style={white, line width=0pt},
        xtick style={draw=none},
        ytick style={draw=none},
        ]
            \addplot [style={softred, fill=softred}] coordinates {
                (1,0.78) 
                (2,0.80) 
                (3,0.88) 
                (4,0.95) 
            };
            \addplot [style={softblue, fill=softblue}] coordinates {
                (1,0.31) 
                (2,0.50) 
                (3,0.22) 
                (4,0.22) 
            };
            \addplot [style={softorange, fill=softorange}] coordinates {
                (1, 0.19) 
                (2, 0.15) 
                (3,0.20) 
                (4,0.20) 
            };
            \addplot [style={ForestGreen, fill=ForestGreen}] coordinates {
                (1, 0) 
                (2, 0) 
                (3,0) 
                (4,0) 
            };
            \addplot [style={softpurple, fill=softpurple}] coordinates {
                (1,0.36) 
                (2,0.34)
                (3,0.45) 
                (4,0.45) 
            };
            \end{axis}
            \end{tikzpicture}
            \centering
        \caption{\hspace{5pt}GIN}
    \end{subfigure}%
    \hfill
    \begin{subfigure}{0.24\textwidth}
        \centering 
        \begin{tikzpicture}
            \begin{axis}
        [ybar=0.01cm, 
        bar width=0.08cm,
        width=0.85\textwidth,  
        height=0.65\textwidth,  
        enlarge x limits={abs=0.35},  
        enlarge y limits=false,
        ylabel near ticks,
        xlabel near ticks,
        scale only axis,
        ymin=0,
        ymax=1,
        yticklabel style={font=\tiny},
        xtick=data,
        xticklabels={LastFM, EN, Flight, Dgraph},
        xticklabel style={rotate=0, xshift=0pt, yshift=4pt, anchor=north , font=\tiny},
        x label style={at={(axis description cs:0.5,-0.1)}, anchor=north},
        y label style={font=\small, at={(axis description cs:-0.1,0.5)}},
        major x tick style={opacity=0},
        minor x tick num=1,
        minor tick length=1 ex,
        ymajorgrids=true,
        xmajorgrids=true,
        grid style={white},
        axis background/.style={fill=gray!15},
        axis line style={white, line width=0pt},
        xtick style={draw=none},
        ytick style={draw=none},
        ]
            \addplot [style={softred, fill=softred}] coordinates {
                (1,0.65) 
                (2,0.67) 
                (3,0.88) 
                (4,0.95) 
            };
            \addplot [style={softblue, fill=softblue}] coordinates {
                (1,0.28) 
                (2,0.19)
                (3,0.22) 
                (4,0.21) 
            };
            \addplot [style={softorange, fill=softorange}] coordinates {
                (1, 0.19) 
                (2, 0.17) 
                (3,0.19) 
                (4,0.19) 
            };
            \addplot [style={ForestGreen, fill=ForestGreen}] coordinates {
                (1,0.18) 
                (2,0.20) 
                (3,0.18) 
                (4,0.18) 
            };
            \addplot [style={softpurple, fill=softpurple}] coordinates {
                (1,0.40) 
                (2,0.30) 
                (3,0.44) 
                (4,0.36) 
            };
            \end{axis}
            \end{tikzpicture}
        \caption{\hspace{5pt}SAGE}
    \end{subfigure}
    \caption{F1-score of attacks in dynamic scenarios with negative samples constrained to nodes undergoing structural changes.}
    \label{fig:dynamic:changed_nodes:full}
\end{figure*}

\subsection{Threshold Selection.} 
\label{app:threshold}
\noindent\textbf{More Advanced Threshold Selection.}
As discussed in the main paper, determining the optimal threshold requires detailed knowledge about individual nodes. The adversary's capabilities can include access to a shadow dataset, partial knowledge of the target node's properties, or direct information from the target graph (e.g., knowledge of some connections of the target node to infer additional connections). After getting this knowledge, the adversary needs more methods to use them for threshold selection.

Beyond these capabilities, more advanced threshold selection strategies could be explored:

    \textbullet\ Score Distribution Modeling: By analyzing score distributions from shadow datasets or similar nodes, the adversary can approximate thresholds tailored to each node. This method accounts for variations in node influence patterns caused by factors like degree and neighborhood diversity.

    \textbullet\ Dynamic Threshold Adjustment: In dynamic scenarios, thresholds could be updated iteratively as the graph evolves, ensuring the threshold remains relevant despite structural or feature changes.

    \textbullet\ Hybrid Approaches: Combining knowledge from partial observations (e.g., known neighbors) with learned patterns from similar graphs or datasets could improve threshold accuracy.

    \textbullet\ Robust Statistical Techniques: Leveraging statistical models to identify outliers in the score distribution could help the adversary distinguish true connections from noise, even in cases where individual scores are uncertain or influenced by natural graph dynamics.

\noindent\textbf{Additional result of threshold selection methods.}
Table \ref{app:table:dynamic:different_threshold} presents the attack performance for in-graph threshold selection under varying levels of adversarial knowledge. The "Single" setup represents the scenario where the adversary has no knowledge of the target node's true neighbors and uses the score of their auxiliary node as the threshold. The 20\% and 50\% setups denote the percentage of ground truth scores accessible to the adversary.

In the "Single" setup, the attack demonstrates high precision but low recall, as the limited knowledge leads to the misclassification of many connected nodes as unconnected, particularly when the auxiliary node's score is high within the distribution. 
As the adversary's knowledge level increases, the attack achieves more balanced metrics, benefiting from a more informed understanding of the optimal threshold. 

Table \ref{app:table:dynamic:different_threshold:twitch} shows the performance of threshold selection on the Twitch-EN dataset.

\begin{table}[!h]
    \centering
    \caption{Performance of $INF\text{--}MAG$ (static scenario) and $INF\text{--}DIR^*$ (dynamic scenario) with different threshold selections on the LastFM dataset.}
    \label{app:table:dynamic:different_threshold}
    \scriptsize
    \setlength{\tabcolsep}{6pt}  
    \begin{tabular}{llSSSS}  
        \toprule
        \multirow{2}{*}{\textbf{Threshold}} & \multirow{2}{*}{\textbf{Model}} 
        & \multicolumn{2}{c}{\textbf{Static}} & \multicolumn{2}{c}{\textbf{Dynamic}} \\ 
        \cmidrule(lr){3-4} \cmidrule(lr){5-6}
        & & {\textbf{Recall}} & {\textbf{Precision}} & {\textbf{Recall}} & {\textbf{Precision}} \\
        \midrule
        \multirow{4}{*}{In-graph(Single)}   
        & GCN  & 0.61 & 0.99 & 0.58 & 0.62  \\  
        & SAGE & 0.62 & 0.99 & 0.58 & 0.54   \\ 
        & GAT  & 0.61 & 0.95 & 0.62 & 0.69 \\ 
        & GIN  & 0.60 & 0.99 & 0.66 & 0.79    \\ 
        \midrule
        \multirow{4}{*}{In-graph(20\%)}     
        & GCN  & 0.65 & 0.99 & 0.67 & 0.63  \\  
        & GCN  & 0.65 & 0.99 & 0.67 & 0.63  \\  
        & SAGE & 0.68 & 0.99 & 0.66 & 0.54   \\ 
        & GAT  & 0.68 & 0.94 & 0.67 & 0.54  \\ 
        & GIN  & 0.70 & 0.99 & 0.65 & 0.80    \\ 
        \midrule
        \multirow{4}{*}{In-graph(50\%)}     
        & GCN  & 0.84 & 0.99 & 0.84 & 0.58  \\  
        & SAGE & 0.85 & 0.99 & 0.80 & 0.48   \\ 
        & GAT  & 0.84 & 0.93 & 0.85 & 0.50 \\ 
        & GIN  & 0.83 & 0.99 & 0.84 & 0.76    \\ 
        \midrule
        \bottomrule
    \end{tabular}
\end{table}

\begin{table}[!h]
    \centering
    \caption{Performance of $INF\text{--}MAG$ (static scenario) and $INF\text{--}DIR^*$ (dynamic scenario) with different threshold selections on the Twitch-EN dataset.}
    \label{app:table:dynamic:different_threshold:twitch}
    \scriptsize
    \setlength{\tabcolsep}{6pt}  
    \begin{tabular}{llSSSS}  
        \toprule
        \multirow{2}{*}{\textbf{Threshold}} & \multirow{2}{*}{\textbf{Model}} 
        & \multicolumn{2}{c}{\textbf{Static}} & \multicolumn{2}{c}{\textbf{Dynamic}} \\ 
        \cmidrule(lr){3-4} \cmidrule(lr){5-6}
        & & {\textbf{Recall}} & {\textbf{Precision}} & {\textbf{Recall}} & {\textbf{Precision}} \\
        \midrule
        \multirow{4}{*}{In-graph(Single)}   
        & GCN  & 0.58 & 0.99 & 0.67 & 0.57  \\  
        & SAGE & 0.61 & 0.98 & 0.63 & 0.53   \\ 
        & GAT  & 0.62 & 0.94 & 0.71 & 0.57  \\ 
        & GIN  & 0.61 & 0.99 & 0.63 & 0.79    \\ 
        \midrule
        \multirow{4}{*}{In-graph(20\%)}     
        & GCN  & 0.69 & 0.99 & 0.67 & 0.60 \\  
        & SAGE & 0.68 & 0.98 & 0.71 & 0.51   \\ 
        & GAT  & 0.68 & 0.94 & 0.68 & 0.65   \\ 
        & GIN  & 0.70 & 0.99 & 0.69 & 0.78    \\ 
        \midrule
        \multirow{4}{*}{In-graph(50\%)}     
        & GCN  & 0.84 & 0.99 & 0.85 & 0.56  \\  
        & SAGE & 0.85 & 0.96 & 0.84 & 0.52   \\ 
        & GAT  & 0.87 & 0.91 & 0.83 & 0.62 \\ 
        & GIN  & 0.84 & 0.99 & 0.88 & 0.75  \\ 
        \midrule
        \bottomrule
    \end{tabular}
\end{table}

\noindent\textbf{Performance of attacks with deviated threshold.} 
To analyze the impact of threshold selection, we follow the approach in \cite{Inference_Attacks:link_steal:wu2021linkteller} and vary the threshold method. Using top-$k$ selection, where $k=d$ represents the ground truth number of positive samples (i.e., connected nodes), we evaluate performance in the static scenario.
Given the ground truth degree $d$, we incorporate three approximations for evaluation: $0.8d$, $d$, and $1.2d$. This range is selected because, at $k=0.8d$, the precision reaches 1, and at $k=1.2d$, the recall reaches 1, making further exploration less meaningful. Table \ref{tab:attack_performance:inf3:degree} illustrates the $INF\text{--}MAG$ attack's efficacy with different estimated degrees. We observe that with precise degree estimation (where \( \hat{d} = d \)), $INF\text{--}MAG$ consistently exhibits high precision and recall. When the density estimation deviates, the attack remains robust. It's important to highlight that with an estimated degree of $1.2d$, the attack still shows excellent precision on the GIN model because we avoid labeling nodes with an LPS of 0, which is common in this scenario.

\noindent\textbf{Performance of attacks on target node with different degree.} 
In our study, we categorize target nodes across various datasets based on their degree of connectivity:\emph{Unconstrained Subset:} We include nodes from the complete testing set. \emph{Specific Degree Subsets:} For LastFM, nodes with a degree of 5 or less are considered low degree, whereas those with a degree of 10 or more are deemed high degree. For Flickr, we categorize nodes with up to 15 connections as low degrees and those with 30 or more as high degrees. 

Table \ref{tab:attack_performance:inf3:degree} illustrates the efficacy of the $INF\text{--}MAG$ attack under varying conditions. 
We observe no significant performance change across different node degree distributions.

\begin{table}[!h]
    \centering
    \caption{{Performance of $INF\text{--}MAG$ on target nodes with different degree distributions and threshold estimation.}}
    \label{tab:attack_performance:inf3:degree}
    \scriptsize
    \setlength{\tabcolsep}{4pt}
    \begin{subtable}{\linewidth}
            \centering
            \begin{tabular}{lllrrcrrcrr}
                \toprule
                \multirow{2}{*}{\textbf{$\hat{d}$}} & \multirow{2}{*}{\textbf{Model}} &  \multicolumn{2}{c}{\textbf{Low Degree}}  & \multicolumn{2}{c}{\textbf{Unconstrained}} & \multicolumn{2}{c}{\textbf{High Degree}}  \\
                \cmidrule(lr){3-4} \cmidrule(lr){5-6} \cmidrule(lr){7-8} &&  \textbf{Recall}  &   \textbf{Precision}     &   \textbf{Recall}  &   \textbf{Precision}    &   \textbf{Recall}  &   \textbf{Precision}   \\
                \midrule
                \multirow{4}{*}{$\lfloor{0.8d}\rfloor$}   
                & GCN  & 0.686 & 0.952 & 0.717 & 0.966 & 0.796 & 0.995 \\  
                & SAGE & 0.686 & 0.952 & 0.717 & 0.966 & 0.798 & 0.998 \\ 
                & GAT  & 0.676 & 0.940 & 0.704 & 0.949 & 0.783 & 0.98  \\ 
                & GIN  & 0.686 & 0.952 & 0.717 & 0.966 & 0.798 & 0.998 \\ 
                \midrule
                \multirow{4}{*}{$d$}     
                & GCN  & 0.966 & 0.966 & 0.976 & 0.976 & 0.992 & 0.992 \\  
                & SAGE & 0.966 & 0.966 & 0.976 & 0.976 & 0.998 & 0.998 \\ 
                & GAT  & 0.934 & 0.934 & 0.938 & 0.938 & 0.966 & 0.966 \\ 
                & GIN  & 0.966 & 0.966 & 0.976 & 0.976 & 0.998 & 0.998 \\ 
                \midrule
                \multirow{4}{*}{$\lceil{1.2d}\rceil$}   
                & GCN  & 1.0   & 0.789 & 1.0   & 0.801 & 1.0   & 0.833 \\  
                & SAGE & 1.0   & 0.782 & 1.0   & 0.795 & 1.0   & 0.833 \\ 
                & GAT  & 0.986 & 0.771 & 0.989 & 0.787 & 0.998 & 0.831 \\ 
                & GIN  & 1.0   & 0.782 & 1.0   & 0.795 & 1.0   & 0.833 \\ 
                \bottomrule
            \end{tabular}
            \caption{LastFM}
        \end{subtable}

        \begin{subtable}{\linewidth}
            \centering
            \begin{tabular}{lllrrcrrcrr}
                \toprule
                \multirow{2}{*}{\textbf{$\hat{d}$}} & \multirow{2}{*}{\textbf{Model}} &  \multicolumn{2}{c}{\textbf{Low Degree}}  & \multicolumn{2}{c}{\textbf{Unconstrained}} & \multicolumn{2}{c}{\textbf{High Degree}}  \\
                \cmidrule(lr){3-4} \cmidrule(lr){5-6} \cmidrule(lr){7-8} &&  \textbf{Recall}  &   \textbf{Precision}     &   \textbf{Recall}  &   \textbf{Precision}    &   \textbf{Recall}  &   \textbf{Precision}   \\
                \midrule
                \multirow{4}{*}{$\lfloor{0.8d}\rfloor$}   
                & GCN & 0.757 & 0.999 & 0.758 & 0.999 & 0.789 & 0.993 \\  
                & SAGE & 0.757 & 1.0 & 0.757 & 0.999 & 0.792 & 0.990 \\ 
                & GAT & 0.757 & 1.0 & 0.757 & 0.998 & 0.785 & 0.981 \\ 
                & GIN & 0.757 & 1.0 & 758 & 1.0 & 0.80 & 1.0 \\ 
                \midrule
                
                \multirow{4}{*}{$d$}   
                & GCN & 0.999 & 0.999 & 0.999 & 0.999 & 0.985 & 0.985 \\  
                & SAGE & 0.996 & 0.996 & 0.992 & 0.992 & 0.986 & 0.986 \\ 
                & GAT & 0.999 & 0.999 & 0.999 & 0.999 & 0.979 & 0.979 \\ 
                & GIN & 1.0 & 1.0  & 1.0  & 1.0  & 1.0  & 1.0\\ 
                \midrule
            
                \multirow{4}{*}{$\lceil{1.2d}\rceil$}   
                & GCN & 1.0 & 0.825 & 1.0 & 0.829 & 0.999 & 0.888 \\  
                & SAGE & 1.0 & 0.830 & 1.0 & 0.839 & 1.0 & 0.903 \\ 
                & GAT & 1.0  & 0.825 & 1.0 & 0.829 & 1.0 & 0.891 \\ 
                & GIN & 1.0 & 1.0 & 1.0 & 1.0 & 1.0 & 1.0 \\                 
                \bottomrule
            \end{tabular}
            \caption{Flickr}
        \end{subtable}

    \end{table}

\subsection{Model Utility With Differential Privacy}
\label{app:dp:model_utility}
We train and test our model with the implementation of DP. For EdgeRand, we varied the privacy budget $\epsilon$ between 5 and 10. And for LapGraph, we varied the privacy budget $\epsilon$ between 2 and 10. The model utility is depicted in Fig. \ref{fig:def:model_performance}. The findings indicate that a privacy budget exceeding 8 enables the model to surpass the performance of MLP, which makes predictions without utilizing edge information.]

\subsection{Imperceptibility of Malicious Behaviors}
\label{app:imperceptibility}
Our attack includes perturbing the graph during the inference phase, emphasizing the importance of maintaining the stealth of the adversary’s actions. 
We evaluate the imperceptibility of our attacks on two fronts: changes to nodes/edges and alterations to node features. 

\noindent\textbf{Node/Edge Imperceptibility.} 
Previous graph attack methods rely on fixed perturbation budgets to limit changes and maintain stealth, assuming actions like node insertion are imperceptible below a threshold \cite{mu2021hard, bojchevski2019adversarial}. However, these metrics are inadequate in dynamic environments where graphs continuously evolve. To address this, we leverage model explanation techniques \cite{agarwal2022probing, ying2019gnnexplainer, luo2020parameterized, schlichtkrull2020interpreting} to assess the noticeability of adversarial behavior by analyzing each node’s influence on the model’s decisions. Using tools like GNNExplainer \cite{ying2019gnnexplainer}, PGExplainer \cite{luo2020parameterized}, and GraphMask \cite{schlichtkrull2020interpreting}, we find that adversarially inserted nodes rarely rank among the top influencers for any target node, demonstrating their stealth. The result is depicted in Table \ref{tab:attack_performance:explanation}.

\noindent\textbf{Feature Change Imperceptibility.} 
Our experiments indicate that adversaries typically need to introduce perturbations that surpass the graph's natural evolution rate to achieve effective influence. To mitigate this, adversaries could adopt a subgraph-based approach instead of targeting a single node. By distributing modifications across multiple nodes, their activities become better concealed. In this study, we focus on analyzing the potential of influence-based attacks, while minimizing the necessary perturbation rate is reserved for future work. 

\begin{table}  
    \centering
    \caption{{Top 5 influencer rate of auxiliary nodes for target nodes according to explanation methods on LastFM}}
    \label{tab:attack_performance:explanation}
    \setlength\tabcolsep{4.25pt} 
    \begin{tabularx}{\linewidth}{
        >{\centering\arraybackslash}X
        S[table-format=1.2]
        S[table-format=1.2]
        S[table-format=1.2]
        S[table-format=1.2]
        S[table-format=1.2]
    }
    \toprule
    \textbf{Explainer} & \textbf{GCN} & \textbf{SAGE} & \textbf{GAT} & \textbf{GIN} \\ 
    \midrule
        GNNExplainer       &           0.01       & 0.01    & 0.01  & 0.04 \\ 
        PGExplainer   &     0.01       & 0.01    & 0.01  & 0.05 \\ 
        GraphMask    &   0.01       & 0.01    & 0.01  & 0.04 \\ 
    \bottomrule
    \end{tabularx}  
\end{table}

\subsection{Attack Query Numbers}
\label{app:number_queries}
The result can be found in Figure \ref{fig:dynamic:n_queries}.

\begin{figure}[ht]
    \centering
    \begin{tikzpicture}
        \begin{axis}[
            width=0.40\textwidth,  
            height=0.15\textwidth,
            enlarge x limits=0.15,
            enlarge y limits=false,
            ylabel near ticks,
            xlabel near ticks,
            scale only axis,
            ymin=0,
            ymax=1,
            yticklabel style={font=\small},
            xtick=data,
            xticklabels={1, 2, 3,4 ,5, 6, 7, 8, 9, 10, 11, 12, 13, 14, 15},
            xticklabel style={yshift=0ex, rotate=0, font=\tiny}, 
            xlabel={Number of Queries},
            x label style={at={(axis description cs:0.5,-0.1)},anchor=north, font=\small},
            ylabel={F1-score},
            major x tick style={opacity=0},
            minor x tick num=1,
            minor tick length=1 ex,
            ymajorgrids= true,
            xmajorgrids= true,
            grid style={white},
            axis background/.style={fill=gray!15},
            axis line style={white, line width=0pt},
            xtick style={draw=none},
            ytick style={draw=none},
            legend style={
              text=black,
              draw=gray, fill=gray!10,
              legend columns=4,
              at={(1,0)}, 
              anchor=south east, 
              cells={anchor=west},
              font=\tiny\itshape,
              /tikz/every even column/.append style={column sep=0.5cm}
            }
        ]
        \addlegendentry{GAT}
        \addplot [softred, mark=*, line width=1pt] coordinates {
            (1,0.47) (2,0.54) (3,0.52) (4,0.52) (5,0.52) (6,0.47) (7,0.53) (8,0.52) (9,0.50) (10,0.50) (11,0.49) (12,0.53) (13,0.53) (14,0.53) (15,0.54)
        };
        \addlegendentry{GCN} 
        \addplot [softblue, mark=*, line width=1pt] coordinates {
            (1,0.71) (2,0.72) (3,0.73) (4,0.71) (5,0.74) (6,0.74) (7,0.79) (8,0.83) (9,0.82) (10,0.81) (11,0.82) (12,0.82) (13, 0.84) (14,0.85) (15,0.85) };  
        \addlegendentry{GIN} 
        \addplot [softorange, mark=*, line width=1pt] coordinates {
            (1,0.39) (2,0.46) (3,0.48) (4,0.54) (5,0.59) (6,0.61) (7,0.62) (8,0.66) (9,0.64) (10,0.68) (11,0.66) (12,0.68) (13,0.71) (14,0.72) (15,0.76)};
        \addlegendentry{SAGE} 
        \addplot [ForestGreen, mark=*, line width=1pt] coordinates {
            (1,0.40) (2,0.40) (3,0.42) (4,0.42) (5,0.42) (6,0.48) (7,0.50) (8,0.53) (9,0.56) (10,0.56) (11,0.60) (12,0.62) (13,0.63) (14,0.65) (15,0.65)        };
        \end{axis}
        \end{tikzpicture}
     \caption{F1-score of $INF\text{--}DIR^*$ on LastFM with different number of queries.}
     \label{fig:dynamic:n_queries}
 \end{figure}

\begin{addedenv}
{
\section{Theoretical Justification}

\subsection{Why do influence-based attacks work?}
\label{app:justification:intuition}
This section formally presents the ideas of our work, showing how both the \textbf{magnitude} of influence and the \textbf{pattern} of influence can encode distance information between two nodes $i$ and $j$ in a multi-layer GNN-style aggregator. 


\subsubsection{Magnitude-Based Distance Inference}

\paragraph{Notation \& Setup}
\begin{itemize}
    \item Let $G=(V,E)$ be an undirected graph with $n = |V|$ nodes.
    \item Let $A \in \mathbb{R}^{n \times n}$ be the adjacency matrix (possibly with self-loops $A_{ii}=1$).
    \item Let $\hat{A} = D^{-\tfrac12} A\, D^{-\tfrac12}$, where $D$ is the diagonal degree matrix of $A$.
    \item In a \textbf{$K$-layer GCN} (Kipf \& Welling, 2017), each layer can be viewed (in simplified form) as
    \[
      H^{(\ell+1)}
      \;=\;
      \sigma\bigl(\hat{A}\,H^{(\ell)}\,W^{(\ell)}\bigr),
      \quad \ell=0,\dots,K-1,
    \]
    with $H^{(0)} = X$ (initial node features), weight matrices $W^{(\ell)}$, and nonlinearities $\sigma$.
    \item After $K$ layers, node $i$'s embedding is $\mathbf{h}_i \in \mathbb{R}^{d_K}$.
\end{itemize}

We define the \emph{(i, j) ``aggregator coefficient''} (or ``influence weight'') to be approximately:
\[
  \hat{A}^K_{ij}
  \;=\;
  \bigl(\underbrace{\hat{A}\,\hat{A}\,\cdots\,\hat{A}}_{K\text{ times}}\bigr)_{ij}.
\]
In practice, the GCN also includes nonlinearity and weights $W^{(\ell)}$, but $\hat{A}^K_{ij}$ suffices to illustrate how \emph{distance} affects \emph{magnitude}.

\paragraph{Proposition 1 (Magnitude Grows with Proximity)}

\noindent \textbf{Claim}: If the \textbf{shortest path distance} $d(i,j)$ between nodes $i$ and $j$ is \textbf{smaller}, then $\hat{A}^K_{ij}$ (and thus the \textbf{magnitude} of $j$'s contribution to $i$) is \textbf{larger} on average than if $d(i,j)$ is bigger.

\medskip
\noindent \textbf{Arguments Sketch}:
\begin{enumerate}
  \item \textbf{Distance and Powers of $\hat{A}$}:
    \begin{itemize}
      \item Recall $d(i,j)$ is the minimum number of edges in a path from $i$ to $j$. If $d(i,j) \le K$, then there exists a path of length $\le K$. In $\hat{A}^K$, each path contributes a product of terms $\hat{A}_{uv}$ along that path.
    \end{itemize}

  \item \textbf{$\hat{A}_{uv}$ Is Positive for Edges}:
    \begin{itemize}
      \item If $(u,v)\in E$, then $\hat{A}_{uv} = \tfrac{1}{\sqrt{\deg(u)\deg(v)}} > 0$. A single path of length $\ell \le K$ from $j$ to $i$ yields a positive term in $\hat{A}^\ell_{ij}$.
    \end{itemize}

  \item \textbf{Shorter Paths $\implies$ Fewer Normalizing Factors}:
    \begin{itemize}
      \item Each hop in the path multiplies by $\tfrac{1}{\sqrt{\deg(\cdot)\deg(\cdot)}}$. A \textbf{shorter} path means fewer such factors, tending to yield \textbf{larger} overall product. Conversely, a longer path accumulates more dividing factors $\le 1$.
    \end{itemize}

  \item \textbf{Comparing Distances}:
    \begin{itemize}
      \item If $d(i,j) = 1$, then $\hat{A}_{ij}$ itself is typically $\frac{1}{\sqrt{\deg(i)\deg(j)}}$.
      \item If $d(i,j) = 2$, then $\hat{A}^2_{ij} = \sum_{u} \hat{A}_{iu}\hat{A}_{uj}$. Each term introduces two normalizing denominators.
      \item The more hops, the smaller (in aggregate) $\hat{A}^K_{ij}$ tends to be because of repeated division (though multiple parallel paths can partially offset this, it still rarely exceeds the single-hop scenario on average).
    \end{itemize}
\end{enumerate}

Hence, \emph{nodes with fewer hops between them (smaller $d(i,j)$) tend to have larger $\hat{A}^K_{ij}$.} In a GCN, that translates directly into a \textbf{larger magnitude} of the feature vector from $j$ that arrives at node $i$.

\noindent \textbf{Take Away (Magnitude)}:
\begin{quote}
By measuring or estimating $\hat{A}^K_{ij}$---or equivalently, how large $\|\Delta \mathbf{h}_i\|$ is when node $j$'s feature changes---we can distinguish smaller from larger distances. If the magnitude of influence from $j$ to $i$ is large, $j$ is likely topologically closer.
\end{quote}

\subsubsection{Pattern-Based Distance Inference}

In addition to magnitude, we can also examine \textbf{which nodes or edges} are involved in conveying information from $j$ to $i$. This can be seen as the \textbf{``pattern'' of multi-hop aggregation}.

\paragraph{Formalizing ``Pattern''}
\begin{itemize}
    \item \textbf{Expanding $\hat{A}^K$}: One can write
    \[
      (\hat{A}^K)_{ij}
      \;=\;
      \sum_{\pi \in \mathcal{P}_{i\to j}(K)}
      \prod_{\ell=1}^{|\pi|\le K} \hat{A}_{\pi_\ell,\pi_{\ell+1}},
    \]
    where $\mathcal{P}_{i\to j}(K)$ is the set of all walks (paths, possibly repeating nodes) of length up to $K$ from $j$ to $i$. Each walk $\pi$ is a sequence of nodes $\pi_1=j,\pi_2,\dots,\pi_m=i$.
    \item The \textbf{pattern} here refers to \emph{which} nodes or edges appear in these walks and with \emph{what weight}. If a path $\pi$ does not exist (i.e., no edge connecting certain nodes), that walk contributes 0. If it does exist, we see a product of adjacency factors.
\end{itemize}

\paragraph{Proposition 2 (Pattern Discloses Hop Distance)}

\noindent \textbf{Claim}: If $d(i,j) \le K$, then the set of aggregator paths from $j$ to $i$ has an identifiable \textbf{``short-path signature''} that does \textbf{not} appear for distant nodes. Observing the aggregator pattern (which neighbors or edges contribute) lets us distinguish smaller distances from larger ones.

\medskip
\noindent \textbf{Argument Sketch}:
\begin{enumerate}
  \item \textbf{Presence/Absence of Certain Walks}:
    \begin{itemize}
      \item If $d(i,j) = 1$, there is a direct edge $(i,j)$. Hence in the final aggregator, we see a ``direct neighbor'' contribution from $j$ to $i$.
      \item If $d(i,j) = 2$, the aggregator from $j$ to $i$ must pass through some intermediate $u$. So node $u$ is \emph{common} in every path from $j$ to $i$.
      \item For \textbf{larger} $d(i,j)$, the aggregator paths require more intermediates.
    \end{itemize}
  \item \textbf{Graph Convolutional Aggregation}:
    \begin{itemize}
      \item In a standard 2-layer GCN, node $i$'s embedding is
        \[
        \mathbf{h}_i^{(2)} =
        \sigma\Biggl(\sum_{u \in \mathcal{N}(i)} \frac{1}{\sqrt{\deg(i)\deg(u)}}
        \sigma\Biggl(\sum_{w\in\mathcal{N}(u)} \frac{\mathbf{x}_w W^{(1)}}{\sqrt{\deg(u)\deg(w)}} \Biggr) W^{(2)}\Biggr).
        \]
        
        The aggregator pattern \textbf{varies} depending on whether node \( j \) is a direct neighbor (\( j \in \mathcal{N}(i) \)) or a 2-hop neighbor (via some \( u \)), affecting summation terms and \( j \)'s role in the nested sums.

    \end{itemize}
  \item \textbf{Distinct Subgraph Structures}:
    \begin{itemize}
      \item If $j$ is truly 3 or more hops away, there is no 1- or 2-hop aggregator route. The pattern of summation (which nodes feed into $i$) differs from if $j$ were 1 or 2 hops away.
      \item Thus, the set of edges used in the aggregator for node $i$ (the ``pattern of influence'') encodes how short or direct the paths from $j$ to $i$ are.
    \end{itemize}
\end{enumerate}

\smallskip
\noindent \textbf{Take Away (Pattern)}:
\begin{quote}
Observing which nodes or edges actively relay $j$'s features into $i$ (i.e., the aggregator path) can \textbf{reconstruct} whether $j$ is 1 hop, 2 hops, or more. Hence, the pattern alone can identify approximate distance---even ignoring the numerical magnitude of those aggregator coefficients.
\end{quote}

\subsubsection{Take away}
\begin{itemize}
    \item \textbf{Magnitude}: 
    Shorter $d(i,j)$ $\implies$ fewer normalizing factors $\tfrac{1}{\sqrt{\deg(\cdot)\deg(\cdot)}}$, higher aggregator coefficients, and a \emph{larger} overall impact on $\mathbf{h}_i$.
    \item \textbf{Pattern}: 
    The presence or absence of certain ``neighbor routes'' from $j$ to $i$ in the aggregator clearly distinguishes small hop distances from large ones. This is a \textbf{structural} property: you can read off the local subgraph that contributed to node $i$'s update.
\end{itemize}

By \textbf{separately} examining \textbf{(A)} the \textbf{magnitude} (how large $\mathbf{h}_i$ changes due to $j$) and \textbf{(B)} the \textbf{pattern} (which edges/neighbors are used in the aggregator from $j$ to $i$), we obtain two complementary indicators of \textbf{distance} between nodes $i$ and $j$.

\begin{itemize}
    \item \textbf{Magnitude} is a \textbf{quantitative} measure, often correlated with the number of hops and how heavily normalized edges are.
    \item \textbf{Pattern} is a more \textbf{qualitative} or \textbf{structural} signature of which specific paths carry information from $j$ to $i$.
\end{itemize}

Hence, \textbf{both} provide theoretically sound signals that can reveal node distances in GNNs.

\subsection{How to design a good Magnitude-Based link inference}

Many link or distance inference attacks use a threshold on the \textbf{magnitude} of how much a perturbation to node \(i\)'s features changes node \(j\)'s final embedding. Concretely, they measure something like:
\[
  \|\Delta_{j}(i)\|
  \;=\;
  \left\lVert
    \mathbf{h}_j^{(L)}(\mathbf{x}_i + \delta) 
    \;-\;
    \mathbf{h}_j^{(L)}(\mathbf{x}_i)
  \right\rVert
  \quad
  \text{or}
  \quad
  \left\lVert
    \frac{\partial \mathbf{h}_j^{(L)}}{\partial \mathbf{x}_i}
  \right\rVert.
\]
\textbf{Idea:} A \emph{large} magnitude indicates that \(i\) heavily influences \(j\), suggesting a short distance (e.g., a direct connection). A \emph{small} (or zero) magnitude suggests that \(j\) is far from \(i\).

\subsubsection{How GNN Weights Affect the Magnitude}

\paragraph{Aggregator Weights and Attention Coefficients}

Most GNNs (GCN, GraphSAGE, GAT, etc.) include weight-like factors in their message-passing. For instance:
\begin{itemize}
  \item \textbf{Normalized Adjacency:} \(\beta_{u,v} = \hat{A}_{u,v}\) (e.g., \(\hat{D}^{-1/2}A\hat{D}^{-1/2}\) in GCN).
  \item \textbf{Mean Aggregation:} \(\beta_{u,v} = \frac{1}{|\mathcal{N}(u)|}\).
  \item \textbf{GAT Attention:} \(\beta_{u,v} = \alpha_{u,v}\), which can vary widely among edges.
\end{itemize}
These \(\beta_{u,v}\) can differ significantly across edges, even if the nodes are equally close in graph distance. For example, one edge might have high attention (\(\alpha=0.9\)), while another similar edge might have low attention (\(\alpha=0.1\)).

\paragraph{Layer Weight Matrices and Activation Non-Linearities}

Each GNN layer has a weight matrix \(W^{(\ell)}\) and a non-linear activation \(\sigma(\cdot)\). The operator norm \(\|W^{(\ell)}\|\) can scale or shrink vectors, and saturating activations can dampen or vanish gradients. Over multiple layers, these factors multiply in complicated ways that may overshadow the simple “distance = number of hops” effect.

Hence, two equally distant node pairs (\(i\)--\(j\) vs. \(i\)--\(j'\)) could yield very different magnitudes if, for example, the aggregator or attention assigns a high weight to \(j\) but not to \(j'\). This can cause an inference method based purely on magnitude to misjudge which pair is truly “closer” in the underlying graph.

\subsubsection{Why We Need to \emph{Mitigate} or \emph{Normalize} Weight Effects}

\begin{enumerate}
  \item \textbf{Magnitude Can Reflect Both Distance \emph{and} Large Weights:}
  \begin{itemize}
    \item If an edge \((i,j)\) is assigned a very large aggregator or attention weight, the partial derivative \(\|\Delta_j(i)\|\) may spike, even if \(j\) is 2 or 3 hops away (via a path that has large weights at each hop).
    \item Conversely, if an edge \((i', j')\) has low aggregator weights---even if \(i'\) and \(j'\) are in the same or fewer hops---its overall gradient magnitude may appear smaller.
  \end{itemize}
  \item \textbf{Misidentification of True Distance:}
  \begin{itemize}
    \item A node pair with artificially high aggregator weights might “look closer” than it actually is.
    \item A pair with the same topological distance but smaller aggregator weights might “look farther.”
  \end{itemize}
\end{enumerate}
Thus, if the goal is purely to \emph{infer the structural closeness or distance} (e.g., distinguishing 1-hop vs. 2-hop neighbors), the raw magnitude combines both the “distance effect” and the “scaling effect” from the GNN's learned or fixed weights.




\subsubsection{Take Away}

In short, \textbf{it is necessary to mitigate or normalize out the effect of GNN weights} (aggregator factors, learned attention, large or small parameter norms) \textbf{when using \emph{magnitude-based} analysis} to infer true node-to-node distance. Otherwise, large aggregator or model weights can amplify the partial derivative and \emph{confound} the simple question of whether these nodes are close in the graph.

A \textbf{raw gradient magnitude} often \textbf{combines} both the “distance effect” and the “scaling effect,” so normalizing or factoring out the aggregator weighting is crucial to obtain a more direct measure of topological proximity.

\subsection{Why experiment with 2-hops neighbors}
\label{app:2hopselection}
Below is a higher-level theoretical explanation of \emph{why} the link inference attack (based on how node feature perturbations affect other nodes’ embeddings) is \emph{especially compelling} for distinguishing \textbf{1-hop} vs. \textbf{2-hop} neighbors, and how the same argument \emph{generalizes} to nodes that are \(k\)-hops away.

\subsubsection{Why 1-Hop (Edges) vs. 2-Hop Is a Crucial Test Case}

\paragraph{Theoretical Chain Rule Through \(L\) Layers}

Consider an \(L\)-layer GNN. The representation of node \(j\) at the final layer, \(\mathbf{h}_j^{(L)}\), can be written as
\[
   \mathbf{h}_j^{(L)} = F_j\!\Bigl(\{\mathbf{h}_v^{(L-1)} : v \in \mathcal{N}(j)\}\Bigr),
\]
with each \(\mathbf{h}_v^{(\ell)}\) itself depending on the representations of its neighbors in the previous layer. Ultimately, one obtains
\[
   \frac{\partial \mathbf{h}_j^{(L)}}{\partial \mathbf{x}_i} 
   = \sum_{\text{paths } i \rightarrow \dots \rightarrow j} \Bigl(\text{product of aggregator and layer weights}\Bigr).
\]

\noindent\textbf{1-Hop (Direct Edge \((i,j)\)):}\\[1mm]
A 1-hop path \(i \rightarrow j\) contributes directly to the sum. This typically yields \textbf{larger} partial derivatives because there is only one aggregator step from \(i\) to \(j\).

\noindent\textbf{2-Hop (Via an Intermediate Node \(w\)):}\\[1mm]
There is no direct 1-hop path, but there exists a 2-hop path \(i \rightarrow w \rightarrow j\). The partial derivative now involves two aggregator multiplications (one from \(i\) to \(w\) and one from \(w\) to \(j\)). Each step can scale or dampen the influence, so the overall derivative from \(i\) to \(j\) is typically smaller than that from a direct edge, though still nonzero.

Thus, from a pure chain-rule perspective, 1-hop neighbors exert the strongest direct influence, whereas 2-hop neighbors exhibit a weaker (but nonzero) influence. Distances greater than 2 usually further diminish the influence since each extra hop introduces another multiplicative factor.

\subsubsection{2-Hop Negatives Are Harder (But More Revealing)}

\begin{itemize}[leftmargin=*, noitemsep]
    \item \textbf{Negative Samples with Distance \(\geq 3\):}\\  
    If negative samples are chosen from node pairs that are 3 hops or more apart, the partial derivative (or influence) is often extremely small --- or even zero if the nodes lie outside the GNN’s receptive field.
    
    \item \textbf{2-Hop Pairs:}\\  
    Since 2-hop pairs are the closest non-adjacent nodes, they serve as a “hard negative” test set, making it more challenging (and more revealing) to discriminate between direct adjacency and near adjacency.
\end{itemize}

In summary, 2-hop vs.\ 1-hop is used in our experiments because:
\begin{itemize}[leftmargin=*, noitemsep]
    \item These cases are commonly confused (since the nodes are “close” in the graph).
    \item They require the method to demonstrate fine-grained discrimination between direct connections and connections through an intermediate node.
\end{itemize}

\subsubsection{Generalization to Larger Distances}

\paragraph{Negative Pairs with Distance \(> 1\)}

If negative samples consist of any pair of nodes with distance greater than 1, the same reasoning applies:
\begin{itemize}[leftmargin=*, noitemsep]
    \item If \(\operatorname{dist}(i,j)=k\) (with \(k \ge 2\) and \(k \le L\)), the partial derivative 
    \[
      \left\|\frac{\partial \mathbf{h}_j^{(L)}}{\partial \mathbf{x}_i}\right\|
    \]
    is the product of \(k\) aggregator/activation factors, each typically \(\leq 1\). This yields a smaller influence than that from direct (1-hop) neighbors.
    \item If \(k > L\), then in many standard message-passing GNNs there is no computational path from \(i\) to \(j\) (i.e., \(j\) does not aggregate any information from \(i\)); hence, the partial derivative vanishes.
\end{itemize}

\paragraph{Attack ``Easier'' for Larger Distances}

When the distance is larger (e.g., 3, 4, or more hops), the difference in influence relative to 1-hop neighbors typically increases:
\begin{itemize}[leftmargin=*, noitemsep]
    \item For distances \(\geq L+1\), the partial derivative becomes exactly zero in an \(L\)-layer GNN, since node \(j\) does not aggregate any information from nodes beyond \(L\) hops.
    \item Hence, from a theoretical viewpoint, it is easier for the attack to classify such pairs as not being neighbors.
\end{itemize}

While negative samples could be defined as any pair with distance \(\geq 2\), choosing 2-hop pairs is particularly interesting because they represent the borderline case where some influence might still exist (given typical GNN depths). Distinguishing 1-hop from 2-hop adjacency is therefore the core challenge in local edge detection.

\subsubsection{Take Away}

\begin{itemize}[leftmargin=*, noitemsep]
    \item \textbf{1-hop vs.\ 2-hop} is the most challenging local classification problem; pairs with distance \(\geq 3\) are generally easier to detect as negatives because their mutual influence is much weaker or zero.
    \item The same chain-rule argument --- tracking how perturbations to \(\mathbf{x}_i\) propagate through the aggregator layers --- applies to any pair of nodes with distance greater than 1. More hops typically result in a smaller or zero derivative, simplifying classification.
    \item Thus, while the approach generalizes naturally, focusing on 2-hop negative samples serves as a standard benchmark to ensure the attack can truly distinguish between direct adjacency and near adjacency, which is central to link inference attacks.
\end{itemize}

\subsection{Why node centric threshold selection}
\label{app:justification:threshold}
Below is a conceptual and (sketch of a) theoretical explanation for \textbf{why the distribution of “influence scores”} \textbf{can vary \emph{significantly} across different “source” nodes} \(i\), and thus \textbf{why a single \emph{global} threshold for deciding whether \(i\) is connected to \(j\) may be suboptimal}. In other words, each node \(i\) can exhibit its own scale or pattern of scores toward other nodes \(j\), implying we might need a \emph{node-specific threshold} for link inference.

\subsubsection{Setting}

We consider a GNN with \(L\) layers on a graph \(G\). Each node \(u\) has initial features \(\mathbf{x}_u\). The final layer’s embedding for node \(u\) is \(\mathbf{h}_u^{(L)}\). An \emph{influence score} or \emph{sensitivity} of node \(j\)’s embedding with respect to node \(i\)’s features is measured by:
\[
   S(i \to j)
   \;=\;
   \left\lVert 
     \frac{\partial \mathbf{h}_j^{(L)}}{\partial \mathbf{x}_i}
   \right\rVert.
\]
A \textbf{link inference attack} uses \(S(i \to j)\) to decide whether \(i\) and \(j\) share an edge, or are “very close,” etc. Concretely, an attacker might say “If \(S(i \to j)\) is above some threshold \(\tau\), guess that \((i,j)\) is an edge.”

\subsubsection{Why a Single Global Threshold Might Be Inappropriate}

\noindent\textbf{Different Nodes Have Different “Local GNN Dynamics”}

\ding{114} \textbf{Neighborhood Size / Degree Effects.}\\
If node \(i\) has a large degree, the aggregator (e.g., in a standard GCN or GraphSAGE) may \emph{average} or \emph{normalize} across many neighbors. This tends to \emph{dampen} the partial derivatives from \(i\) to each neighbor. Conversely, a low-degree node might yield higher per-edge sensitivities. Hence, even for \emph{true edges}, the magnitude of \(S(i \to j)\) can differ drastically depending on node \(i\) and how the aggregator normalizes or distributes attention across neighbors.

\ding{114} \textbf{Attention Coefficients.}\\
In a GAT-style model, each node \(i\) learns a distinct set of attention weights \(\{\alpha_{i,k}\}\) over its neighbors \(k\). If node \(i\) is ``dominant'' in certain substructures, the GAT might assign large attention to some edges. Another node \(i'\) might have a completely different attention pattern. Thus, the overall scale of partial derivatives from \(i\) to others can be larger or smaller than from \(i'\) to others.

\ding{114} \textbf{Node-Specific Learned Weights / Features.}\\
Although GNN layers typically share parameters globally (e.g., a weight matrix \(W^{(\ell)}\) or an attention transform), the \emph{effective} transformation depends on each node’s features. For example, if the activation function saturates for certain node feature values, the gradient can vanish or saturate. Different nodes can thus systematically exhibit different gradient magnitudes to their neighbors.

\emph{Consequence:} Because each node \(i\) (1) may have a different local degree, (2) can have different adjacency or attention patterns, and (3) might lie in regions of the graph where the learned function saturates or amplifies differently, the distribution 
\[
\{S(i \to j)\}_{j \neq i}
\]
can vary significantly from that of another node \(i'\).

\subsubsection{Theoretical Sketch: Chain Rule Bounds and Node-Specific Scaling}

Below is a \emph{bounding} argument that suggests \emph{per-node scaling factors} can appear in the partial derivatives.

\paragraph{A Simple Layer-Wise Expression}

Consider an \(L\)-layer GNN. The representation of node \(j\) at the final layer, \(\mathbf{h}_j^{(L)}\), can be written as
\[
   \mathbf{h}_j^{(L)} = F_j\!\Bigl(\{\mathbf{h}_v^{(L-1)} : v \in \mathcal{N}(j)\}\Bigr),
\]
with each \(\mathbf{h}_v^{(\ell)}\) itself depending on the representations of its neighbors in the previous layer. Ultimately, one obtains
\[
   \frac{\partial \mathbf{h}_j^{(L)}}{\partial \mathbf{x}_i} 
   = \sum_{\text{paths } i \rightarrow \dots \rightarrow j} \Bigl(\text{product of aggregator and layer weights}\Bigr).
\]

\paragraph{Per-Node Aggregation \& Normalization Terms}

Assume, for simplicity, that each layer’s aggregator uses a factor \(\beta_{u,v}^{(\ell)}\) to scale neighbor \(v\)’s embedding in node \(u\)’s update. For example:
\[
\begin{array}{rcl}
\text{GCN:} & \beta_{u,v}^{(\ell)} &= \tilde{A}_{u,v}, \\[1mm]
\text{GraphSAGE (mean):} & \beta_{u,v}^{(\ell)} &= \frac{1}{|\mathcal{N}(u)|}, \\[1mm]
\text{GAT:} & \beta_{u,v}^{(\ell)} &= \alpha_{u,v}^{(\ell)}.
\end{array}
\]
Then the partial derivative can be bounded by:
\[
   \left\lVert 
     \frac{\partial \mathbf{h}_j^{(L)}}{\partial \mathbf{x}_i}
   \right\rVert
   \;\le\;
   \sum_{p \in \Pi_{i \to j}} 
      \prod_{\ell \in p} 
         \left\lVert W^{(\ell)}\right\rVert
         \,\cdot\,
         \max_{(u,v)\in p} \{\beta_{u,v}^{(\ell)}\}
         \,\cdot\,
         \max \{\sigma'\},
\]
where \(\Pi_{i \to j}\) denotes the set of all layer-by-layer paths from \(i\) to \(j\). Notice that \(\beta_{u,v}^{(\ell)}\) depends on the node/edge pair \((u,v)\). Therefore, for node \(i\) the magnitude of \(\beta_{i,\cdot}^{(\ell)}\) and the aggregator structure may differ from that of another node \(i'\).

\paragraph{Node-Specific “Scaling” Factor}

Define, for example,
\[
  \Gamma(i) 
  \;=\; 
  \max_{\substack{v \in \mathcal{N}(i) \\ \ell=1,\ldots,L}} \;\beta_{i,v}^{(\ell)},
\]
which represents the maximum aggregator weight from \(i\) to any of its neighbors (across all layers). Since degrees, attention distributions, and normalization factors differ, \(\Gamma(i)\) can vary significantly from node to node. Consequently, node \(i\) might produce systematically larger or smaller partial derivatives \(\left\|\frac{\partial \mathbf{h}_j^{(L)}}{\partial \mathbf{x}_i}\right\|\). In other words:
\begin{enumerate}
    \item The distribution \(\{S(i \to j)\}_{j}\) may be shifted or scaled by \(\Gamma(i)\).
    \item Another node \(i'\) may have a very different \(\Gamma(i')\), shifting its score distribution accordingly.
\end{enumerate}
Thus, a global threshold \(\tau\) applied uniformly to all \(\left\|\frac{\partial \mathbf{h}_j^{(L)}}{\partial \mathbf{x}_i}\right\|\) might misclassify edges for nodes with particularly high or low \(\Gamma\) values.

\subsubsection{Empirical \& Practical Consequences}

\begin{itemize}
    \item \textbf{Different Neighborhood Sizes \(\to\) Different Score Ranges:}\\
    A node with 50 neighbors may have its gradient “diluted” due to normalization, so even true edges produce a smaller partial derivative magnitude. Conversely, a node with only 2 neighbors may yield a larger derivative for each edge. Thus, using a single global threshold \(\tau\) may cause many false negatives for high-degree nodes or false positives for low-degree nodes.
    
    \item \textbf{Different Learned Attention Patterns:}\\
    If a node’s neighborhood is heterogeneous (with some edges receiving high \(\alpha_{u,v}\) and others low), the range of partial derivative magnitudes may be wide. Another node may exhibit a more uniform distribution. Each node, therefore, can have a different mean and variance in its \(\{S(i \to j)\}\) distribution.
    
    \item \textbf{Nonlinear Activations:}\\
    A node’s hidden representation might fall in the saturated region of a nonlinearity, yielding a small gradient flow, while another node may operate in a near-linear region, yielding a larger gradient. This again results in node-specific differences in influence scores.
\end{itemize}

For link inference attacks (or any gradient-based adjacency detection), a single universal threshold is often too crude. A \emph{per-node threshold} — for example, one based on normalizing or standardizing the distribution \(\{S(i \to j)\}\) for each node \(i\) — may better separate true edges from non-edges.

\subsubsection{Take Away}
Analyzing the chain rule in message-passing GNNs shows that aggregator weights \(\beta_{u,v}^{(\ell)}\) and activation functions can induce \emph{node-specific scaling effects} in the partial derivatives from node \(i\) to node \(j\). 
Each node \(i\) ends up with its own distribution of influence scores \(\{S(i \to j)\}_j\). This distribution may have a different mean and variance from that of another node \(i'\), so a single global threshold \(\tau\) may be too high for some nodes and too low for others.
To improve the accuracy of link inference attacks (or any gradient-based edge detection), it may be beneficial to adopt a \emph{node-specific threshold} --- for instance, by normalizing the scores \(\{S(i \to j)\}\) per node --- in order to better distinguish true edges from non-edges.

}
\end{addedenv}

\end{document}